\documentclass[a4paper]{article}
\usepackage{graphicx}

\usepackage{amsmath,amssymb,amsfonts}
\usepackage{physics}
\usepackage{tensor}
\usepackage{comment}

\usepackage{mathtools}

\usepackage{caption}
\usepackage{subcaption}

\usepackage{tikz}
\usetikzlibrary{plotmarks}

\usepackage{makecell}
\usepackage{multirow}

\usepackage{booktabs}
\usepackage{bigstrut}

\usepackage{hyperref}
\usepackage{cleveref}

\usepackage{soul}

\usepackage{makecell}

\usepackage[title]{appendix}

\newcommand{\paren}[1]{\left(#1\right)}
\newcommand{\bracket}[1]{\left[#1\right]}
\newcommand{\bracee}[1]{\left\lbrace#1\right\rbrace}

\numberwithin{equation}{section}

\newcolumntype{M}[1]{>{\centering\arraybackslash}m{#1}}
\newcolumntype{N}{@{}m{0pt}@{}}

\begin{document}

\begin{titlepage}

\vskip 2.2cm

\begin{center}

{\large \bf Simulating Quantum Mechanics with a $\theta$-term \\ and an 't Hooft Anomaly on a Synthetic Dimension}

\vskip 1.4cm

{{Jiayu Shen}$^{a,b,e,}$\footnote{jiayus3@illinois.edu},  {Di Luo}$^{a,c,d,e,f,g}$, {Chenxi Huang}$^{a,e}$,\\{Bryan K. Clark}$^{a,c,d,e}$, {Aida X. El-Khadra}$^{a,b,e}$,\\ {Bryce Gadway}$^{a,e}$, and  {Patrick Draper}$^{a,b,e}$}
\vskip 1cm
{{\it $^{a}$Illinois Quantum Information Science and Technology Center}}\\
{{\it $^{b}$Illinois Center for Advanced Studies of the Universe}}\\
{{\it $^{c}$Institute for Condensed Matter Theory}}\\
{{\it $^{d}$NCSA Center for Artificial Intelligence Innovation}}\\
{{\it $^{e}$Department of Physics, University of Illinois at Urbana-Champaign, Urbana, IL 61801}}\\
{{\it $^{f}$Center for Theoretical Physics, Massachusetts Institute of Technology, Cambridge, MA 02139}}\\
{{\it $^{g}$The NSF AI Institute for Artificial Intelligence and Fundamental Interactions, Cambridge, MA 02139}}\\

\vspace{0.3cm}

\vskip 1.5cm

\begin{abstract}
A topological $\theta$-term in gauge theories, including quantum chromodynamics in 3+1 dimensions, gives rise to a sign problem that makes classical Monte Carlo simulations impractical. Quantum simulations are not subject to such sign problems and are a promising approach to studying these theories in the future. In the near term, it is interesting to study simpler models that retain some of the physical phenomena of interest and their implementation on quantum hardware. For example, dimensionally-reducing gauge theories on small spatial tori produces quantum mechanical models which, despite being relatively simple to solve, retain interesting vacuum and symmetry structures from the parent gauge theories. Here we consider quantum mechanical particle-on-a-circle models, related by dimensional reduction to the 1+1d Schwinger model, that possess a $\theta$-term and realize an 't Hooft anomaly or global inconsistency  at $\theta = \pi$. These models also exhibit the related phenomena of spontaneous symmetry breaking and instanton-anti-instanton interference in real time. We propose an experimental scheme for the real-time simulation of a particle on a circle with a $\theta$-term and a $\mathbb{Z}_n$ potential using a synthetic dimension encoded in a Rydberg atom. Simulating the Rydberg atom with realistic experimental parameters, we demonstrate that the essential physics can be well-captured by the experiment, with expected behavior in the tunneling rate as a function of $\theta$. Similar phenomena and observables can also arise in more complex quantum mechanical models connected to  higher-dimensional nonabelian gauge theories by dimensional reduction.
\end{abstract}

\end{center}

\vskip 1.0 cm

\end{titlepage}

\setcounter{footnote}{0} 
\setcounter{page}{1}
\setcounter{section}{0} \setcounter{subsection}{0}
\setcounter{subsubsection}{0}
\setcounter{figure}{0}

\section{Introduction}

Euclidean Lattice Field Theory is the most powerful approach currently available for studying the nonperturbative dynamics of a large class of gauge theories, including quantum chromodynamics (QCD). However, lattice field theory computations generically give rise to sign problems which keep important phenomena out of reach, for example, real-time dynamics, finite-density systems, or theories with topological terms. 
In principle, quantum computers and quantum simulators can be used to study such theories and phenomena, but these techniques are at the early stages of development.
In the near term, it is useful to consider simpler theories in lower dimensions retaining some aspects of the real higher dimensional physics of interest. The phenomena exhibited by these theories offer milestone physics targets, and benchmarks for comparison with other types of computations, along the road towards full, fault-tolerant quantum  simulations of realistic gauge theories relevant for high energy physics. For reviews of lattice gauge theory (LGT) simulations on quantum devices, see~\cite{Banuls:2019bmf,klco2021standard,doi:10.1080/00107514.2016.1151199,Zohar_2015}, and for recent work on fault-tolerant algorithms for LGT simulations see~\cite{Shaw:2020udc,tong2021provably,kan2021lattice}.

The quantum mechanical (QM) particle on the circle is one of the simplest theories with instantons and a $\theta$-term. In that sense it represents one such benchmark theory for quantum simulation. It is also of interest because it is the low-energy effective theory describing the dimensional reduction of 1+1d U(1) gauge theory on a spatial circle. Dimensional reduction often leads to simpler theories that retain some of the interesting physics  of the parent gauge theory, including symmetry structures and topological aspects. The particle on the circle example can be generalized to include an $n$-fold periodic potential, in which case it is the dimensional reduction of the weakly-coupled massive charge-$n$ Schwinger model~\cite{PhysRev.128.2425}. The QM circle-valued degree of freedom corresponds to the Wilson loop around the spatial circle in the gauge theory and the $\mathbb{Z}_n$ potential is dynamically generated by integrating out the charged matter. Recently, digital quantum computations of the phase structure of the Schwinger model with a $\theta$-term have been studied in~\cite{chakraborty2020digital,honda2021digital}.

The most straightforward questions to ask in a quantum simulation involve real time evolution, so we are naturally led to consider phenomena with two sign problems: real-time dynamics associated with the $\theta$-term.  An interesting class of observables is the tunneling rate between classical vacua with a $\mathbb{Z}_n$ symmetric potential on the circle. In continuum Euclidean language, the presence of the $\theta$-term generates interference between instanton amplitudes in different directions around the circle. For example, with $n=2$ and $\theta=\pi$, the tunneling rate between the two wells vanishes due to perfect destructive interference between the two instantons. This is in fact related to a fundamental structural property of the theory, the presence of an 't Hooft anomaly or a global inconsistency between two discrete symmetries, which is matched in the infrared by an exactly twofold degenerate ground state~\cite{Gaiotto:2014kfa,Gaiotto_2017}. In the dimensional reduction picture, the anomaly structure in the QM theory is dictated by the same structure in the Schwinger model. A similar anomaly or global inconsistency also arises in 4d Yang-Mills theory, occurring  between time reversal symmetry and a 1-form $\mathbb{Z}_n$ center symmetry, and  which is responsible for the spontaneous breaking of charge conjugation times parity (CP) at $\theta=\pi$~\cite{Gaiotto_2017}. 
As such, studies of the QM particle on the circle can also be used as a toy model for exploring methods and observables probing the anomaly structure of higher-dimensional gauge theories in quantum simulations.

An 't Hooft anomaly~\cite{Hooft1980} is a quantum mechanical obstruction to gauging certain global symmetries. Intuitively, it either means that a single symmetry cannot be gauged, or that two symmetries cannot be gauged together, because gauging one explicitly breaks the other. If neither are gauged, then neither is broken, but the existence of the anomaly still implies a nontrivial infrared limit of the theory due to 't Hooft's anomaly matching conditions. For example, one possible realization of an 't Hooft anomaly is a degeneracy of ground states. If the ground states are related by an exchange symmetry $S_1$ which has an anomaly with another symmetry $S_2$, gauging $S_2$ may eliminate some of the ground states, thereby explicitly breaking $S_1$ and ``saturating" the anomaly. Global inconsistencies are similar symmetry properties constraining the behavior of a given theory at two different values of its couplings, for example, at $\theta = 0$ and $\theta = \pi$~\cite{Gaiotto_2017}. Global inconsistencies also imply obstructions to gauging global symmetries, but they are slightly milder than an 't Hooft anomaly: the gauging is possible at one value of the couplings, but not both. 

The QM $\mathbb{Z}_n$ particle on a circle model can realize either an 't Hooft anomaly or a global inconsistency at $\theta=\pi$, depending on the parity of $n$~\cite{10.1093/ptep/ptx148}. In both cases, these symmetry structures result in ground state degeneracy and spontaneous symmetry breaking  at $\theta = \pi$. The spectral properties inferred from the anomaly/inconsistency govern the slow dynamics of the system, providing a complementary explanation to the picture of instanton/anti-instanton interference.

In this work we  study  encodings of  $\mathbb{Z}_n$ symmetric particle-on-a-circle models on quantum simulators, utilizing a synthetic dimension mapped to the states of a Rydberg atom. We compute the tunneling rate between different wells as a function of $n$, $\theta$, and the potential frequency, first in the continuum instanton gas approximation, then by direct solution of the discrete time-dependent Schr\"odinger equation (including the effects of spatial discretization), and finally with models of synthetic Rydberg lattice implementations that include realistic experimental non-idealities. In this way we infer how properties of the experimental platform and encoding affect the continuum prediction. 

This paper is organized as follows. In Sec.~\ref{sec:continuumtheory} we discuss  slow dynamics in the continuum particle on a circle model with a $\mathbb{Z}_n$ symmetric cosine potential. Using the instanton gas approximation to determine the low-lying states, we compute the time-dependent rate to hop between vacua as a function of the topological angle $\theta$. In Sec.~\ref{sec:discrete} we latticize the circle, mapping the theory to a tight-binding model with a finite Hilbert space. In this theory the $\theta$-angle is realized by a complex hopping parameter that cannot be gauged away. We study this theory with exact diagonalization (ED) and compare to the continuum model to assess the effects of discretization. In Sec.~\ref{sec:experiment} we map the tight-binding model to a synthetic dimension of a Rydberg atom. In this numerically investigated experimental model, different Rydberg levels of the atom can be used to encode the discretized spatial circle, and the amplitudes and phases of incident microwaves control the amplitude and phase of the hopping parameter. The $\mathbb{Z}_n$ potential is encoded by tuning the microwaves off-resonance. We run real-time simulations of the Rydberg encoding, solving the time-dependent Schr\"odinger equation including realistic experimental effects, for $n=2,3$ on $4$, $6$, $8$, and $12$ lattice sites. We compare to the idealized tight-binding model and find that the Rydberg encoding should be able to realize the dynamical phenomena associated with the $\theta$-term and anomaly/inconsistency structure.

\section{Continuum theory}\label{sec:continuumtheory}

The continuum Euclidean action for a particle on a circle with an $n$-well potential is
\begin{equation}
	S_{\theta} \bracket{\tilde{x}} = \int d \tilde{\tau} \, \bracket{\frac{1}{2} m \paren{\frac{d \tilde{x}}{d \tilde{\tau}}}^2 + \tilde{\lambda} \paren{1 - \cos \paren{\frac{n \tilde{x}}{R}}} - \frac{i \theta}{2 \pi R} \frac{d \tilde{x}}{d \tilde{\tau}}}
	,
\label{eq:theory_dim}
\end{equation}
where $\tilde{x}$ is the position of the particle on a circle of radius $R$, $\tilde{x} \sim \tilde{x} + 2 \pi R$. $\tilde{\tau} \equiv i \tilde{t}$ is the Euclidean time,  and $\theta$ is the topological angle. This action can be nondimensionalized with $I \equiv m R^2$, the moment of inertia, serving as a unit of length or time. We define dimensionless quantities $\tau \equiv \tilde{\tau} / I$, $x \equiv \tilde{x} / R$, $\lambda \equiv \tilde{\lambda} I$, in terms of which the action is
\begin{equation}
	S_{\theta} \bracket{x} = \int d \tau \, \bracket{\frac{1}{2} \paren{\frac{d x}{d \tau}}^2 + V \paren{n x} - \frac{i \theta}{2 \pi} \frac{d x}{d \tau}}
	,
\label{eq:theory_1}
\end{equation}
where $V (n x) = \lambda ( 1 - \cos (n x))$. The $l$-th well is centered at $x = 2\pi l / n$. On a compactified Euclidean time circle, $x (\tau+\beta) - x (\tau) =  2\pi \mathbb{Z}$. Then the $\theta$-term has the property $\exp [- S_{\theta} [x]] =\exp [- S_{\theta + 2 \pi} [x]]$,
so $\theta \sim \theta + 2\pi$.

At $\theta=0$ or $\pi$ and $\lambda = 0$, there are two global symmetries of this theory: the time-reversal symmetry $\tau \mapsto -\tau$ and the $U(1)$-rotation global symmetry $x \mapsto x + \alpha$ where $\alpha$ is independent of $\tau$. Both can be lifted to symmetries of 2D electrodynamics; in the latter case, it is identified with the electric 1-form center symmetry. At $\theta = \pi$, there is a mixed 't Hooft anomaly between these symmetries~\cite{Gaiotto_2017,10.1093/ptep/ptx148}.
This anomaly implies a nontrivial vacuum structure, which in this case manifests as a degenerate ground-state doublet. The same result is obtained by taking the spacetime symmetry to be the parity operation $x \mapsto -x$~\cite{Gaiotto_2017}.

With $\lambda > 0$, the rotation symmetry breaks to a discrete subgroup, $U(1) \rightarrow \mathbb{Z}_n$. For even $n$, at $\theta=\pi$, there is also a mixed 't Hooft anomaly between the parity symmetry and the $\mathbb{Z}_n$ rotation symmetry, again saturated by a degenerate ground-state doublet. For odd $n$, there is no mixed anomaly. Instead, there is a similar but slightly weaker phenomenon, a global inconsistency \cite{10.1093/ptep/ptx148}. That is, it is possible to add a counterterm to restore CP symmetry if the U(1) symmetry is gauged, but only at $\theta=0$ or $\theta=\pi$, not both. If we demand that there is no anomaly at $\theta=0$, then the result is the same, a degenerate ground-state doublet at $\theta = \pi$.

\begin{figure}[!ht]
\centering
		\centering
		\includegraphics[width=0.5\textwidth]{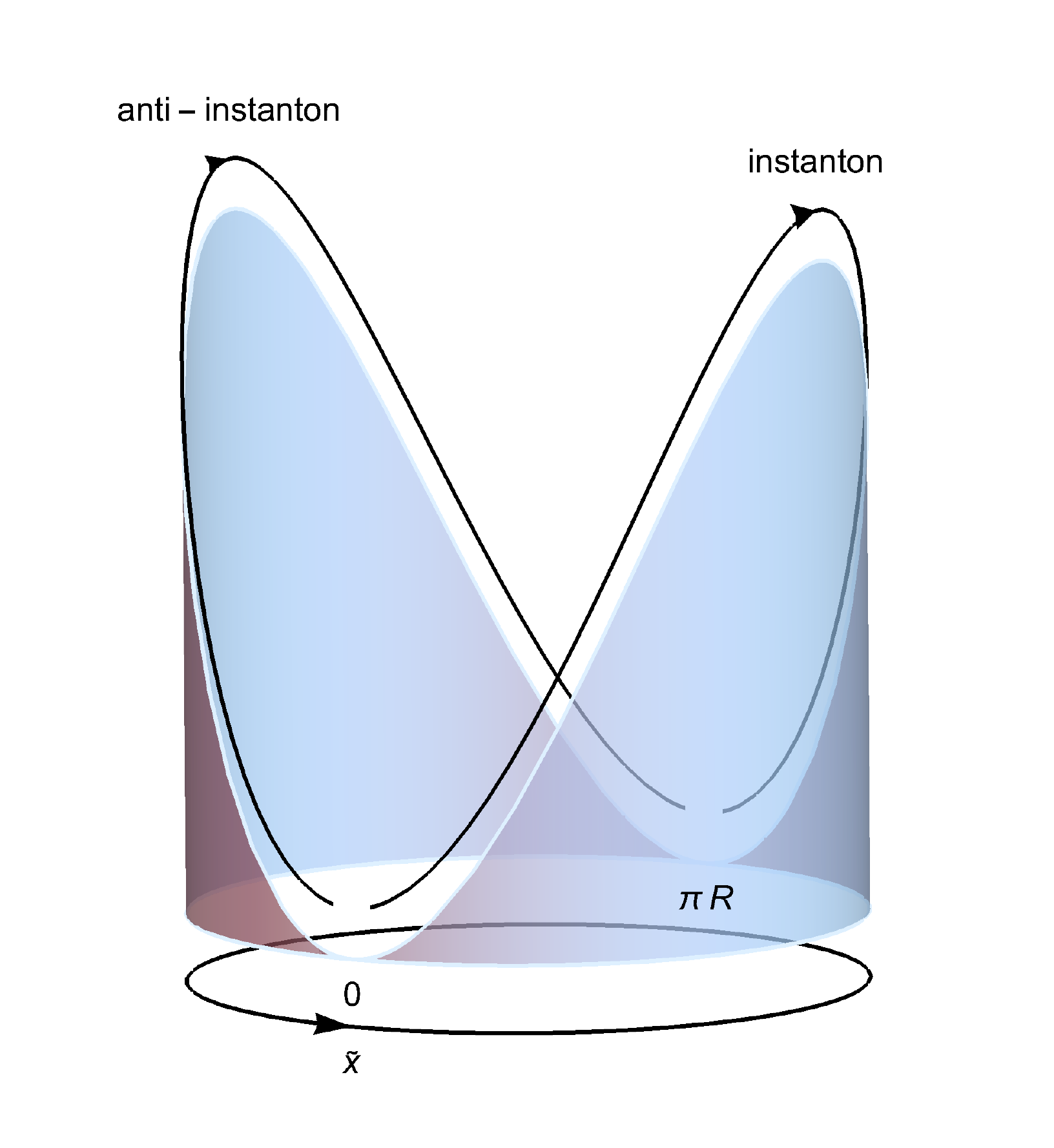}
		\caption{An illustration of the instanton and the anti-instanton in the example of the $n=2$ potential on the circle.}
		\label{fig:cartoon_potential}
\end{figure}

For real-time simulation of this theory, we can use the  Hamiltonian
\begin{equation}
	H = \frac{1}{2} \paren{p - \frac{\theta}{2 \pi}}^2 + V \paren{n x}
	,
\label{eq:continuum_ham_1}
\end{equation}
where $p \equiv \partial L / \partial \dot x = d x / d \tau +\theta/2 \pi$ is the conjugate momentum. If the initial state is  localized near the $l$-th well, then quantum tunneling can occur, and the tunneling dynamics probes the vacuum structure of the theory. Semiclassically and in the continuum, tunneling is described by instantons. The instanton/anti-instanton is the classical solution that takes the particle from $x = 0$ to $x = \pi$ counterclockwise/clockwise, as is shown in Fig~\ref{fig:cartoon_potential}. The contribution to the tunneling amplitude from $x = 0$ to $x = \pi$ from an instanton is proportional to $e^{- S_{I}  + i \theta / 2}$, and that from an anti-instanton is proportional to $e^{- S_{I}  - i \theta / 2}$ where $S_I$ is the real part of the Euclidean instanton action. 
The sum of these two contributions (and quantum fluctuations around them) is proportional to $2 e^{- S_{I}} \cos (\theta / 2)$. At $\theta = \pi$, the sum of the amplitudes vanishes. Contributions from multiple instantons/anti-instantons are discussed in Appendix \ref{sec:diga}.

We perform a semiclassical analysis in the dilute instanton gas approximation (DIGA) and including the one loop fluctuation determinant in Appendix \ref{sec:diga} (see, \textit{e.g.}, \cite{Vainshtein:1981wh}). The DIGA analysis shows that the lowest $n$ energies can be described by an $n$-dimensional tight-binding effective Hamiltonian with eigenvalues
\begin{equation}
	E_k \paren{\theta} = \frac{\omega}{2} - 2 \omega d \cos \paren{\frac{2 \pi k + \theta}{n}}
	,
\label{eq:diga_spectrum}
\end{equation}
where $k = 0, 1, 2, \cdots, n-1\, (\mathrm{mod} \, n)$, $\omega = n \sqrt{\lambda}$,
and $d$ is the instanton density, $d = (4/n) e^{- 8 \omega / n^2} \sqrt{\omega/\pi}$. The semiclassical limit corresponds to a large instanton action, of which the real part is $S_I = 8 \omega / n^2 \gg 1$.

A curious property of Eq.~(\ref{eq:diga_spectrum}) is that the $2\pi$ periodicity of $\theta$ is realized in a non-minimal way, with an associated shift of energy branch: $\theta\rightarrow\theta+2\pi$, $k\rightarrow k-1$. This ``monodromy" phenomenon is also believed to arise in pure $SU(N)$ Yang-Mills theory, as first conjectured by Witten using large $N$ arguments~\cite{PhysRevLett.81.2862}, and occurs in a number of other related theories, including softly broken supersymmetric QCD~\cite{Dine:2016sgq} and Yang-Mills on $R^3\times S^1$~\cite{Aitken:2018mbb}. Qualitatively, the quantum mechanical particle on the circle can be thought of as analogous to a sphaleron direction in field space in YM~\cite{Gaiotto_2017}. 

We can also read off the topological susceptibility  at $\theta=0$ from the $\theta$-dependence of the ground state energy,
\begin{equation}
	\chi_t = \frac{2 \omega d}{n^2} = \frac{8 \omega}{n^3} \sqrt{\frac{\omega}{\pi}} e^{- 8 \omega / n^2} = \frac{8}{\sqrt{\pi}} \paren{\frac{\omega}{n^2}}^{3/2} e^{- 8 \omega / n^2}
	.
\end{equation}
The low-lying energy levels can be probed by real-time dynamics. Taking $n=2$ as an example, the lowest two eigenstates are
\begin{equation}
	\ket{E_0 \paren{\theta}} = \frac{1}{\sqrt{2}} \paren{\ket{0} + \ket{1}} \quad \text{with } E_0 \paren{\theta} = \frac{\omega}{2} - 2 \omega d \cos \paren{\frac{\theta}{2}}
	,
\end{equation}
\begin{equation}
	\ket{E_{1} \paren{\theta}} = \frac{1}{\sqrt{2}} \paren{\ket{0} - \ket{1}} \quad \text{with } E_1 \paren{\theta} = \frac{\omega}{2} + 2 \omega d \cos \paren{\frac{\theta}{2}}
	,
\end{equation}
where $\ket{0}$ and $\ket{1}$ are localized in the centers of the two potential wells at $x = 0$ and $x = \pi$ respectively.
If the initial state is  $\ket{0}$, then the time-evolution will exhibit an angular frequency of $\abs{E_1(\theta) - E_0 (\theta)} = 4 \omega d \cos(\theta / 2)$. At $\theta = \pi$, the two energies become degenerate and there is  no tunneling from $\ket{0}$ to $\ket{1}$. This is a result of the mixed 't Hooft anomaly,\footnote{The state $\ket{E_1(\pi)}$ is odd under the $\mathbb{Z}_2$ rotation symmetry ($\ket{0} \mapsto \ket{1}$, $\ket{1} \mapsto \ket{0}$), $\ket{E_1(\pi)} \mapsto -\ket{E_1(\pi)}$, analogous to one-form center symmetries in higher dimensional gauge theories, so it disappears from the gauge-invariant spectrum if this symmetry is gauged. The state $\ket{E_0 (\pi)}$ is invariant under the symmetry and survives if the symmetry is gauged. Under the global $\mathbb{Z}_2$ time-reversal symmetry, $\ket{E_1(\pi)}$ is  exchanged with $\ket{E_0(\pi)}$. Therefore, if the $\mathbb{Z}_2$ ``center'' symmetry is gauged, the time-reversal symmetry is explicitly broken. There is thus a mixed 't Hooft anomaly between the two symmetries, which implies a nontrivial structure in the low-energy part of the original theory (before gauging). In this paper we are only concerned with the ungauged theory, so we do not explicitly introduce a gauge field.} and at the semiclassical level, it arises via quantum interference between the instanton and the anti-instanton tunneling in opposite directions on the circle.
More generally, the time dependent probabilities to hop from well $0$ to itself or from well $0$ to $1$  are
\begin{equation}
\begin{aligned}
    P_{\theta} \paren{0, 0; t} = \cos^2 \paren{2 \omega d \cos \paren{\frac{\theta}{2}} t }
    ,
\end{aligned}
\label{eq:n_2_p_00}
\end{equation}
\begin{equation}
\begin{aligned}
    P_{\theta} \paren{0, 1; t} = \sin^2 \paren{2 \omega d \cos \paren{\frac{\theta}{2}} t }
    .
\end{aligned}
\label{eq:n_2_p_01}
\end{equation}

A more complicated example is $n = 3$. Eq.~(\ref{eq:diga_spectrum}) shows that the spectra at $\theta = 0$ and $\theta = \pi$ are related to each other by an inversion. Assuming that only the lowest three eigenstates dominate the dynamics of tunneling, then densities starting from a position eigenstate will be the same for $\theta = 0$ and $\theta = \pi$.
This phenomenon in dynamics differs from examples with even $n$.

\section{Discretization}
\label{sec:discrete}
\subsection{The tight-binding Hamiltonian}

The particle on the circle with a topological term can also be thought of as a charged particle on a circle moving in a homogenous transverse magnetic field. The natural discretization is the tight-binding model with complex hopping parameters. 
The discrete Hamiltonian is
\begin{equation}
\begin{aligned}
	 H 
	&=  \sum_{i} \bracket{ w_{i, i+1} b^{\dagger}_{i+1} b_{i} + w_{i, i+1}^{*} b^{\dagger}_{i} b_{i+1} + V_{i} b^{\dagger}_{i} b_{i}} \\
	&=  \sum_{i} \bracket{ - \frac{1}{2 m \xi^2} \paren{ e^{i \frac{\theta}{2 \pi R} \xi} b^{\dagger}_{i+1} b_{i} + e^{- i \frac{\theta}{2 \pi R} \xi}  b^{\dagger}_{i} b_{i+1}} + \paren{V \paren{n \xi i} + \frac{1}{m \xi^2}} b^{\dagger}_{i} b_{i}}
\end{aligned}
\label{eq:tight-binding_ham}
\end{equation}
with periodic boundary conditions. Here $b^{\dagger}_i$ and $b_i$ are bosonic creation and annihilation operators at site $i$, and we have restored the dimensionful parameters.  $\xi$ is the lattice spacing on the circle of radius $R = \xi n_{\mathrm{sites}}$ and $n_{\mathrm{sites}}$ is the number of sites.
(See Appendix \ref{sec:tbh_derivation} for how to discretize the continuum Hamiltonian to the tight-binding model.)
This discrete Hamiltonian can be diagonalized exactly, and it shown to have the same spectrum
in the continuum limit $\xi \rightarrow 0$ as
\begin{equation}
	H = \frac{1}{2m} \paren{p - \frac{\theta}{2 \pi R}}^2 + V \paren{n x}
	.
\end{equation}
Note that $\sum_i b_i^\dagger b_i / (m \xi^2) = 1 / (m \xi^2)$ is a constant term.

 Without the circle topology, one can make a diagonal unitary transformation $\ket{x} = e^{i \alpha \paren{x}} \ket{x}'$ to make the hopping parameter real, and then the theory is equivalent to the usual real-hopping tight-binding Hamiltonian. However, this redefinition cannot be performed globally on a circle if $\theta \notin 2 \pi \mathbb{Z}$. With $b^{\dagger}_i = e^{- i \alpha_i} b^{\dagger\prime}_i$, $b_i = e^{i \alpha_i} b_i'$, the hopping parameter transforms as $w_{i, i+1}' = e^{i \paren{\alpha_{i} - \alpha_{i + 1}}} w_{i, i+1} $. To make the hopping parameter real, we would like to choose
\begin{equation}
	\alpha_{i} - \alpha_{i + 1} = \frac{\theta}{2 \pi R} \xi
	.
\end{equation}
Then, periodicity of the $\alpha_i$ implies
\begin{equation}
	2 \pi \mathbb{Z} \ni \alpha_{i = 0} - \alpha_{i = 0} = \alpha_{i = 0} - \alpha_{i = \frac{2 \pi R}{\xi}} = \frac{1}{R} \frac{\theta}{2 \pi} \xi \frac{2 \pi R}{\xi} \implies \theta \in 2 \pi \mathbb{Z}
	.
\end{equation}
Any values of $\theta$ that do not satisfy this condition will not allow a transformation to make all hopping parameters real. A generic theory at a value of $\theta \notin 2 \pi \mathbb{Z}$ is not equivalent to a real-hopping tight-binding model. The $\theta$-parameter that classifies these theories is
\begin{equation}
	\theta = \arg{\prod_i w_{i, i+1}} \quad (\text{mod } 2 \pi)
	,
\label{eq:phase_hopping}
\end{equation}
which is analogous to the Polyakov loop around the spatial direction.

\subsection{Numerical exact diagonalization results}

We use the numerical exact diagonalization (ED) method to analyze the discrete Hamiltonian.

\begin{figure}[!t]
\centering
	\begin{subfigure}[ht]{0.48\textwidth}
		\centering
		\includegraphics[width=\textwidth]{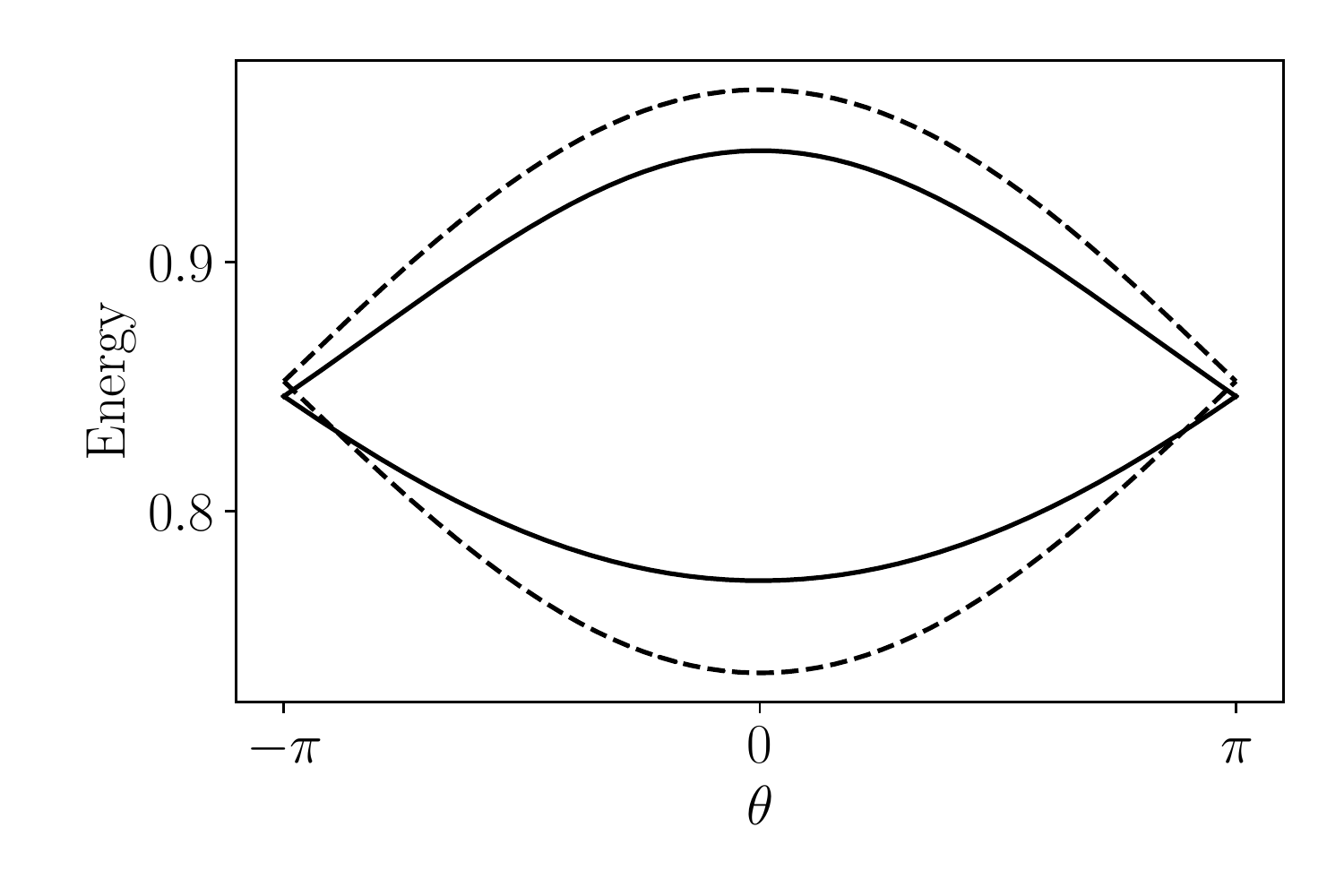}
		\caption{}
		\label{subfig:spectrum_n_2_1}
	\end{subfigure}
	\hfill
	\begin{subfigure}[!htbp]{0.48\textwidth}
		\centering
		\includegraphics[width=\textwidth]{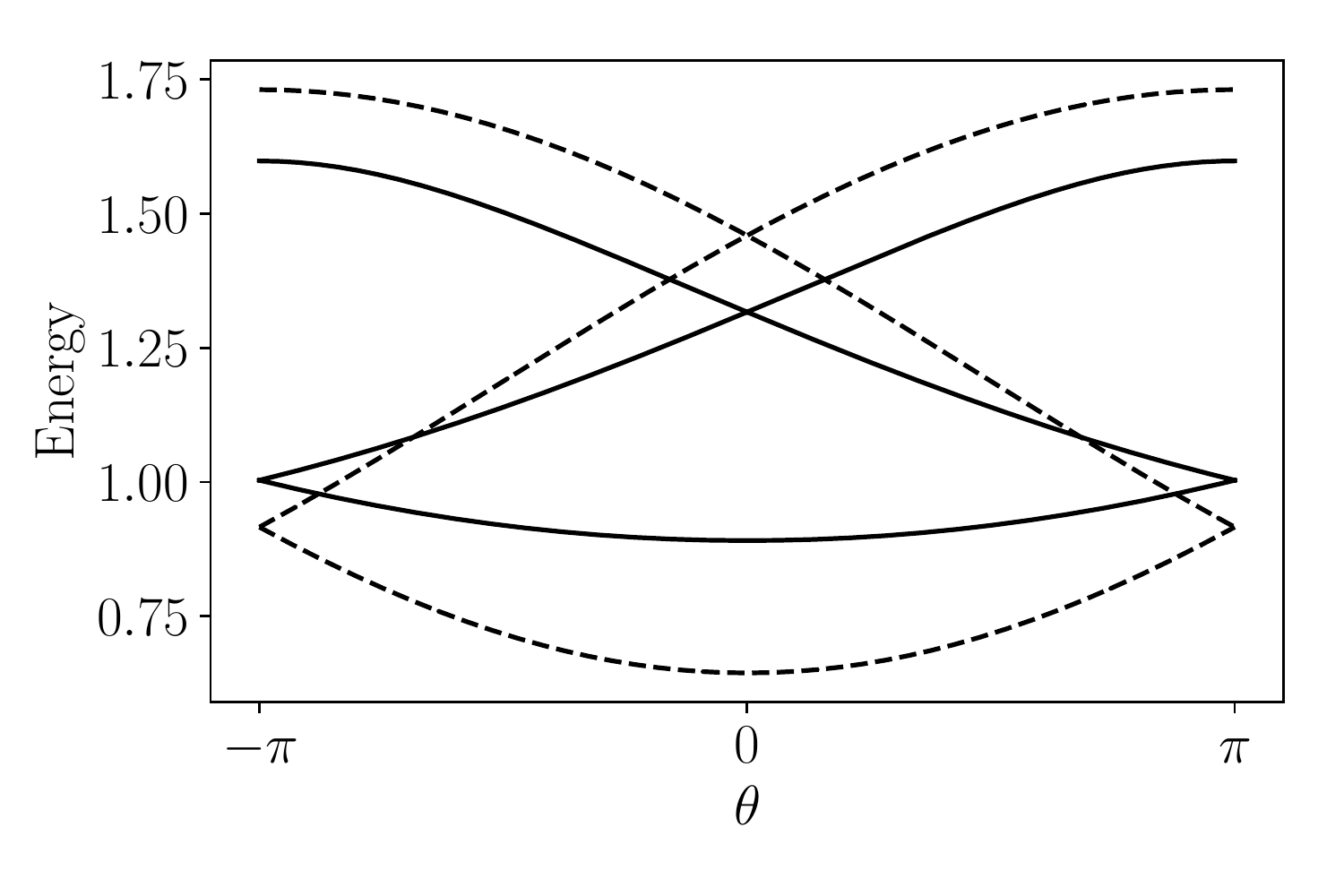}
		\caption{}
		\label{subfig:spectrum_n_3_1}
	\end{subfigure}
	\vfill
	\begin{subfigure}[ht]{0.48\textwidth}
		\centering
		\includegraphics[width=\textwidth]{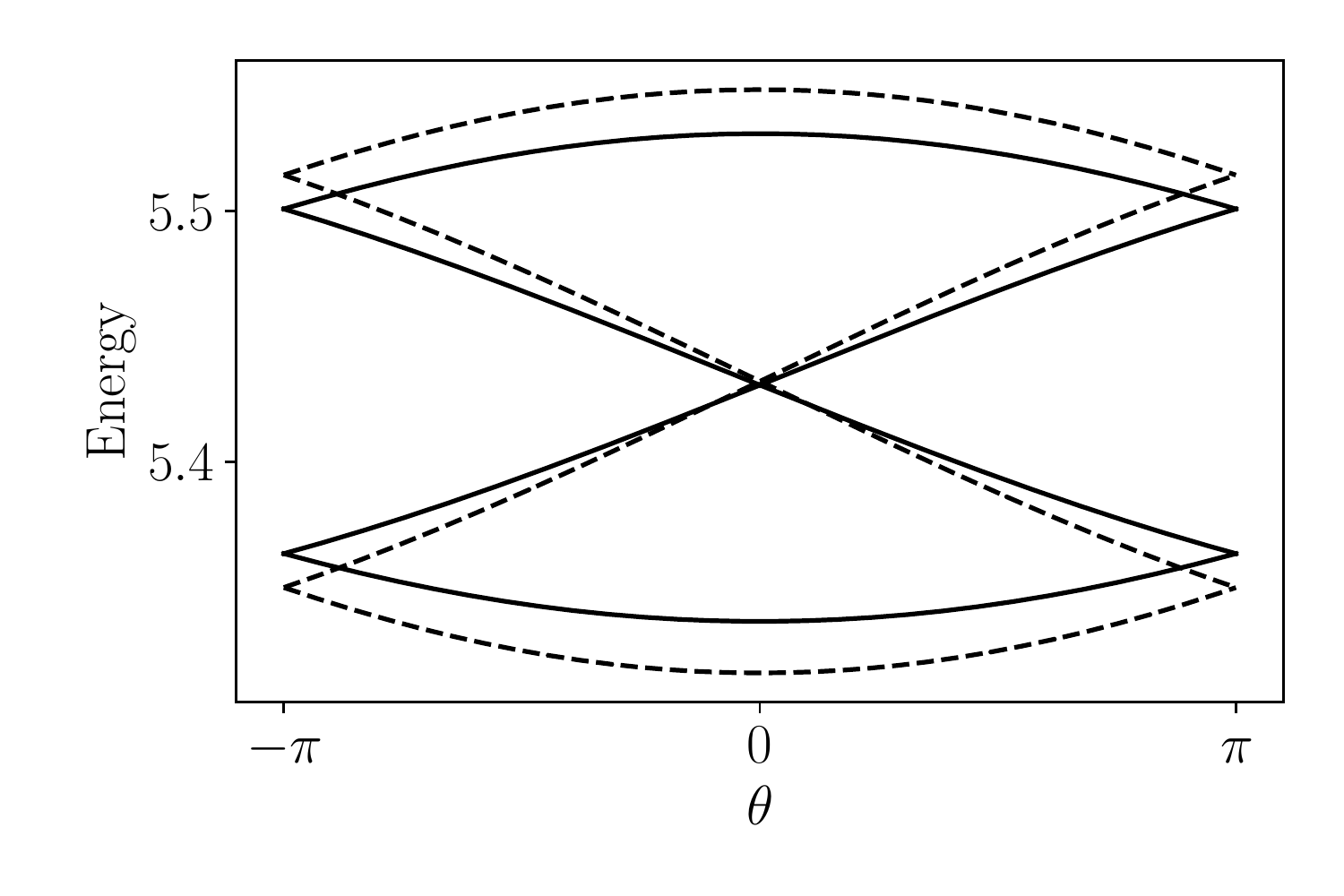}
		\caption{}
		\label{subfig:spectrum_n_4_1}
	\end{subfigure}
	\hfill
	\begin{subfigure}[!htbp]{0.48\textwidth}
		\centering
		\includegraphics[width=\textwidth]{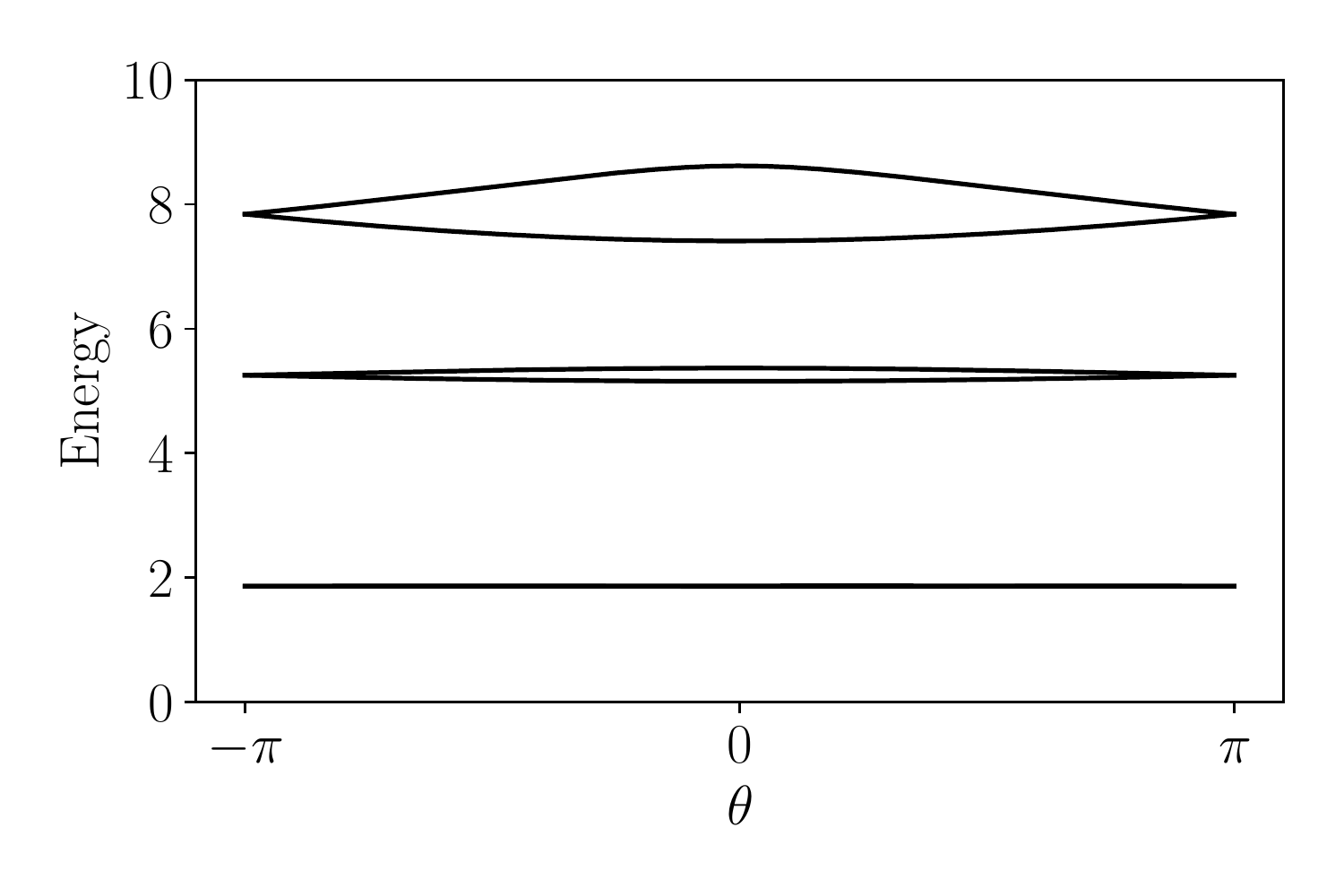}
		\caption{}
		\label{subfig:spectrum_n_2_2}
	\end{subfigure}
\caption{Comparison between the spectra from ED (solid) on $n_s = 120$ lattice sites and DIGA (dashed). We show the lowest $n$ energy branches with (\subref{subfig:spectrum_n_2_1}) $n=2$, $\omega=2$;  (\subref{subfig:spectrum_n_3_1}) $n=3$, $\omega=3$; (\subref{subfig:spectrum_n_4_1}) $n=4$, $\omega=12$.
Only $\theta \in [-\pi, \pi]$ is shown; the $2\pi$-periodicity in $\theta$ is exact in both ED and DIGA.
The qualitative shapes of spectra are well captured by DIGA, indicating reasonable proximity to the continuum semiclassical limit. 
In the semiclassical limit $\omega \rightarrow \infty$, tunneling between different potential wells is exponentially suppressed and the energy levels are dominated by the perturbative spectra of the local potential wells which are approximately harmonic oscillators.
In (\subref{subfig:spectrum_n_2_2}) with $n = 2$, $\omega = 4$, $n_s = 120$, the first 6 branches from ED are plotted to illustrate the general structure of the spectrum including excited states. Every two energy branches are connected at $\theta = \pm \pi$. The lowest two branches are very close and almost overlap with the given plotting scale. The lowest two branches are approximately at the ground state energy level, $\omega / 2 = 2$, of a harmonic oscillator with frequency $\omega$. Similarly, the $(2 s + 1)$-th and $(2 s + 2)$-th branches are approximately at the $s$-th excited harmonic oscillator energy $(s + 1/2 ) \omega$.
}
\label{fig:spectrum}
\end{figure}

The ED method yields the whole spectrum of the tight-binding Hamiltonian on $n_s$ sites, and we can compare the lowest $n$ energies with the continuum DIGA prediction in Eq.~(\ref{eq:diga_spectrum}). We show this comparison in Fig.~\ref{fig:spectrum}. For moderate values of $\omega$, the qualitative features of the lowest $n$ energies are consistent between ED and DIGA. Ground state degeneracy occurs at $\theta=\pi$ for any $n$ and $\omega$ \cite{10.1093/ptep/ptx148} in both computations. The ED and DIGA ground state energies are offset relative to one another, but these offsets are much smaller than the spacing between harmonic oscillator energies $\sim \omega$. To isolate the nonperturbative contribution and suppress the effects of these relative offsets, the differences $\omega / 2 + E_{k} - (1/n) \sum_{k' = 1}^{n} E_{k'}$ as computed with DIGA are plotted in Fig.~\ref{fig:spectrum}. The shapes of the lowest $n$ energies from the ED and DIGA are not of the same size, but this discrepancy, as we show below,  decreases as $\omega$ increases and the semiclassical limit is approached.

\begin{figure}[!t]
\centering
	\begin{subfigure}[ht]{0.48\textwidth}
		\centering
		\includegraphics[width=\textwidth]{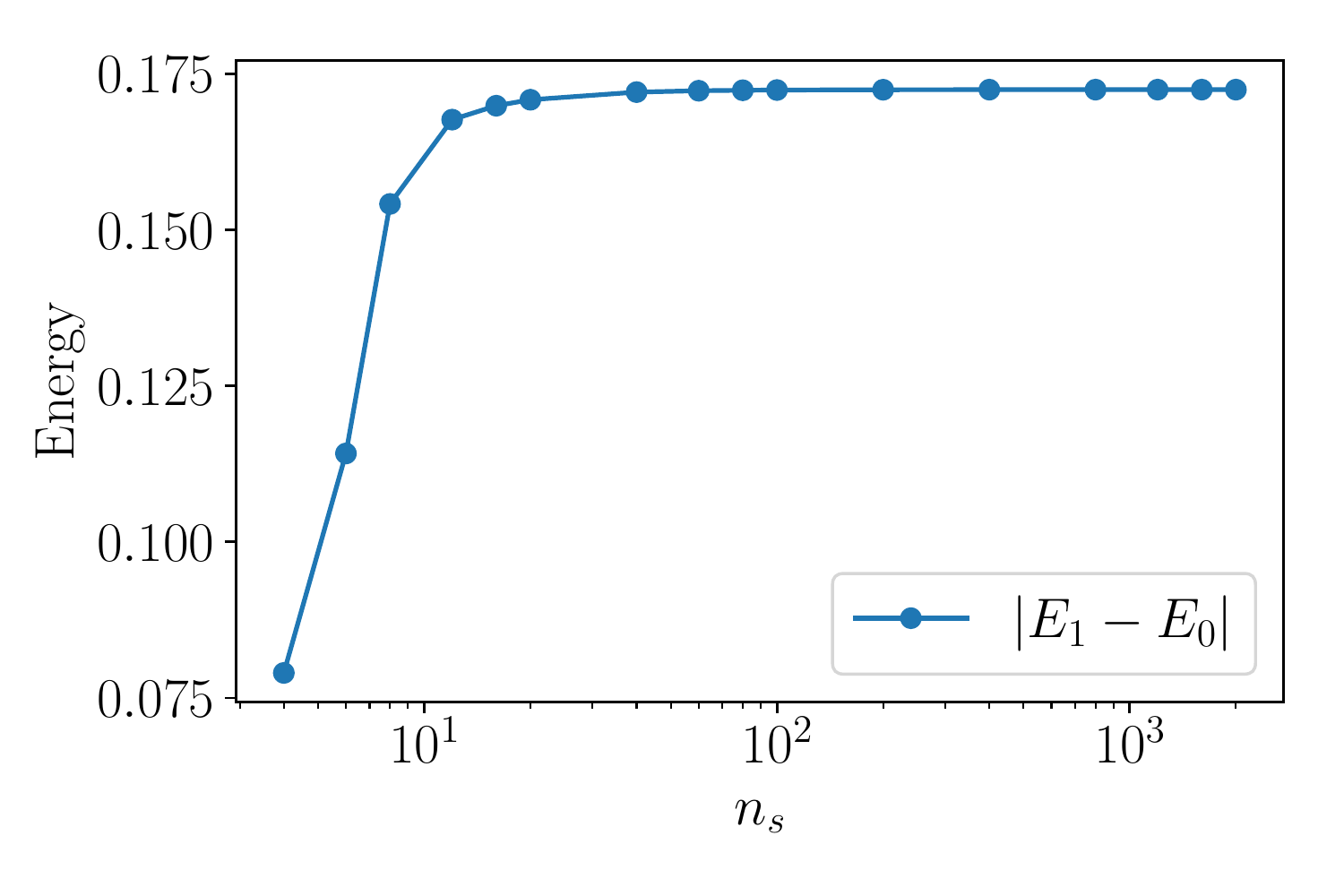}
		\caption{}
		\label{subfig:vary_ns_1}
	\end{subfigure}
	\hfill
	\begin{subfigure}[!htbp]{0.48\textwidth}
		\centering
		\includegraphics[width=\textwidth]{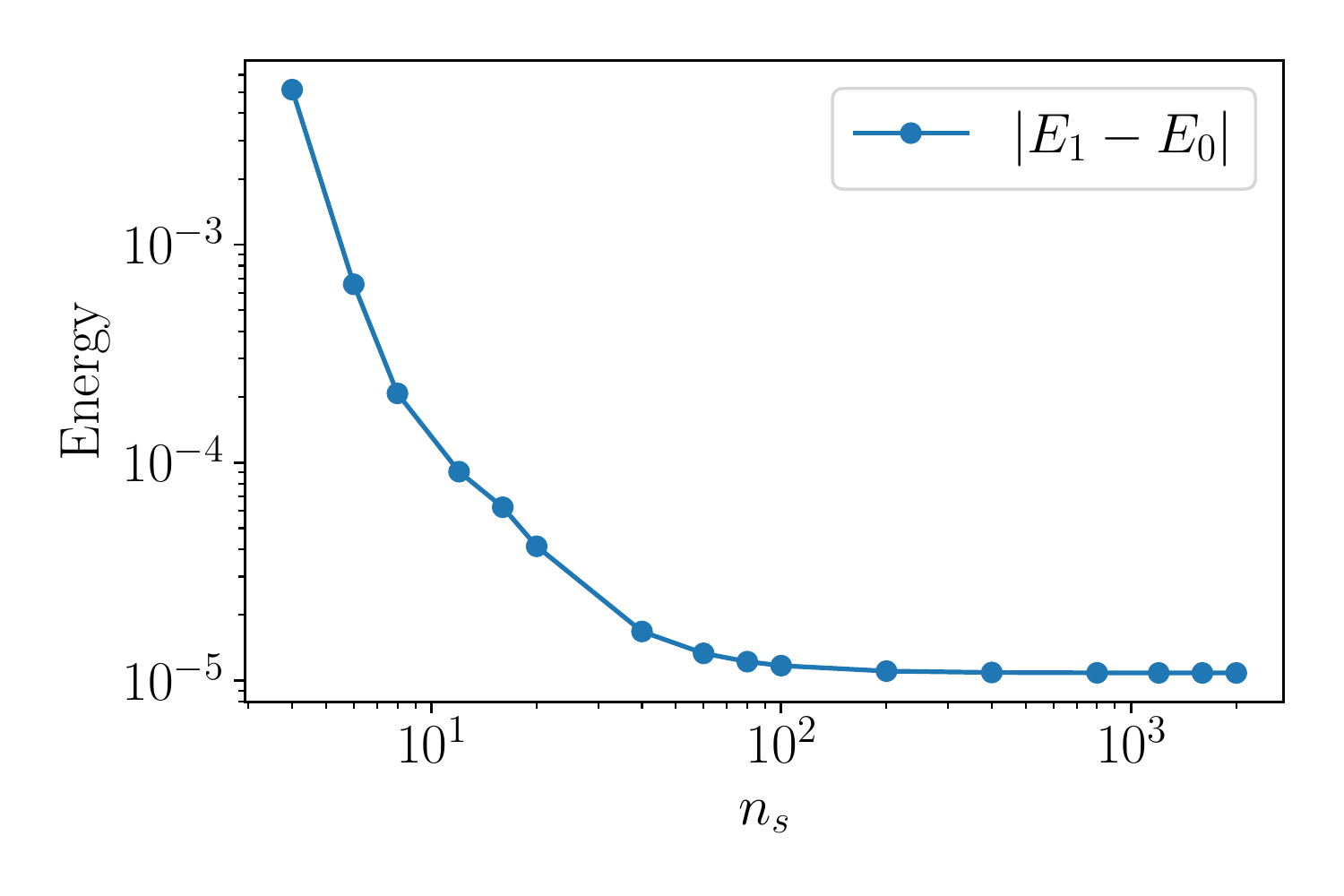}
		\caption{}
		\label{subfig:vary_ns_2}
	\end{subfigure}
\caption{The energy difference between the two lowest states $\abs{E_1 - E_0}$ with varying $n_s$. Parameters used in exact diagonalization are $n = 2$, $\theta = 0$, (\subref{subfig:vary_ns_1}) $\omega = 2$, (\subref{subfig:vary_ns_2}) $\omega = 8$. The energy difference starts to converge from $n_s = O(10)$.}
\label{fig:vary_ns}
\end{figure}

Now let us examine the continuum limit of ED. We take $n=2$, $\theta = 0$ as an example with $\omega = 2, 8$ in the ED shown in Fig.~\ref{fig:vary_ns}. The continuum limit is approached at around $n_s = O(10)$.

\begin{figure}[!t]
\centering
	\begin{subfigure}[ht]{0.48\textwidth}
		\centering
		\includegraphics[width=\textwidth]{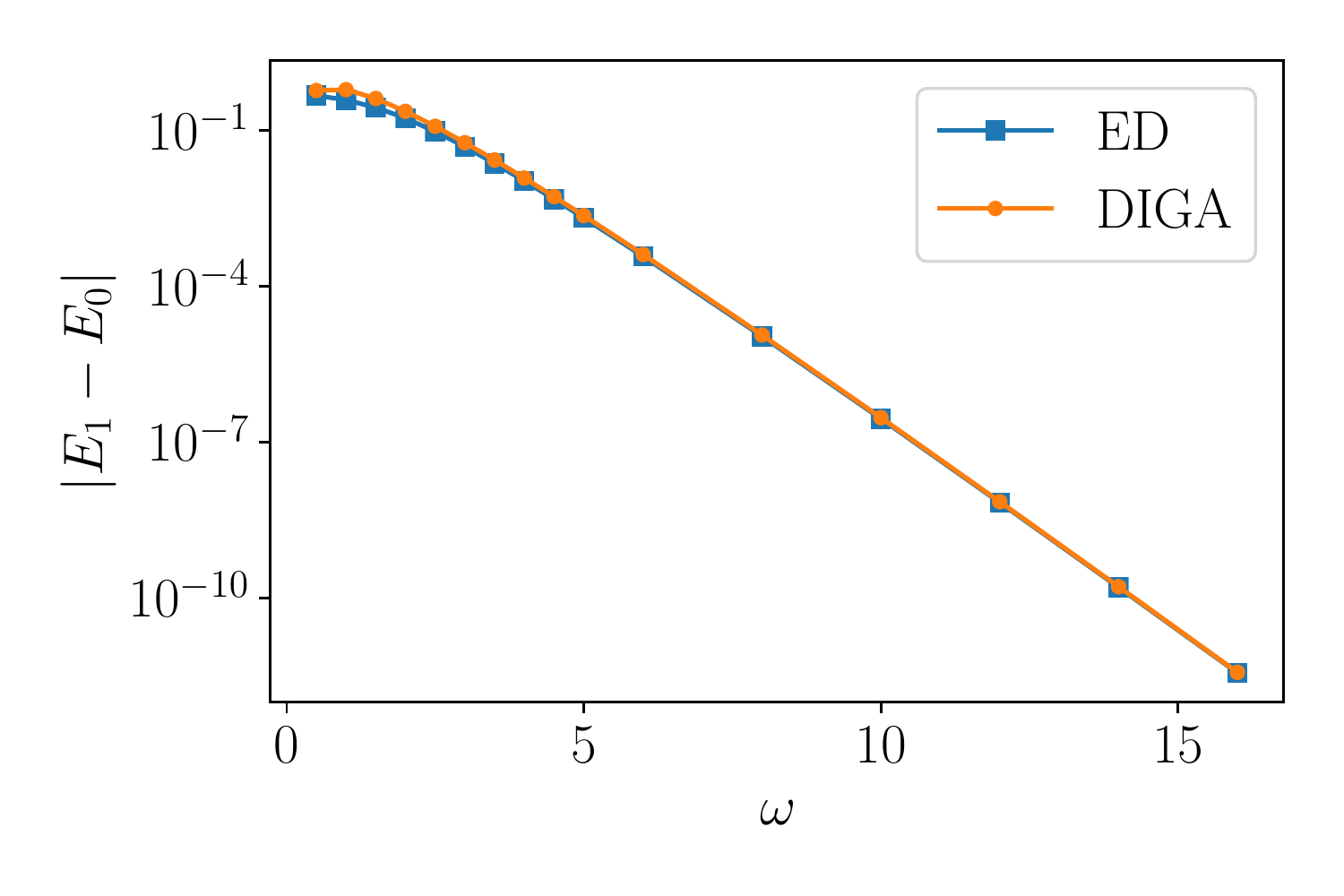}
		\caption{}
		\label{subfig:vary_omega_lo}
	\end{subfigure}
	\hfill
	\begin{subfigure}[!htbp]{0.48\textwidth}
		\centering
		\includegraphics[width=\textwidth]{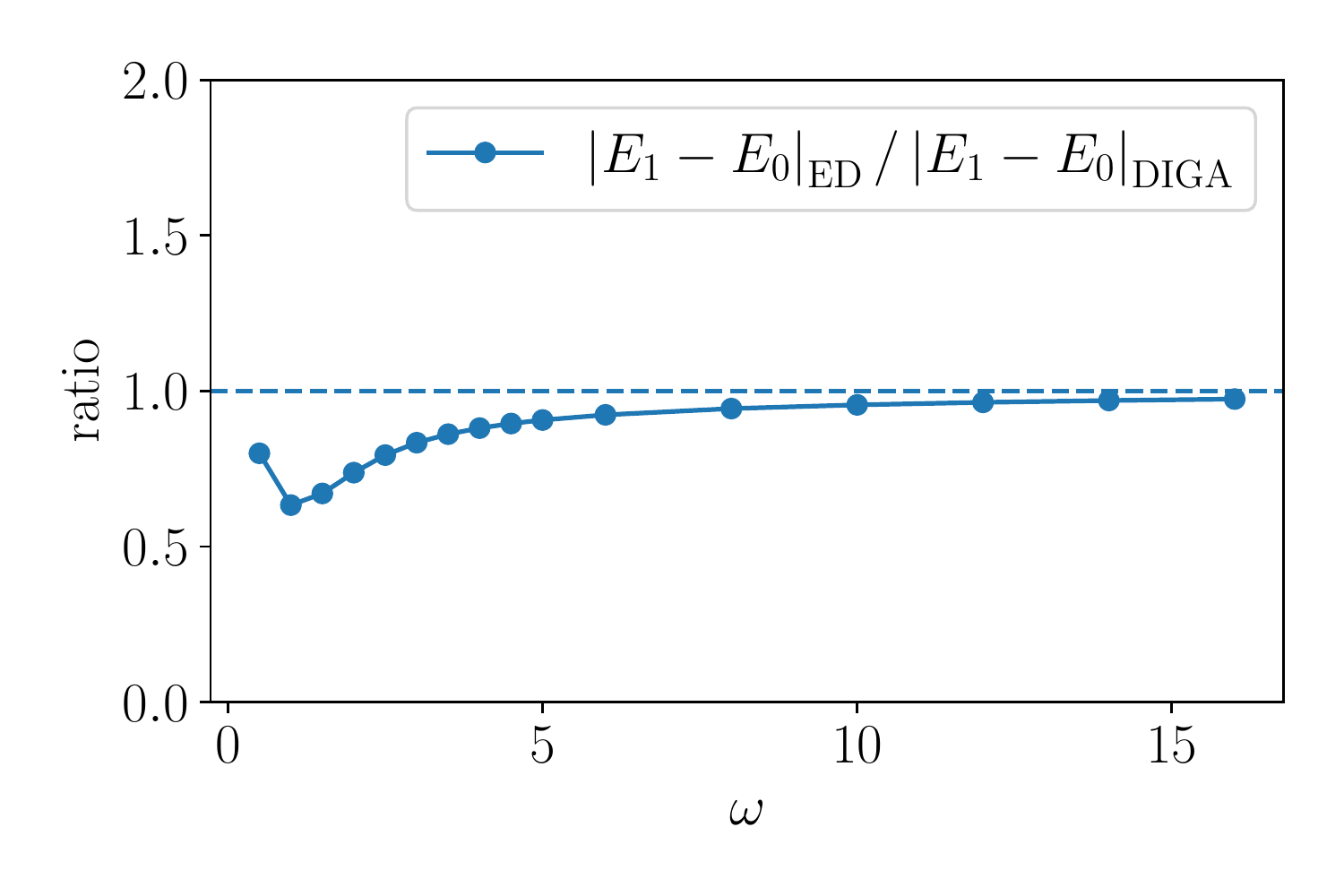}
		\caption{}
		\label{subfig:vary_omega_ratio}
	\end{subfigure}
\caption{For $n=2$, $n_s = 2000$, $\theta=0$, with varying $\omega \in [0.5, 16.0]$, (\subref{subfig:vary_omega_lo}) the energy difference between the two lowest states $\abs{E_1 - E_0}$ from ED and DIGA and (\subref{subfig:vary_omega_ratio}) the ratio between the results of ED to DIGA. The DIGA gives the correct order of magnitude from $\omega = 0.5$, and starts to converge at around $\omega = 2$.}
\label{fig:vary_omega}
\end{figure}

In Fig.~\ref{fig:vary_omega}, we compare the DIGA and ED results for the splitting $\abs{E_1 - E_0}$ between the ground and the first excited states in the two-well potential, using a large number of sites $n_s = 2000$. The results match at the order of magnitude level for all $\omega \geq 0.5$, and converge rapidly for $\omega > 2$.

For $\omega$ not too large, qualitative features of spectra are well-captured with a few lattice sites. $n_s = 2, 4, 6, 8, 12$ lattice sites are studied in Sec.~\ref{sec:experiment} below.

\section{Experimental schemes}
\label{sec:experiment}

\subsection{Encoding using a synthetic dimension}

\begin{figure}[!ht]
\centering
	\begin{subfigure}[ht]{0.32\textwidth}
		\centering
		\includegraphics[width=\textwidth]{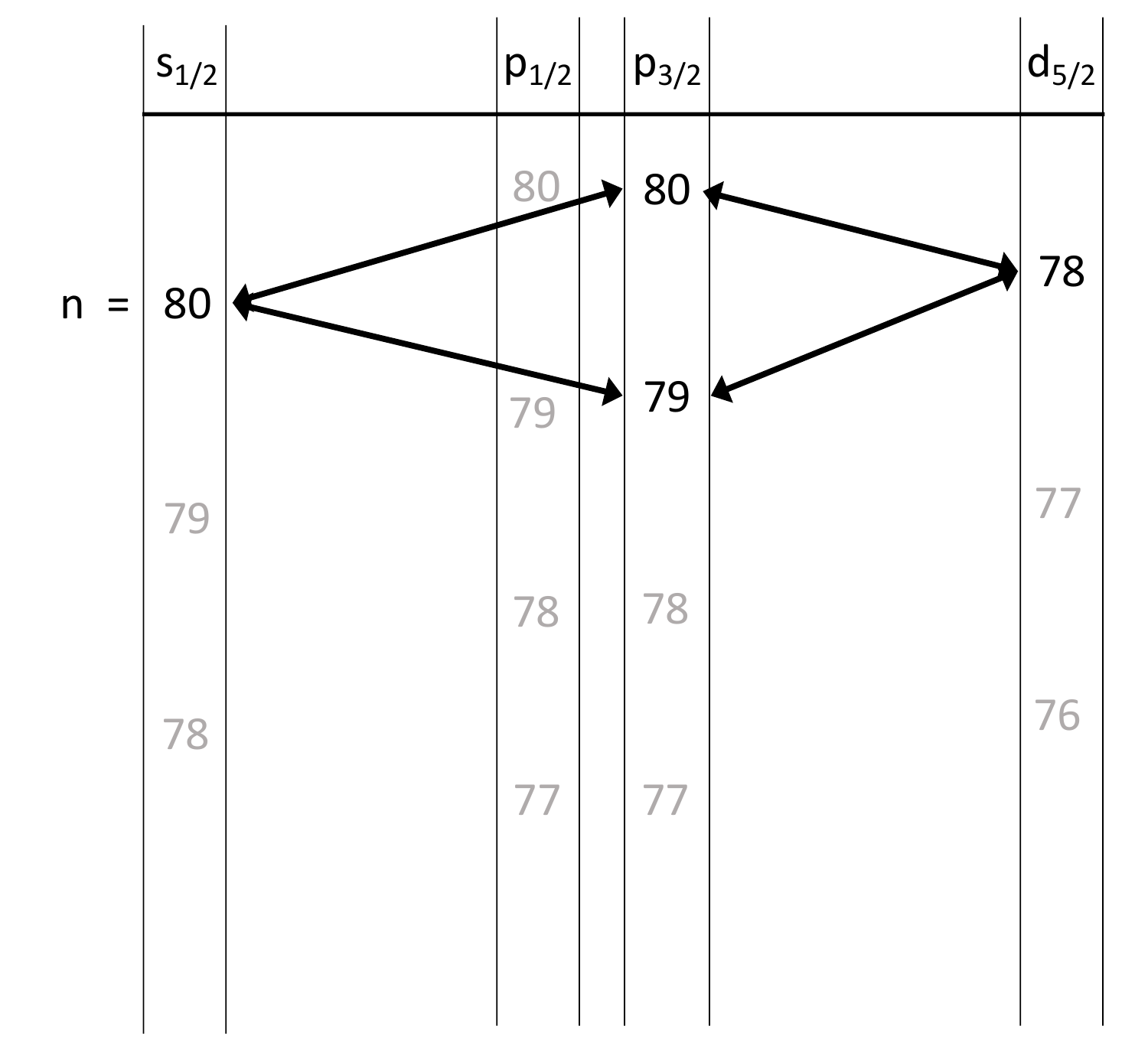}
		\caption{}
		\label{subfig:4-state}
	\end{subfigure}
	\hfill
	\begin{subfigure}[ht]{0.32\textwidth}
		\centering
		\includegraphics[width=\textwidth]{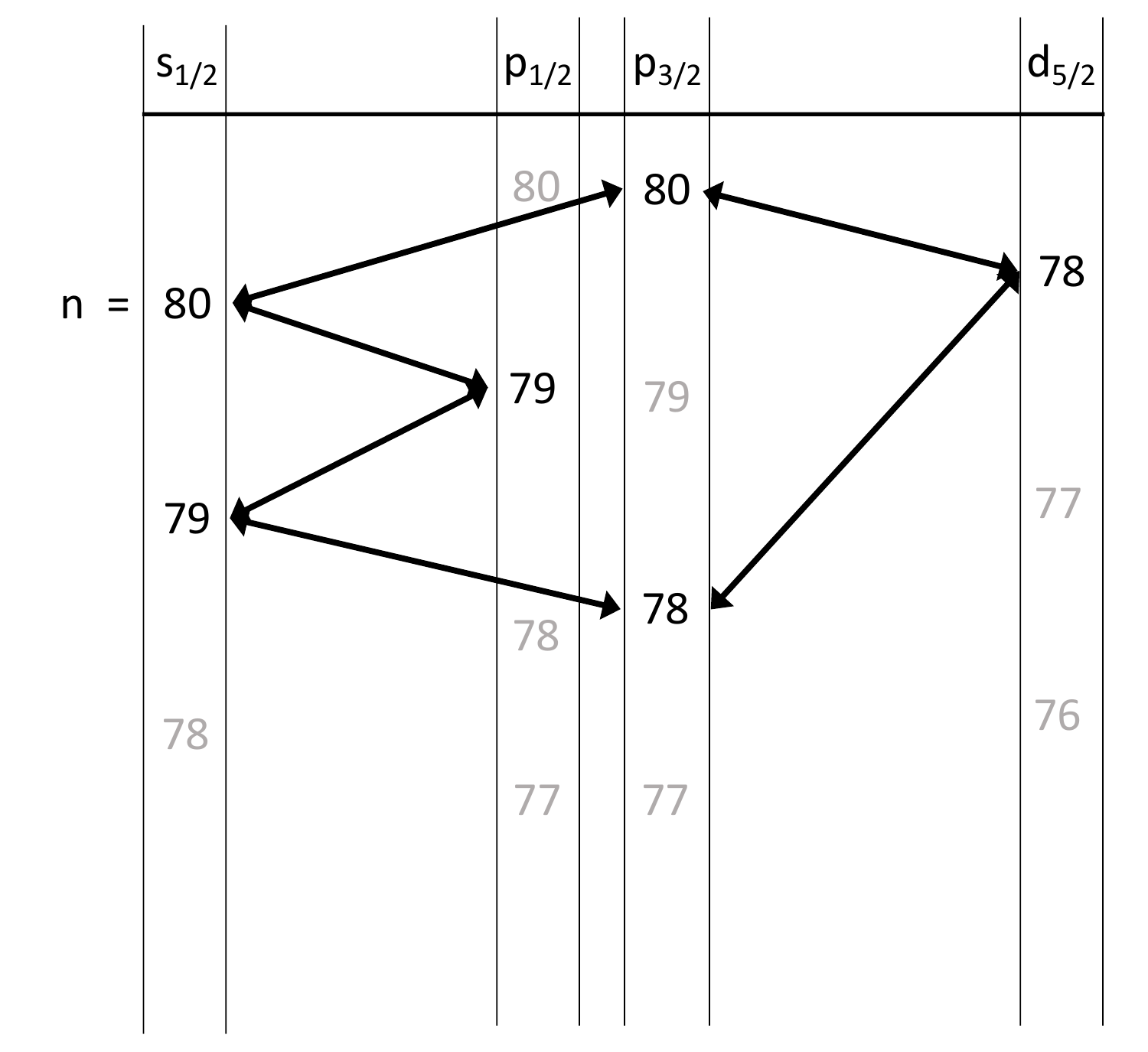}
		\caption{}
		\label{subfig:6-state}
	\end{subfigure}
	\hfill
	\begin{subfigure}[ht]{0.32\textwidth}
		\centering
		\includegraphics[width=\textwidth]{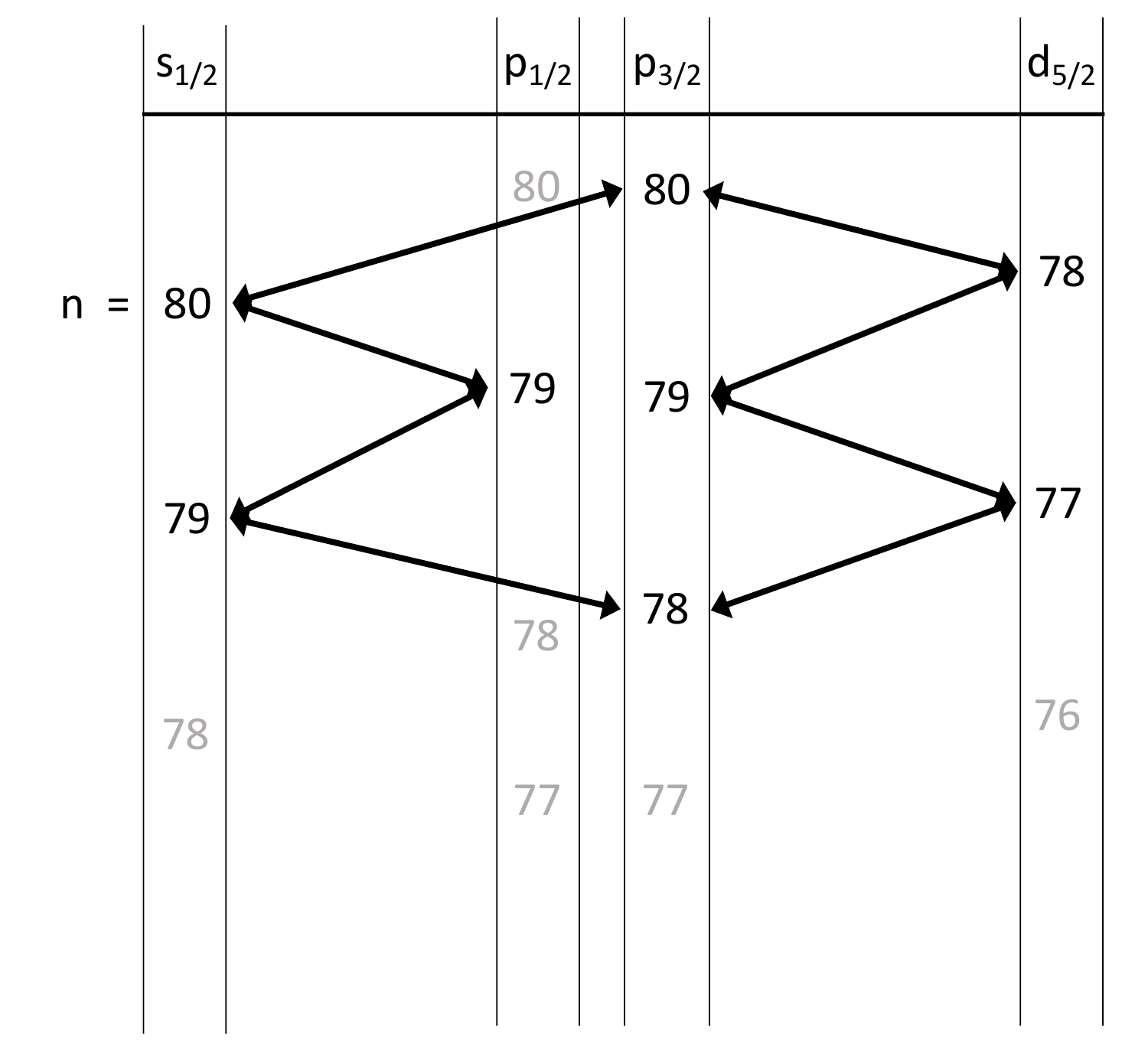}
		\caption{}
		\label{subfig:8-state}
	\end{subfigure}
	\begin{subfigure}[ht]{0.32\textwidth}
		\centering
		\includegraphics[width=\textwidth]{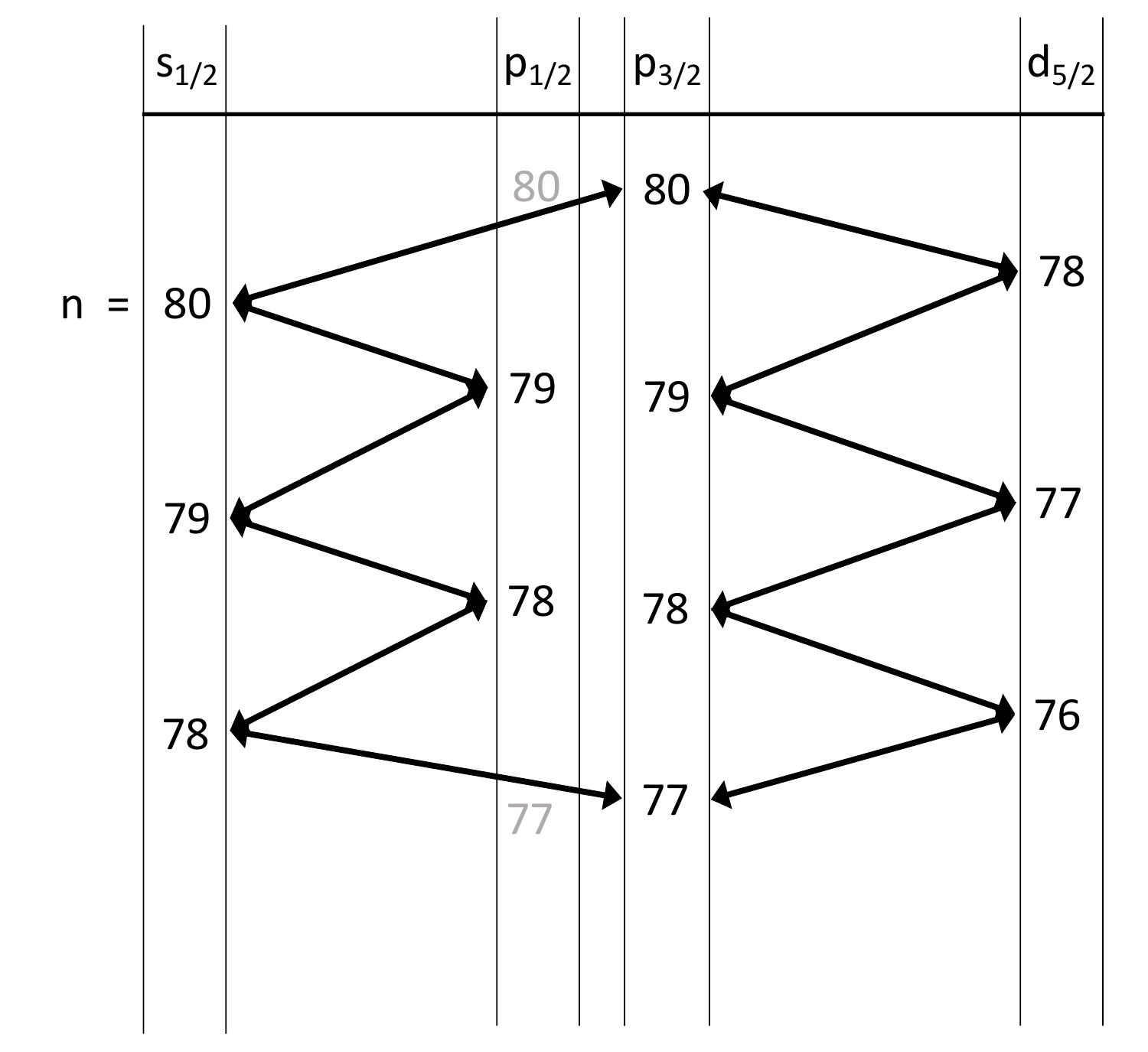}
		\caption{}
		\label{subfig:12-state}
	\end{subfigure}
	\hfill
\caption{A loop-like synthetic dimension encoded in individual Rydberg atoms by connecting (\subref{subfig:4-state}) 4 states, (\subref{subfig:6-state}) 6 states, (\subref{subfig:8-state}) 8 states, and (\subref{subfig:12-state}) 12 states with resonant microwaves. In this approach, transitions between connected states are enabled by applying microwave electric fields at the frequency resonant with (or slightly detuned from, to elicit an effective site energy) the state-to-state frequency splitting. Each state in a loop represents a discretized position on the circle.  $s_{1/2}$, $p_{1/2}$, $p_{3/2}$, and $d_{5/2}$ states are used to encode the position. $d_{3/2}$ states are also included in the numerical simulation.  Gray-colored states lie outside of the subspace of the effective synthetic lattice, and transitions to such states are suppressed by energy differences between unwanted transitions and the desired (driven) transitions.}
\label{fig:synthetic_dimension_1}
\end{figure}

A synthetic dimension~\cite{Review-OzawaPrice} is an effective dimension encoded in, \textit{e.g.}, the internal degrees of freedom of an atom or molecule, which can be used to study a range of phenomena related to dynamics and transport. In the quantum mechanics problem considered here, a single Rydberg atom can be used to encode a discretized spatial circle on which the particle resides: as shown in Fig.~\ref{fig:synthetic_dimension_1}, multiple internal Rydberg states can be used to encode position (site) basis states of a lattice with periodic boundary conditions.
Here, we describe a lattice formed by Rydberg levels (high-lying electronic states of an atom's valence electron), which are plentiful, long-lived, and have recently been utilized for the generation of synthetic lattices~\cite{RydSynth}. 

For the single-particle studies we describe, one may either employ individual Rydberg atoms excited from a bulk gas of laser-cooled atoms~\cite{RydSynth} or, alternatively, from a single atoms trapped in an optical tweezer~\cite{RydSing}.
For both cases, synthetic lattices of Rydberg states may formed by global addressing with multi-frequency microwave fields, \textit{e.g.}, provided by a single high-bandwidth source and horn antenna.
In the latter approach based on tweezer-trapped atoms, the scalability of optical tweezer arrays~\cite{Kaufman2021} provides natural opportunities to introduce strong interparticle interactions for the study of correlated dynamics in synthetic dimensions~\cite{sundar2018synthetic}.

Rydberg synthetic lattices can be formed by coupling different Rydberg levels with resonantly oscillating microwave electric fields. The coupling of these electronic states is governed by electric dipole selection rules, \textit{i.e.}, $\Delta L = \pm 1$, $|\Delta J| \leq 1$, $|\Delta m_J| \leq 1$. Individual tunneling links of the synthetic lattice are formed by spectrally resolving a given state-to-state transition. Zeeman shifts by a bias magnetic field help to separate out different magnetic sub-levels ($m_J$ states) and their transitions. We note that, along with hard constraints due to selection rules, there are some practical limitations on the lattice graphs that can be formed, with practical considerations related to the range of microwave frequencies that need to be applied.

With suitably chosen parameters (frequencies, amplitudes, and phases) of external microwaves, in the rotating wave approximation (RWA) and within the decay time of excited states, one can form a tight-binding model described by the Hamiltonian
\begin{equation}
\begin{aligned}
	 H 	=  \sum_{i} \bracket{ w_{i, i+1} b^{\dagger}_{i+1} b_{i} + w_{i, i+1}^{*} b^{\dagger}_{i} b_{i+1} + V_{i} b^{\dagger}_{i} b_{i}}.
\end{aligned}
\label{eq:tight-binding_ham_dipole}
\end{equation}
Here, the absolute values of hopping parameters $\abs{w_{i, i+1}}$ are tuned via the amplitudes of the microwaves; the complex phases of the hopping parameters $w_{i, i+1}$ are tuned by the phases of the microwaves (and accounting for path-dependent phases from the sources to the atoms); and state-dependent potential terms $V_i$ are controlled by the detuning between the microwave frequencies and the resonance frequency of the relevant transition. Consequently, it is possible to have full spectroscopic control over the parameters in Eq.~(\ref{eq:tight-binding_ham_dipole}).

Order-of-magnitude estimates can be used to establish suitable ranges of parameters viable with realistic experiments based on Rydberg synthetic lattices. The hopping parameter magnitudes $\abs{w}$  generated by resonant microwaves corresponds to the (resonant) Rabi frequency $\Omega$ for a given transition:
\begin{equation}
    \frac{1}{2 m \xi^2}	= \abs{w} = \Omega
	.
\end{equation}
The $\mathbb{Z}_n$ potential is set by detuning as
\begin{align}
	\Delta_i &= \Delta \paren{1 - \cos \paren{\frac{n \xi i}{R}}} \nonumber\\
	\Delta&\equiv\tilde\lambda= \tilde{\omega}^2 I / n^2.
\end{align}
Here $\Delta$ is the detuning scale and $\tilde{\omega} \equiv \omega / I$ is the perturbative fast frequency.
We regard $\Omega$, $\Delta$, the number of sites (states) $n_s = 2 \pi R / \xi$, and a microwave relative phase $\theta$ from Eq.~(\ref{eq:phase_hopping}) as the experimental variables.
The energy scales of the theory are
the fast frequency
\begin{equation}
	\tilde{\omega} = \sqrt{2 \Delta \Omega} \frac{2 \pi n}{n_s}
	,
\end{equation}
and the nonperturbative slow frequency
\begin{align}
	\tilde{\omega}_{\mathrm{DIGA}} = & 2 \tilde{\omega} d\nonumber\\
	= & 2 \tilde{\omega} \times 2 e^{-\frac{8}{n^2} \tilde{\omega} I} \frac{2}{n} \sqrt{\frac{\tilde{\omega} I}{\pi}} \nonumber\\
	= & 8 \paren{2 \Delta^3 \Omega}^{1/4} \sqrt{\frac{2 n}{n_s}} \exp \paren{- 2 \sqrt{\frac{2 \Delta}{\Omega}} \frac{n_s}{\pi n}}
	.
\end{align}
The dimensionless quantity 
\begin{equation}
	\omega = \tilde{\omega} I = \sqrt{\frac{\Delta}{2 \Omega}} \frac{n n_s}{2 \pi}
	,
\label{eq:omegaI}
\end{equation}
 characterizes the theory.
The continuum limit $n_s \rightarrow \infty$ can be understood as $n_s / n \rightarrow \infty$ since a greater number of potential wells $n$ demands a greater number of sites $n_s$ to resolve the shape of the potential. 
For given values of $n$ and $n_s$, the parameters of the theory $\tilde{\omega}$ and $I$ are determined by the experimental parameters $\Omega$ and $\Delta$.

\subsection{Comparison of real-time dyanamics}

For concreteness, we consider the use of potassium ($^{39}$K) Rydberg states in the range of $n \sim 80$, as shown in Fig.~\ref{fig:synthetic_dimension_1}. This choice is without loss of generality, as a change of atomic species or the range of principal quantum numbers will only result in slight modifications of the microwave frequencies and bandwidths required. We incorporate moderate Zeeman shifts of the state energies, assuming applied quantization fields in the range of tens of Gauss.
We select the $m_J = +J$ magnetic sublevels to serve as the relevant states that are part of our synthetic lattice.
All state energies and state-to-state transition frequencies are determined using the open-source Alkali Rydberg Calculator (ARC) platform~\cite{SIBALIC2017319,ROBERTSON2021107814}.
The synthetic lattice technique relies on having the desired transitions (those representing tunneling connections between sites/states of the synthetic lattice) be separated in frequency from those that lead to states outside of the ``synthetic lattice'' subspace. In these explorations, the frequency separation $\Delta_{\mathrm{sep}}$ of desired transitions from unwanted transitions are at the scale of $\sim$10~MHz. To ensure the validity of the rotating wave approximation that underlies our description of the tight-binding synthetic lattice, we are limited to considering Hamiltonian terms, $w$ and $V$, that are small compared to $\Delta_{\mathrm{sep}}$.

\begin{figure}[!ht]
\centering
	\begin{subfigure}[ht]{0.48\textwidth}
		\centering
		\includegraphics[width=\textwidth]{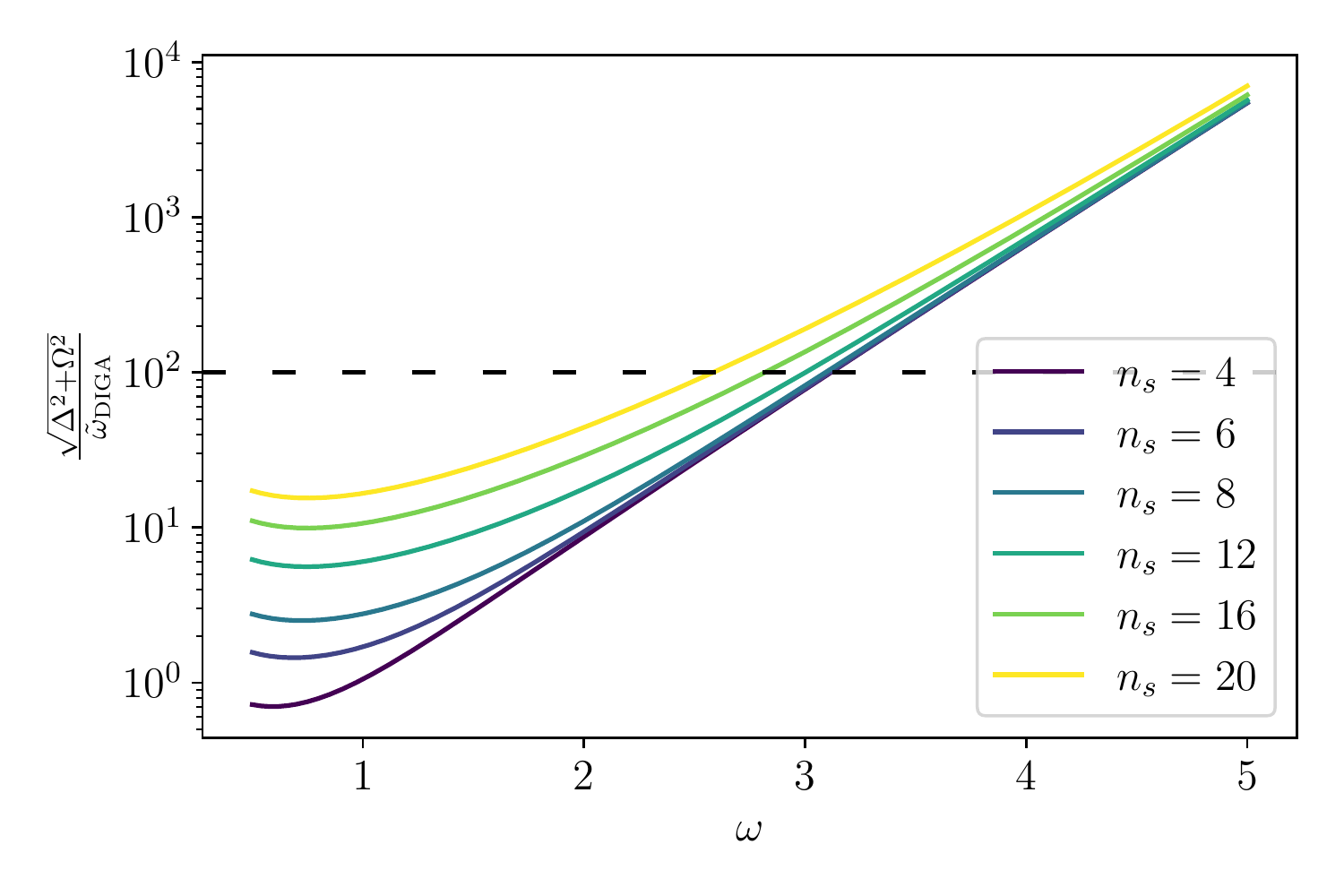}
		\caption{}
		\label{subfig:exp_params_n_2}
	\end{subfigure}
	\hfill
	\begin{subfigure}[!htbp]{0.48\textwidth}
		\centering
		\includegraphics[width=\textwidth]{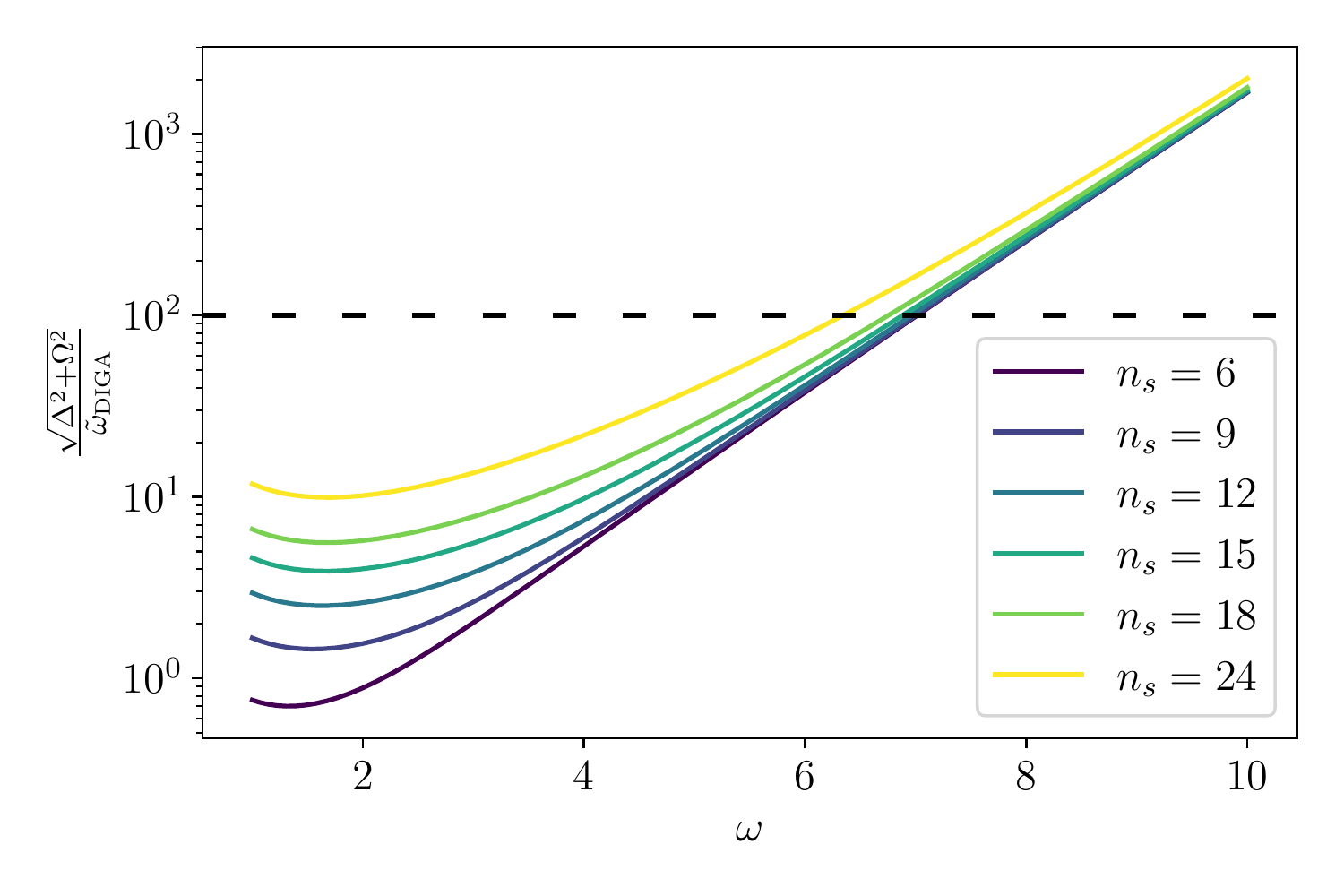}
		\caption{}
		\label{subfig:exp_params_n_3}
	\end{subfigure}
	\caption{The ratio $\sqrt{\Delta^2 + \Omega^2} / \tilde{\omega}_{\mathrm{DIGA}}$ for (\subref{subfig:exp_params_n_2}) $n=2$ (\subref{subfig:exp_params_n_3}) $n=3$. For a realistic experiment for observing a complete tunneling event, the ratio must be well below $O(100)$, which constrains the parameters $\omega$, $n$, $n_s$.}
\label{fig:exp_params}
\end{figure}

The condition $\sqrt{\Delta^2 + \Omega^2} \ll \Delta_{\mathrm{sep}}$ controls the goodness of the RWA. $\Delta_{\mathrm{sep}}$ is on the order of $\sim$10~MHz. For observing the tunneling event, the characteristic tunneling time scale $2\pi / \tilde{\omega}_{\mathrm{DIGA}}$ must not substantially exceed the nominal Rydberg lifetime $\sim$10 microseconds. Therefore, the ratio $\sqrt{\Delta^2 + \Omega^2} / \tilde{\omega}_{\mathrm{DIGA}} \ll 10\text{ MHz} \times 10~\mu\text{s} = 100$ is a constraint for a realistic experiment. If this ratio is close to $O(100)$, then a fraction of the tunneling process  might still be observed. Fig.~\ref{fig:exp_params} demonstrates this ratio for $n = 2$ and $n = 3$ with a few values of $n_s$.

We compare two real-time simulations: a simulation of the ideal tight-binding Hamiltonian, Eq.~(\ref{eq:tight-binding_ham_dipole}), and a realistic simulation of the Rydberg atom. In the latter case we solve the Schr\"{o}dinger equation with time-dependent external fields (microwaves) and including 54 (for 4-state)/126 (for 6-/8-/12-state) nearby (potentially connected)  Rydberg states in the simulation. The initial state is prepared as a position eigenstate at the bottom of the $0$-th well, \textit{i.e.}, a Kronecker delta distribution. This preparation of an atom within a single Rydberg state can be accomplished through a Rydberg laser $\pi$-pulse, an adiabatic sweep, or a STIRAP excitation.

The site-resolved (along the synthetic dimension) population dynamics of the Rydberg system may, \textit{e.g.}, be achieved by performing state-selective field ionization as in Ref.~\cite{RydSynth}. Alternatively, the phenomenology associated with the 't Hooft anomaly, namely a transition to frozen dynamics due to instanton-anti-instanton interference, may be observed even if only measuring the dynamics of a single (initially occupied) Rydberg state. Such a partial measurement scheme, resolving the dynamics in only one or a few Rydberg levels, may be more amenable to experiments based on optical tweezers.

\begin{figure}[!ht]
\centering
	\begin{subfigure}[ht]{0.32\textwidth}
		\centering
		\includegraphics[width=\textwidth]{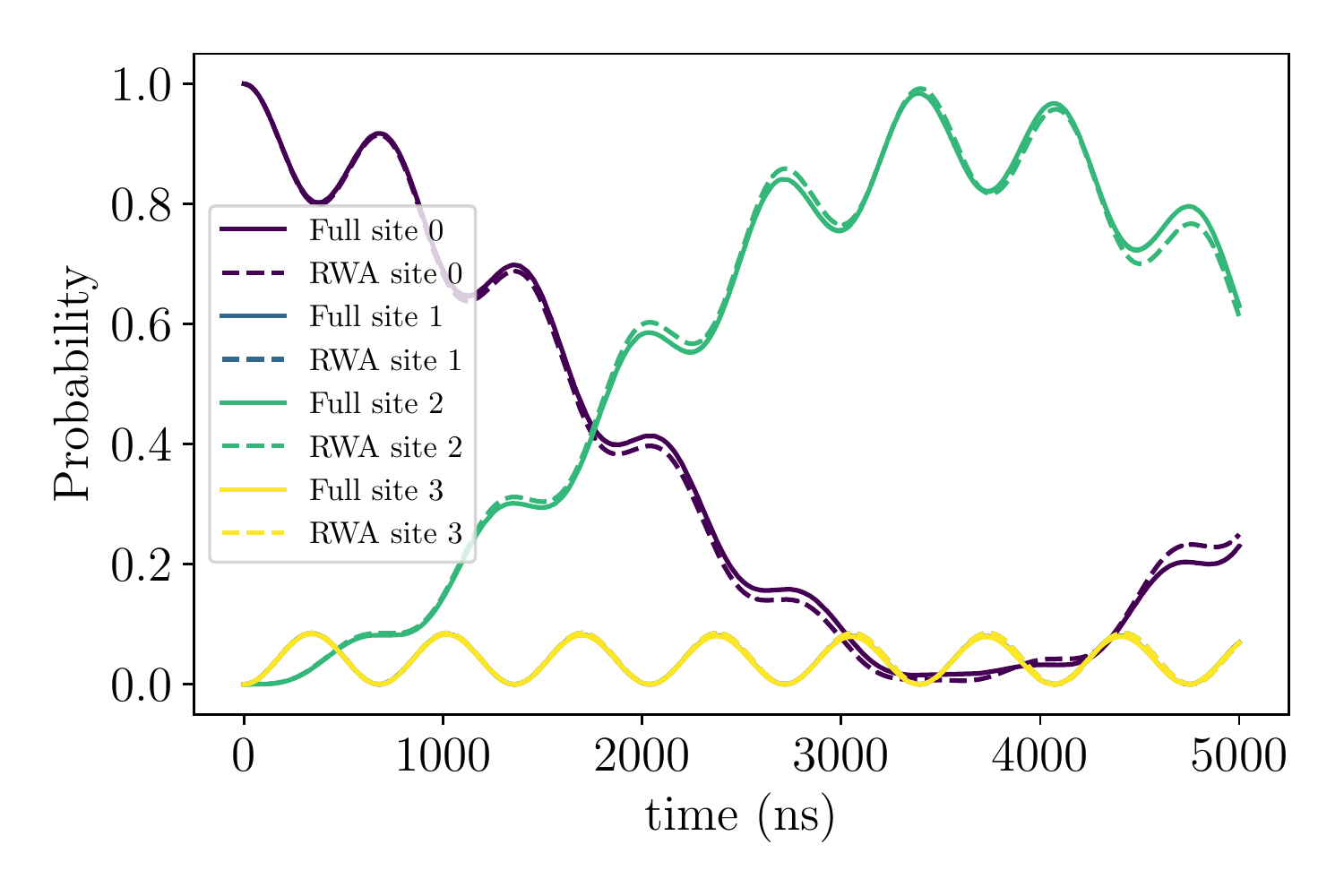}
		\caption{}
		\label{subfig:comparison_exp_theta_0_1}
	\end{subfigure}
	\hfill
	\begin{subfigure}[ht]{0.32\textwidth}
		\centering
		\includegraphics[width=\textwidth]{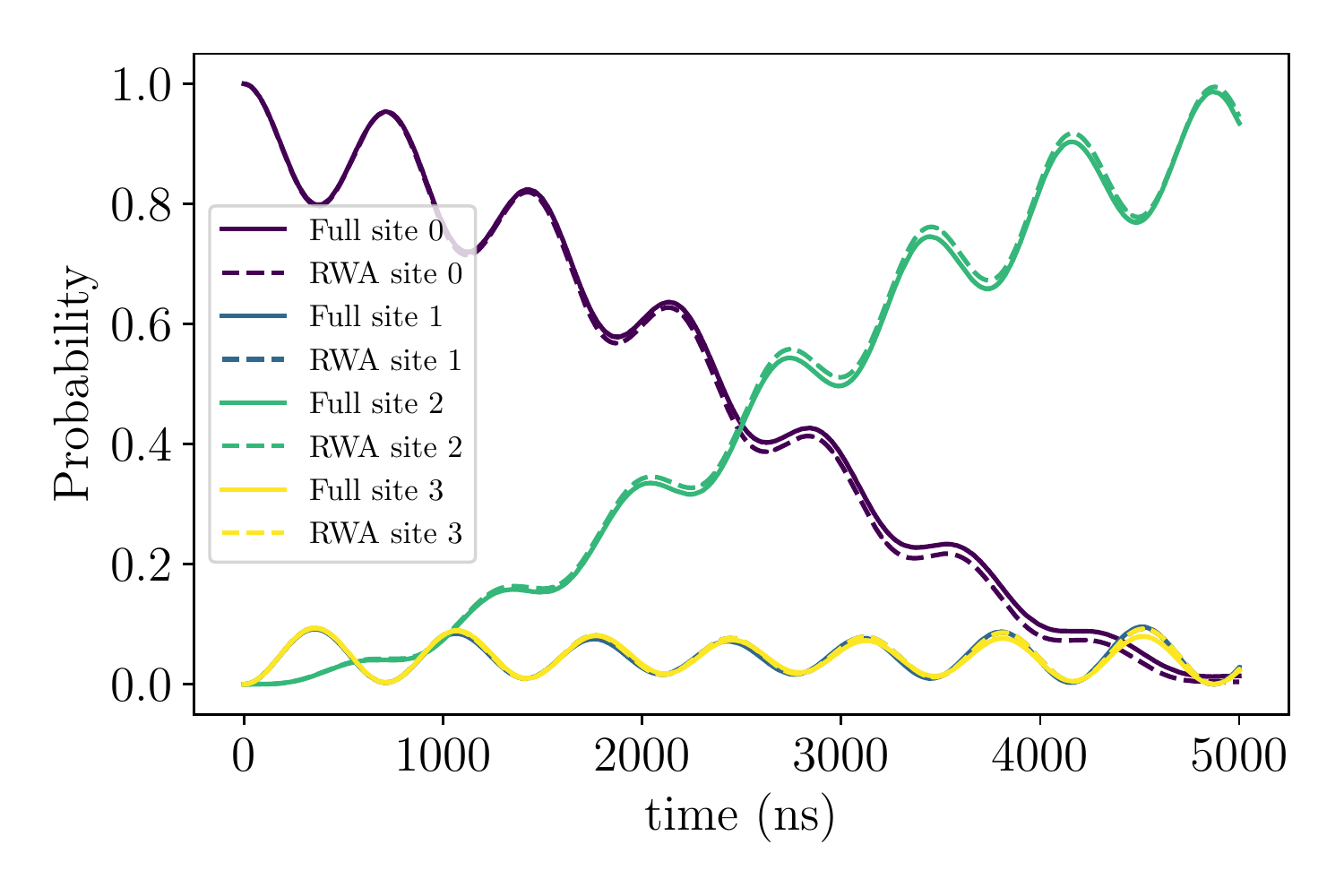}
		\caption{}
		\label{subfig:comparison_exp_theta_0.5pi_1}
	\end{subfigure}
	\hfill
	\begin{subfigure}[!htbp]{0.32\textwidth}
		\centering
		\includegraphics[width=\textwidth]{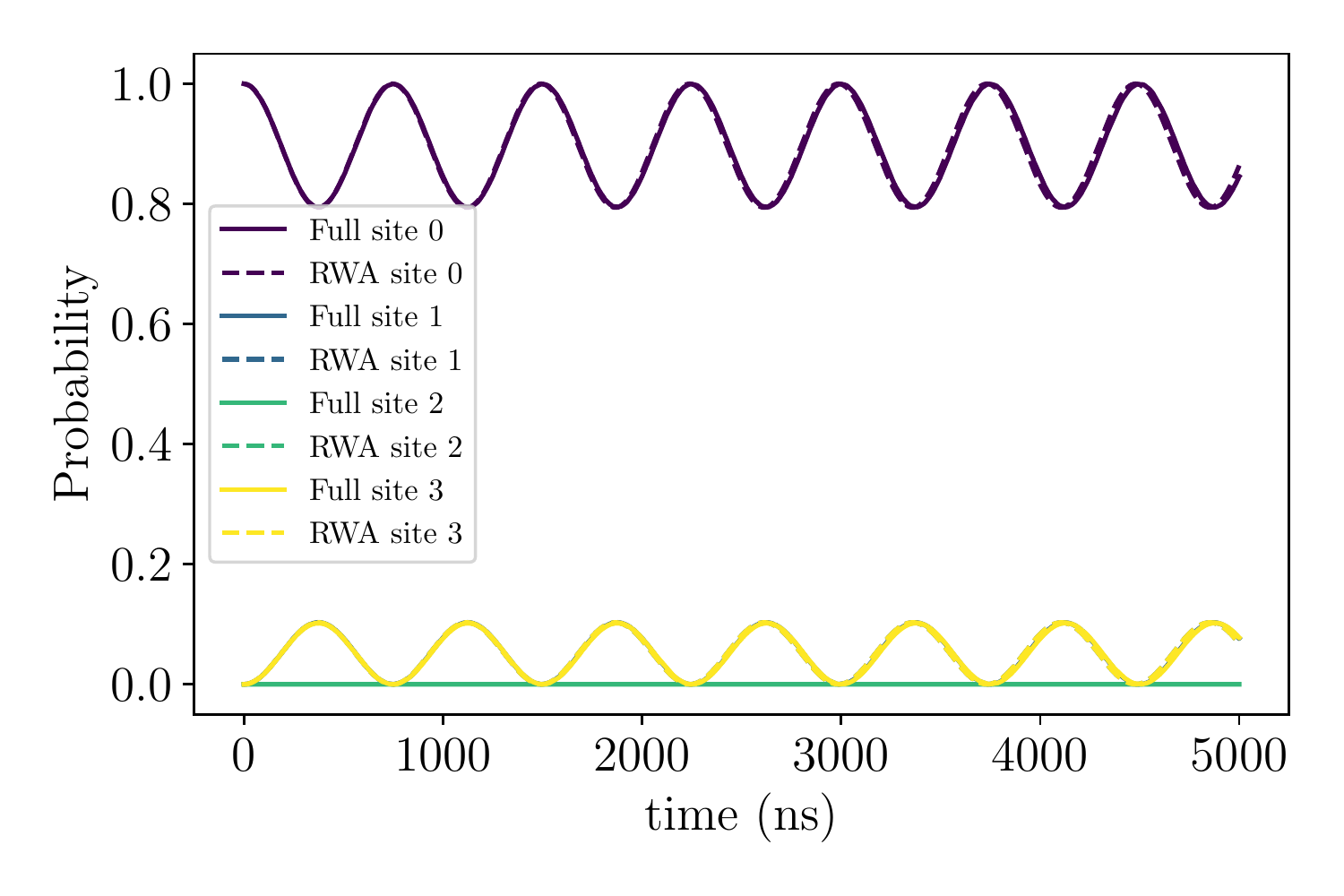}
		\caption{}
		\label{subfig:comparison_exp_theta_pi_1}
	\end{subfigure}
	\begin{subfigure}[ht]{0.32\textwidth}
		\centering
		\includegraphics[width=\textwidth]{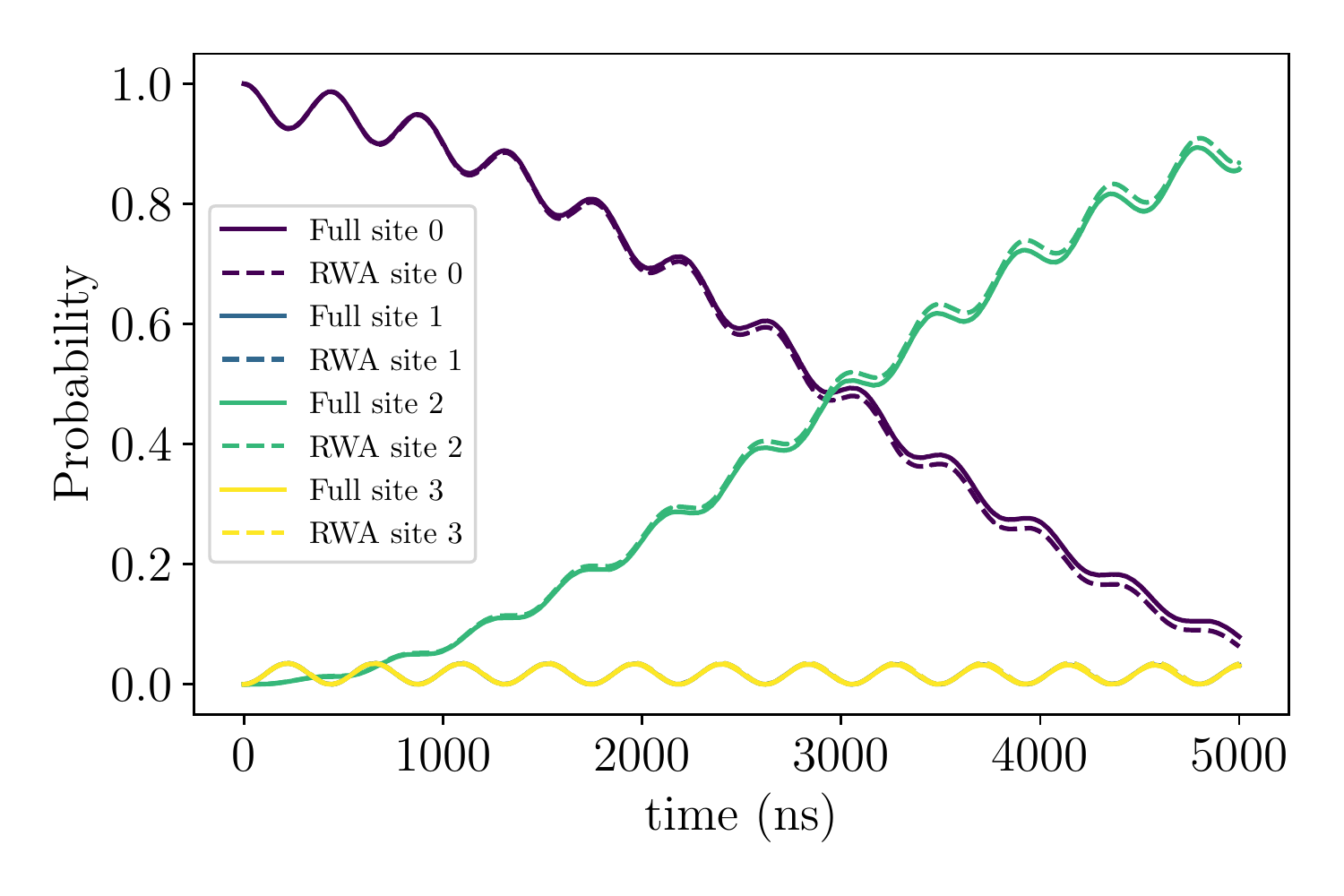}
		\caption{}
		\label{subfig:comparison_exp_theta_0_2}
	\end{subfigure}
	\hfill
	\begin{subfigure}[ht]{0.32\textwidth}
		\centering
		\includegraphics[width=\textwidth]{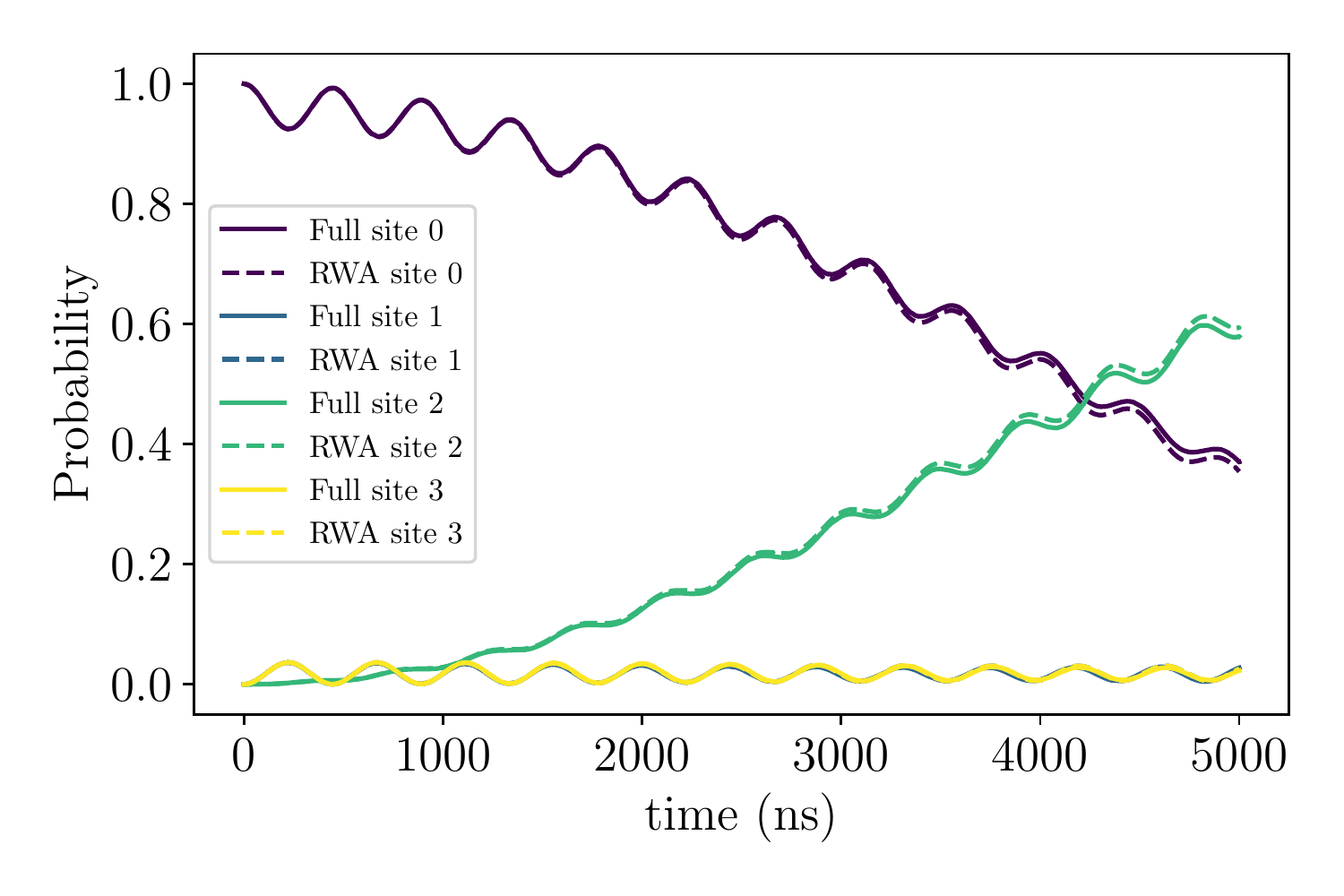}
		\caption{}
		\label{subfig:comparison_exp_theta_0.5pi_2}
	\end{subfigure}
	\hfill
	\begin{subfigure}[!htbp]{0.32\textwidth}
		\centering
		\includegraphics[width=\textwidth]{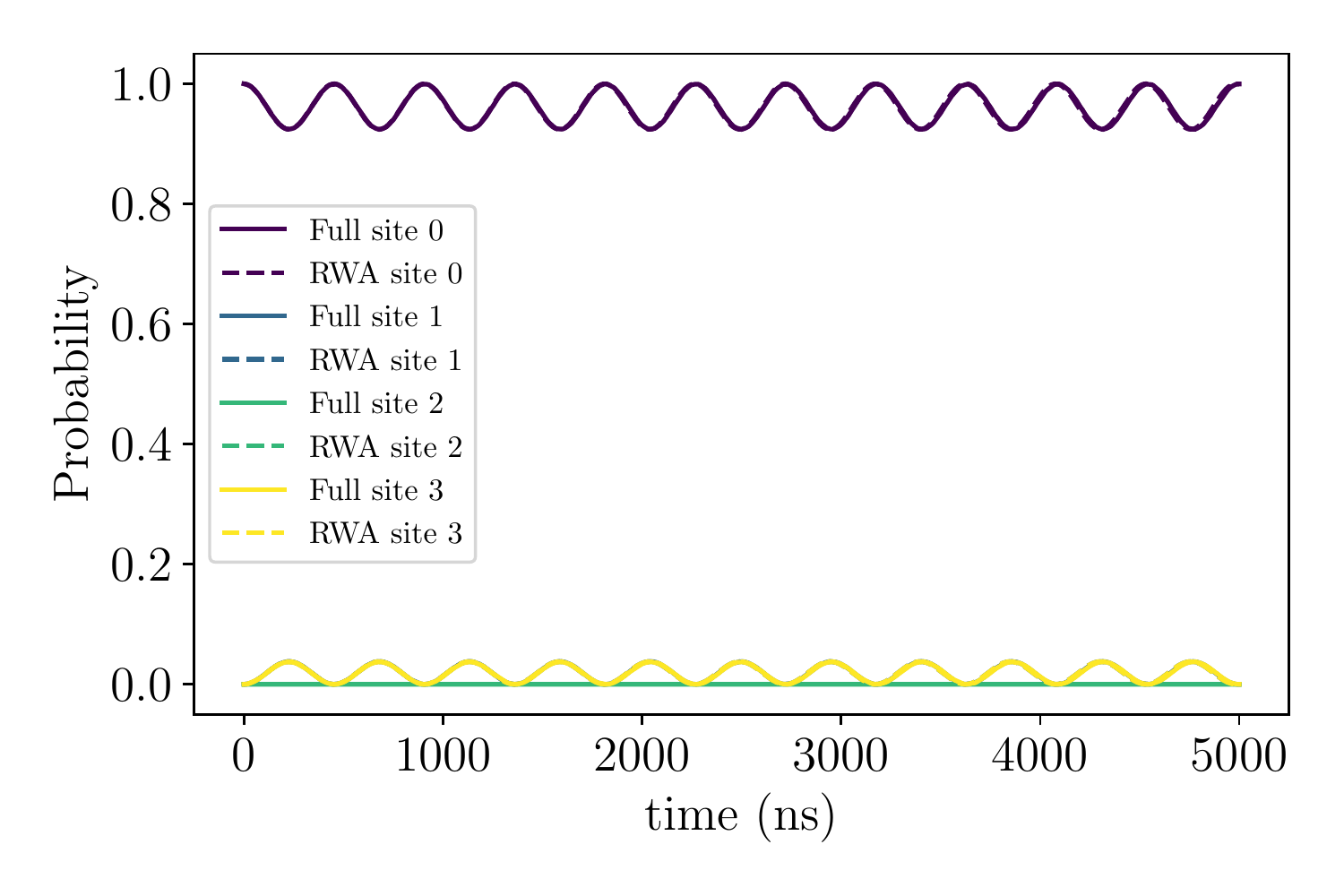}
		\caption{}
		\label{subfig:comparison_exp_theta_pi_2}
	\end{subfigure}
	\begin{subfigure}[ht]{0.32\textwidth}
		\centering
		\includegraphics[width=\textwidth]{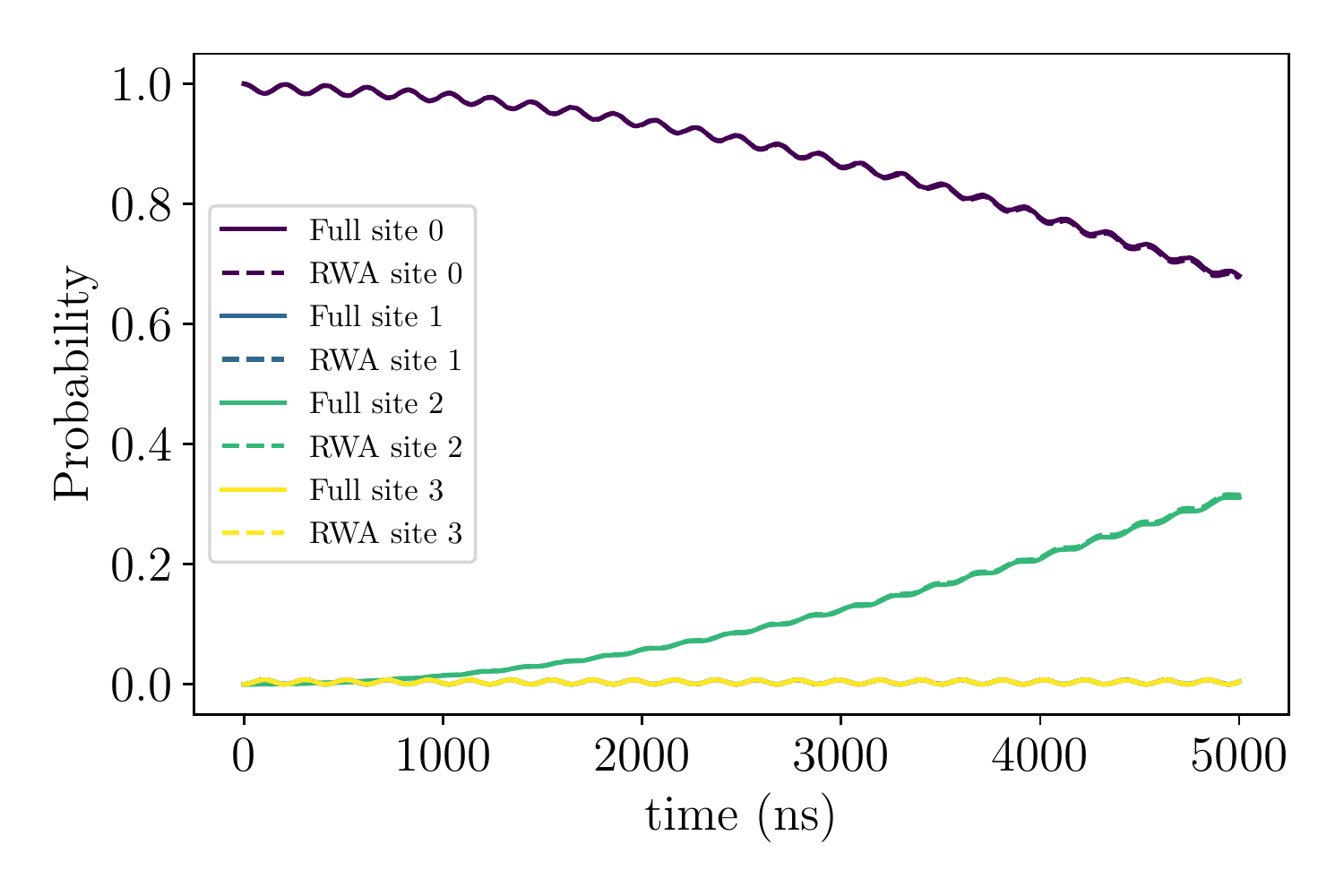}
		\caption{}
		\label{subfig:comparison_exp_theta_0_3}
	\end{subfigure}
	\hfill
	\begin{subfigure}[ht]{0.32\textwidth}
		\centering
		\includegraphics[width=\textwidth]{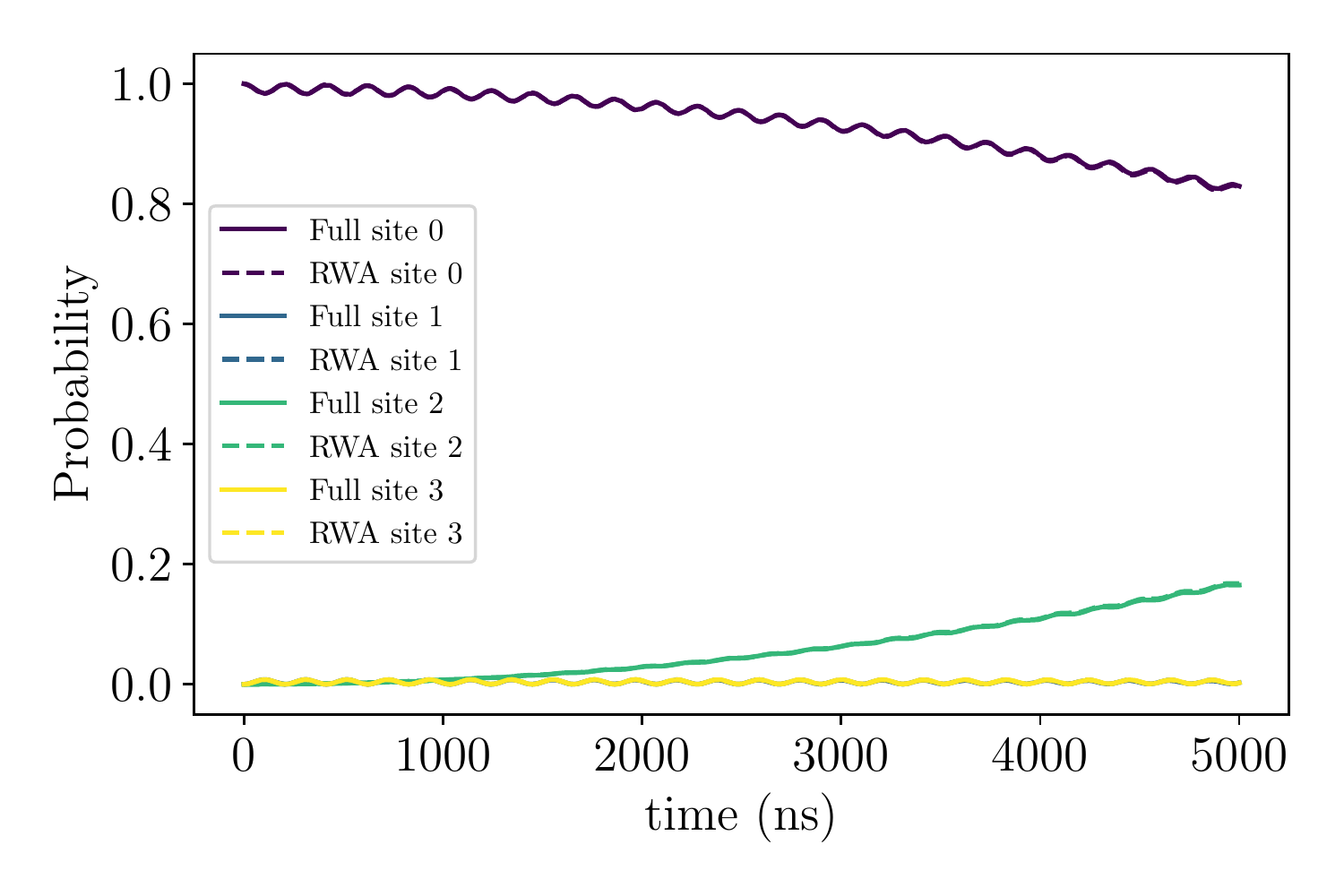}
		\caption{}
		\label{subfig:comparison_exp_theta_0.5pi_3}
	\end{subfigure}
	\hfill
	\begin{subfigure}[!htbp]{0.32\textwidth}
		\centering
		\includegraphics[width=\textwidth]{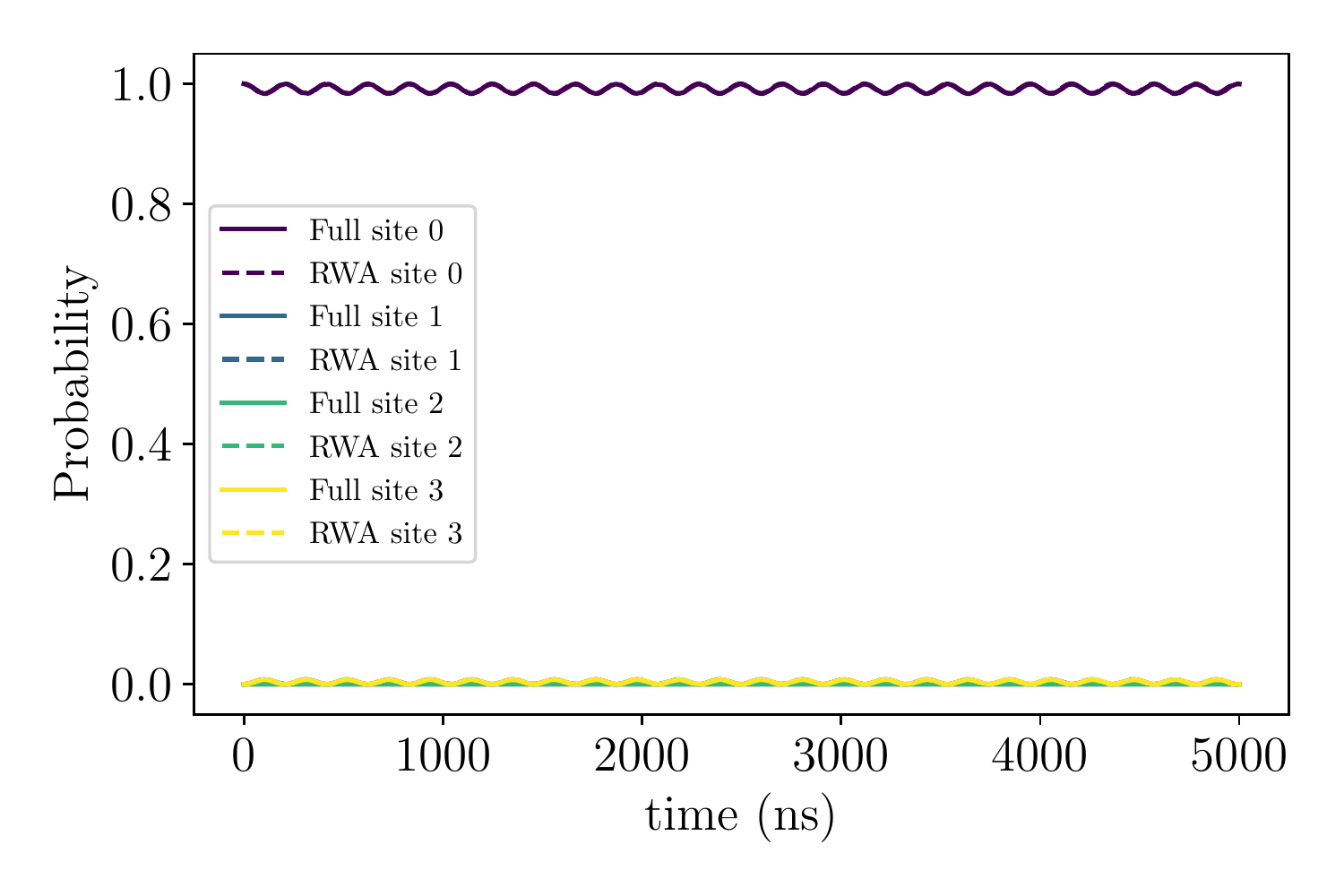}
		\caption{}
		\label{subfig:comparison_exp_theta_pi_3}
	\end{subfigure}
\caption{Comparison in probabilities as functions of experimental time between full Rydberg simulations (solid) and RWA (tight-binding) simulations (dashed) for $n=2$, $n_s = 4$. Sites 0 and 2 are centers of the two potential wells. Sites 1 and 3 are tops of the two potential barriers. Parameters used in simulation are listed in Table~\ref{table:params_n_2_ns_4}.
The 't Hooft anomaly occurs at (\subref{subfig:comparison_exp_theta_pi_1})(\subref{subfig:comparison_exp_theta_pi_2})(\subref{subfig:comparison_exp_theta_pi_3}) $\theta = \pi$.
The moment of inertia $I$ is $\omega / \tilde{\omega} = 150~\mathrm{ns}$ in all panels. The magnetic field is $B = 15~\mathrm{Gauss}$. The probabilities at sites 1 and 3 are suppressed because they are at the two maxima of the potential. The full simulation of the Rydberg atom and the simulation of the idealized RWA limit nearly overlap.}
\label{fig:comparison_exp_1}
\end{figure}

For the minimal case $n = 2$, $n_s = 4$,  Fig.~\ref{fig:comparison_exp_1} shows good agreement between the full simulation of 54 Rydberg states and the tight-binding model corresponding to the RWA. Continuum semiclassical calculations,  Eqs.~(\ref{eq:n_2_p_00}),(\ref{eq:n_2_p_01}), show that the  tunneling frequency scales as $\sim 2 \tilde{\omega} d \cos (\theta / 2)$. With a fixed potential height, the tunneling frequency decreases when $\theta$ increases from $0$ to $\pi$. From left to right, the columns of Fig.~\ref{fig:comparison_exp_1} depict this trend for $\theta =0, \pi/2,\pi$.
At $\theta = \pi$, the slow dynamics result from the exact degeneracy of the two lowest energy eigenstates in the idealized limit, reflecting the prediction of the 't Hooft anomaly.
For a fixed $\theta \neq \pi$ and a fixed $I= 150~\mathrm{ns}$ in Fig.~\ref{fig:comparison_exp_1}, the tunneling frequency decreases exponentially with $\omega$.
Although $n = 2$, $n_s = 4$ is not close to the continuum limit, qualitatively we observe that the tunneling frequency decreases from top to bottom in Fig.~\ref{fig:comparison_exp_1}.
Fitting the frequency to the simulation results (see Table~\ref{table:params_n_2_ns_4}), $\cos(\theta/2)$ behavior of the tunneling frequency is captured well even with the minimal number of sites. The remaining factor $2\tilde{\omega} d$ is not perfectly captured by the $n_s = 4$ simulation, but the decrease in $\tilde{\omega}_{\mathrm{tun}}$ from smaller to larger $\omega$ at a fixed $\theta \neq \pi$ is consistent with the behavior of $2\tilde{\omega} d$.

The simulated experimental time scale is 5 microseconds, which is a reasonable timescale over which one may expect to maintain coherent internal state dynamics, taking into account noise due to stray electric and magnetic fields, any effects of thermal motion~\cite{Brow-Analysis}, state decay, and decoherence due to blackbody radiation. For explicit comparison, the experimental synthetic Rydberg lattice results of Ref.~\cite{RydSynth} are consistent with fully coherent microwave driven dynamics over a time of 5 microseconds.

\begin{table}[t!]
\begin{center}
\small
\begin{tabular}{l|l|l|l|l l l l}
\hline\hline
~ & Parameters & $\theta$ & $\tilde{\omega}_{\mathrm{tun}}$~$(10^{-4}~\mathrm{ns}^{-1})$ & $\tilde{\omega}_{\mathrm{fast}}$~$(\mathrm{ns}^{-1})$ & $2 A_1$ & $A_2$ & $\varphi_{\mathrm{fast}}$\\
\hline
(\subref{subfig:comparison_exp_theta_0_1}) & \multirow{3}{*}{\thead{$\tilde{\omega} = 0.01~\mathrm{ns}^{-1}$, $\omega = 1.5$, \\\textit{i.e.},\\ $\Delta = 0.00375~\mathrm{ns}^{-1} = 2\pi \times 0.597~\mathrm{MHz}$,\\ $\Omega = 0.00135~\mathrm{ns}^{-1} = 2\pi \times 0.215~\mathrm{MHz}$}} & $0$ & $8.56$ & $0.0095$ & $0.91$ & $0.035$ & $-0.78$ \\ \cline{1-1} \cline{3-8}
(\subref{subfig:comparison_exp_theta_0.5pi_1}) &                        & $\pi / 2$ & $5.98 \approx 8.56 \cos(\pi/4)$ & $0.0086$ & $0.90$ & $0.055$ & $-0.085$ \\ \cline{1-1} \cline{3-8} 
(\subref{subfig:comparison_exp_theta_pi_1}) &                        & $\pi$ & $0.013 \approx 8.56 \cos(\pi/2)$ & $0.0084$ & $0.90$ & $0.10$ & $0.000$ \\ \hline
(\subref{subfig:comparison_exp_theta_0_2}) & \multirow{3}{*}{\thead{$\tilde{\omega} = 0.0133~\mathrm{ns}^{-1}$, $\omega = 2.0$,\\ \textit{i.e.},\\ $\Delta = 0.00667~\mathrm{ns}^{-1} = 2\pi \times 1.06~\mathrm{MHz}$,\\ $\Omega = 0.00135~\mathrm{ns}^{-1} = 2\pi \times 0.215~\mathrm{MHz}$}} & $0$ & $5.17$ & $0.0141$ & $0.96$ & $0.026$ & $0.006$  \\ \cline{1-1} \cline{3-8} 
(\subref{subfig:comparison_exp_theta_0.5pi_2}) &                        & $\pi / 2$ & $3.66 \approx 5.17 \cos(\pi/4)$ & $0.0140$ & $0.96$ & $0.029$ & $-0.029$ \\ \cline{1-1} \cline{3-8} 
(\subref{subfig:comparison_exp_theta_pi_2}) &                        & $\pi$ & $0.000 = 5.17 \cos(\pi/2)$ & $0.0138$ & $0.96$ & $0.038$ & $0.000$\\ \hline
(\subref{subfig:comparison_exp_theta_0_3}) & \multirow{3}{*}{\thead{$\tilde{\omega} = 0.02~\mathrm{ns}^{-1}$, $\omega = 3.0$,\\ \textit{i.e.},\\ $\Delta = 0.015~\mathrm{ns}^{-1} = 2\pi \times 2.39~\mathrm{MHz}$,\\ $\Omega = 0.00135~\mathrm{ns}^{-1} = 2\pi \times 0.215~\mathrm{MHz}$}} & $0$ & $2.40$ & $0.0304$ & $0.99$ & $0.007$ & $0.001$  \\ \cline{1-1} \cline{3-8} 
(\subref{subfig:comparison_exp_theta_0.5pi_3}) &                        & $\pi / 2$ & $1.70 \approx 2.40 \cos(\pi/4)$ & $0.0302$ & $0.99$ & $0.007$ & $-0.003$ \\ \cline{1-1} \cline{3-8} 
(\subref{subfig:comparison_exp_theta_pi_3}) &                        & $\pi$  & $0.000 = 2.40 \cos(\pi/2)$ & $0.0302$ & $0.99$ & $0.008$ & $0.000$ \\ \hline\hline
\end{tabular}
\end{center}
\caption{Parameters used in Fig.~\ref{fig:comparison_exp_1}, and results of a fit to the real-time evolution of probabilities. The probability to evolve from site $0$ back to site $0$ is modeled by $P (0, 0; t) = A_1 (1 +  \cos (\tilde{\omega}_{\mathrm{tun}} \tilde{t})) + A_2 \cos(\tilde{\omega}_{\mathrm{fast}} \tilde{t} + \varphi_{\mathrm{fast}})$, which is Eq.~(\ref{eq:n_2_p_00}) modified by an additional fast-oscillation term at frequency $\tilde{\omega}_{\mathrm{fast}} \sim 2 \tilde{\lambda} = 2 \Delta$. The fast frequency $\tilde{\omega}_{\mathrm{fast}}$ deviates from the potential curvature $\tilde{\omega}$ in the continuum limit because of the low resolution of the potential with only $n_s = 4$ sites. Fit parameters are $\tilde{\omega}_{\mathrm{tun}}$, $\tilde{\omega}_{\mathrm{fast}}$, $A_1$, $A_2$, and $\varphi_{\mathrm{fast}}$. $\tilde{\omega}_{\mathrm{tun}}$ corresponds to the DIGA result $2 \tilde{\omega}_{\mathrm{DIGA}} \cos(\theta / 2) = 4 \tilde{\omega} d \cos(\theta / 2)$ in the semiclassical  and continuum limit.
For a given value of $\omega$, with the three values $\theta = 0, \pi/2, \pi$, the fit results for $\tilde{\omega}_{\mathrm{tun}}$ exhibit close proportionality to $\cos(0/2) = 1$, $\cos(\pi/4) = 0.707$, $\cos(\pi/2) = 0$ respectively.
}
\label{table:params_n_2_ns_4}
\end{table}	

In general, for even $n$, the 't Hooft anomaly at $\theta = \pi$ is reflected in the two-fold degeneracy of all  energy eigenstates (doublets)~\cite{10.1093/ptep/ptx148}. Some energy eigenstates at $\theta = 0$, \textit{e.g.}, the ground state, are singlets. Therefore, the spectral structure is quite different between $\theta =0$ and $\theta = \pi$, which results in different real-time probability evolution. In the $n = 2$ example shown in Fig.~\ref{fig:comparison_exp_1}, there is tunneling at $\theta = 0$ but no tunneling at $\theta = \pi$.

When $n$ is odd, a global inconsistency between $\theta = 0$ and $\theta = \pi$ arises. This is a weaker condition that does not require all energy eigenstates to be doublets at $\theta = \pi$. If there is a singlet at one of the $\theta$-values ($0$ or $\pi$), then it cannot be continuously connected to a singlet at the other $\theta$-value~\cite{10.1093/ptep/ptx148}. The global inconsistency can be reflected by a singlet at $\theta = \pi$ being continuously connected to a doublet at $\theta = 0$, which is in fact what happens when $n$ is odd, explicitly shown by the example in Fig.~\ref{subfig:spectrum_n_3_1}.
It is possible for $\theta = 0$ and $\theta = \pi$ to have the same number of singlets, realized by their similar spectra related by an inversion in the energy space. As a result, the dynamics of $\theta = 0$ and $\theta = \pi$ can be very similar for odd $n$. In the case of $n = 3$, since the tunneling is not forbidden at $\theta = 0$, it will not be forbidden at $\theta = \pi$, either.

\begin{figure}[!ht]
\centering
	\begin{subfigure}[ht]{0.32\textwidth}
		\centering
		\includegraphics[width=\textwidth]{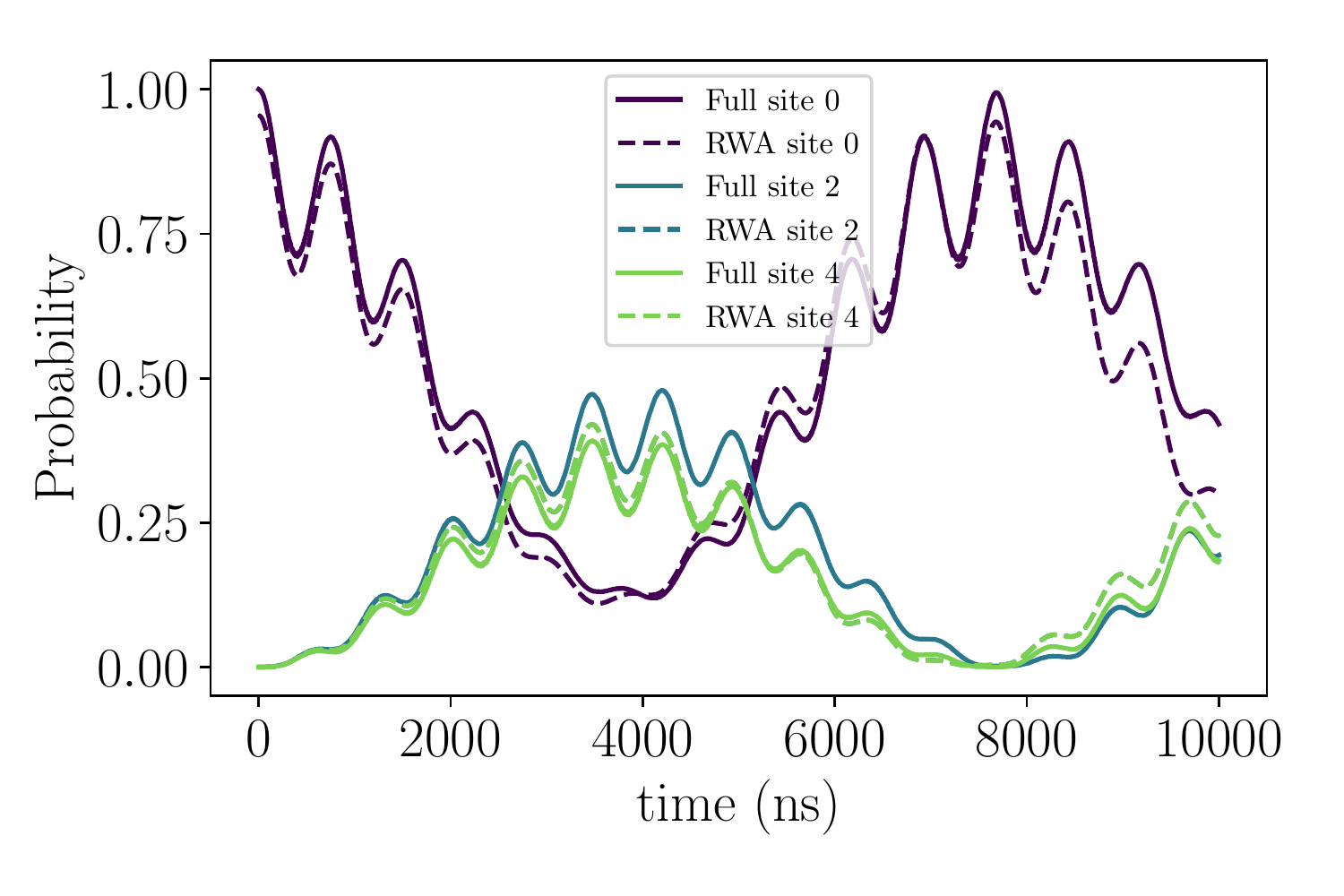}
		\caption{}
		\label{subfig:ns_6_prob_exp_theta_0_1}
	\end{subfigure}
	\hfill
	\begin{subfigure}[ht]{0.32\textwidth}
		\centering
		\includegraphics[width=\textwidth]{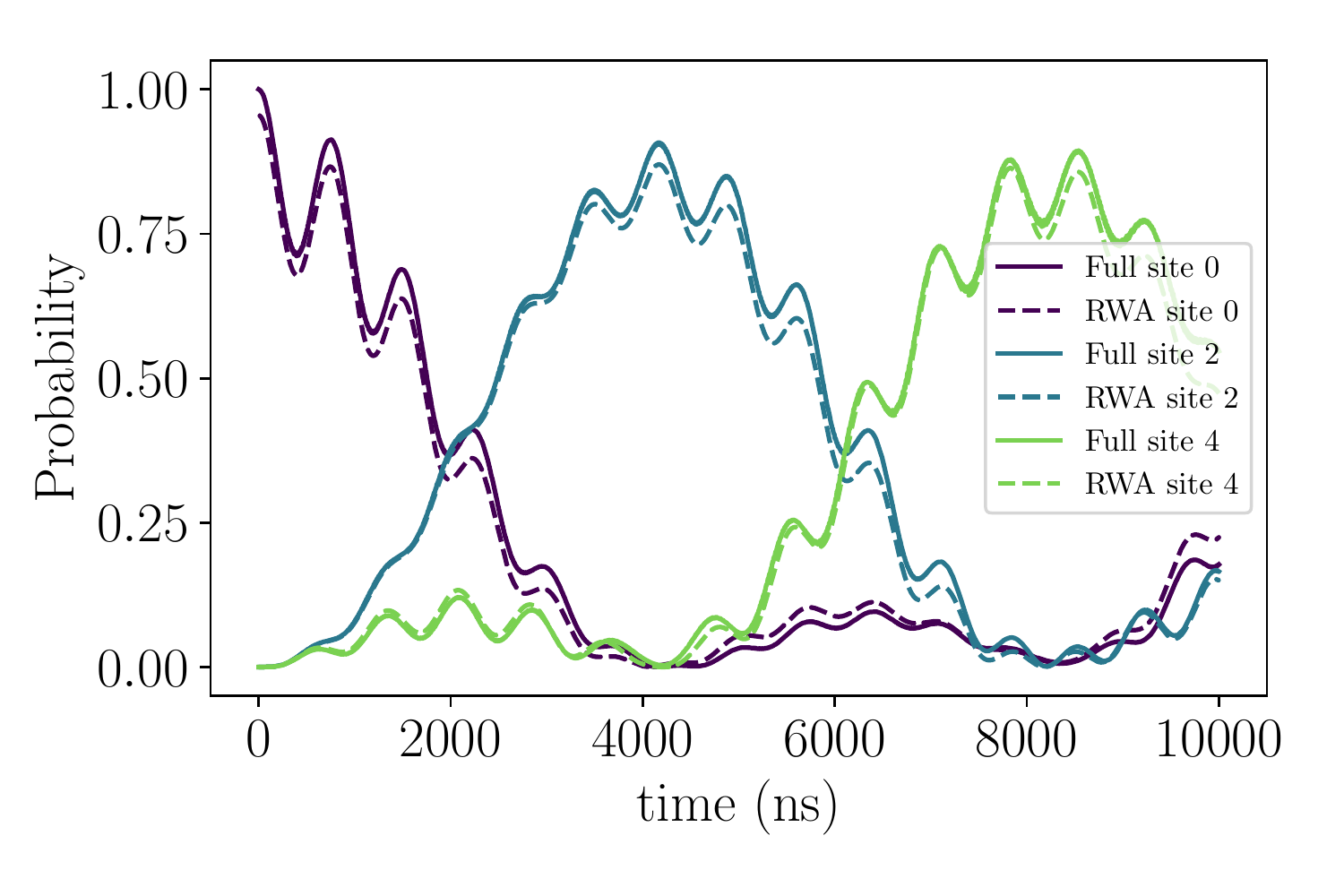}
		\caption{}
		\label{subfig:ns_6_prob_exp_theta_0.5pi_1}
	\end{subfigure}
	\hfill
	\begin{subfigure}[ht]{0.32\textwidth}
		\centering
		\includegraphics[width=\textwidth]{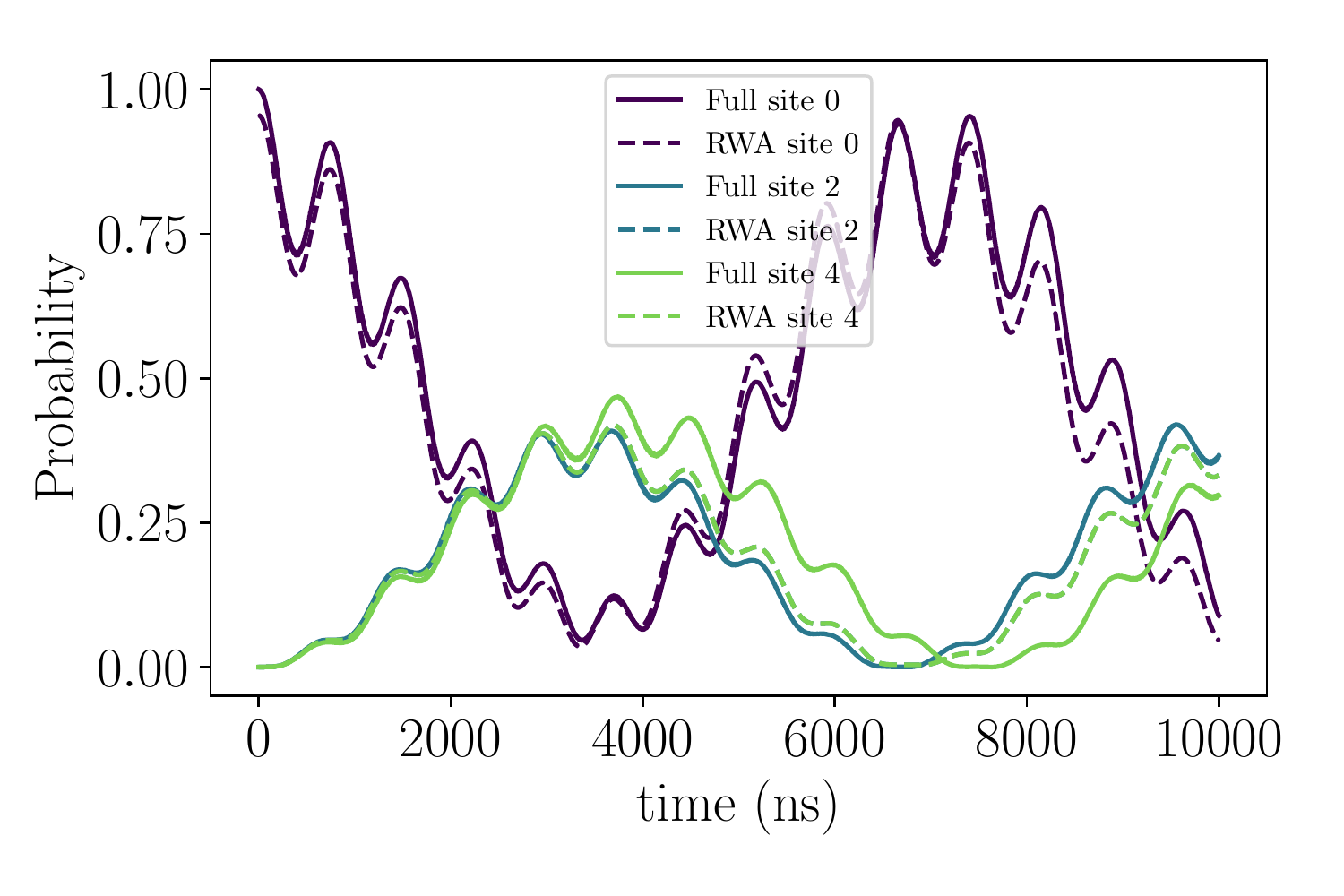}
		\caption{}
		\label{subfig:ns_6_prob_exp_theta_pi_1}
	\end{subfigure}
	\begin{subfigure}[!htbp]{0.32\textwidth}
		\centering
		\includegraphics[width=\textwidth]{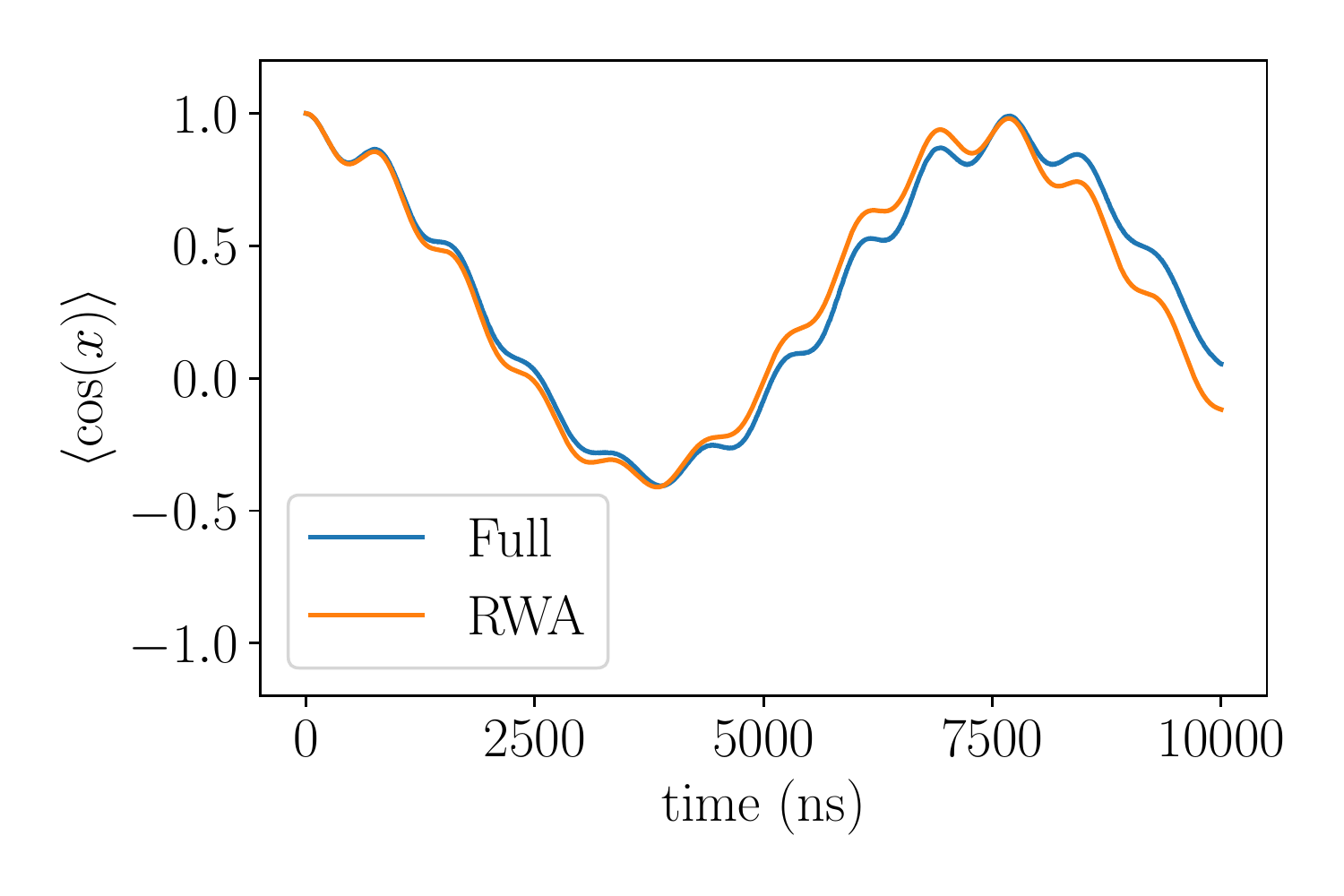}
		\caption{}
		\label{subfig:ns_6_avg_pos_cos_theta_0_1}
	\end{subfigure}
	\hfill
	\begin{subfigure}[ht]{0.32\textwidth}
		\centering
		\includegraphics[width=\textwidth]{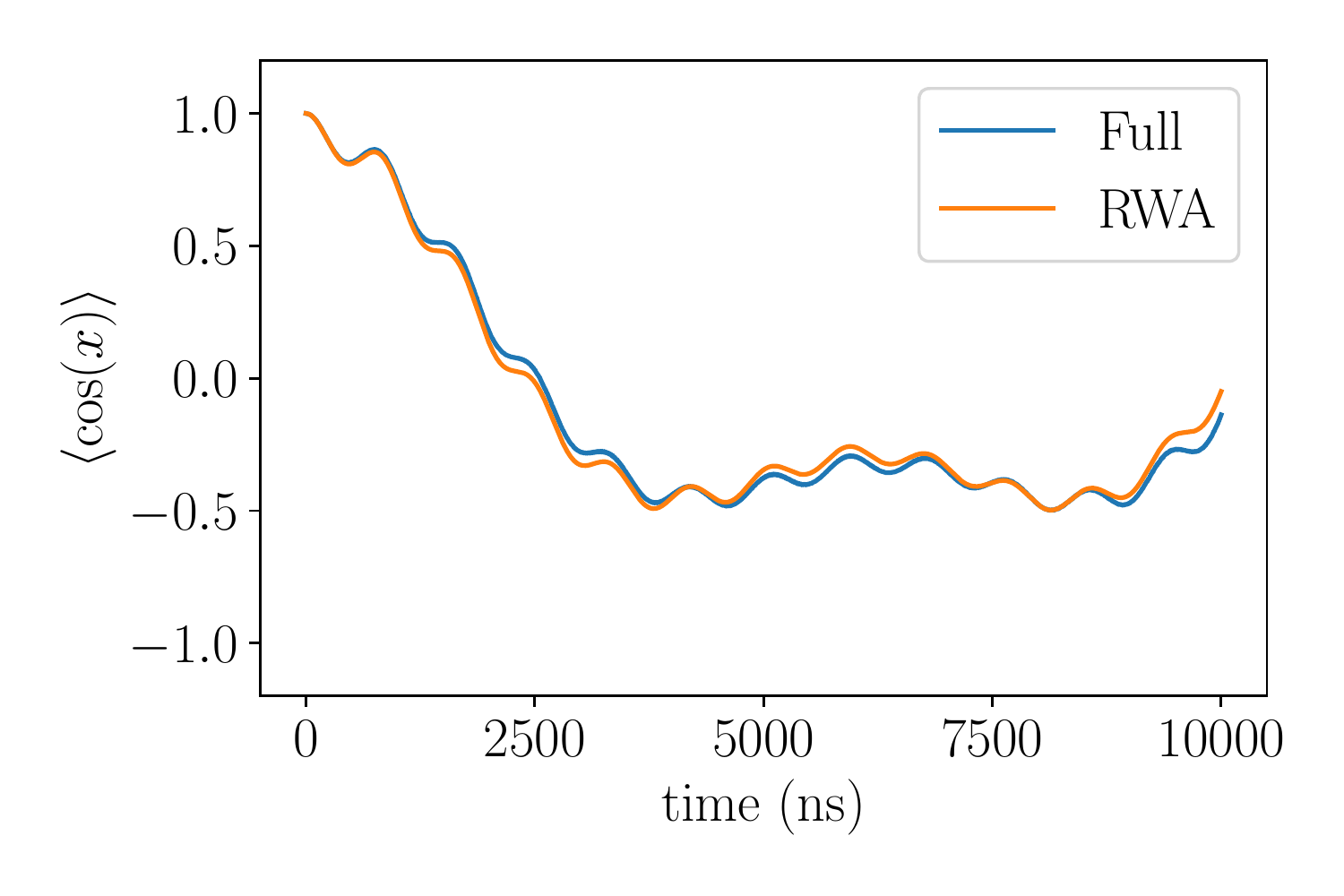}
		\caption{}
		\label{subfig:ns_6_avg_pos_cos_theta_0.5pi_1}
	\end{subfigure}
	\hfill
	\begin{subfigure}[ht]{0.32\textwidth}
		\centering
		\includegraphics[width=\textwidth]{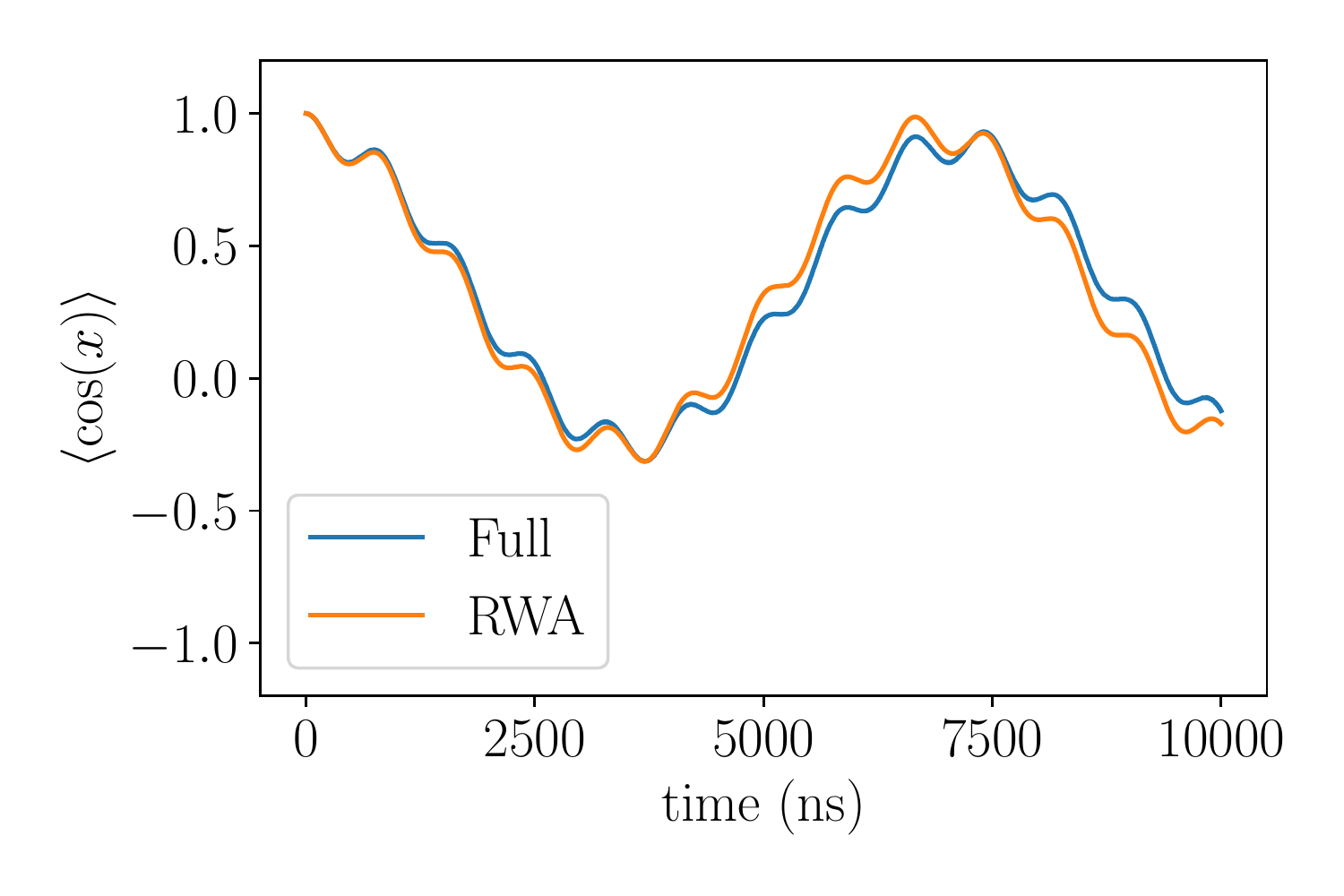}
		\caption{}
		\label{subfig:ns_6_avg_pos_cos_theta_pi_1}
	\end{subfigure}
	\begin{subfigure}[!htbp]{0.32\textwidth}
		\centering
		\includegraphics[width=\textwidth]{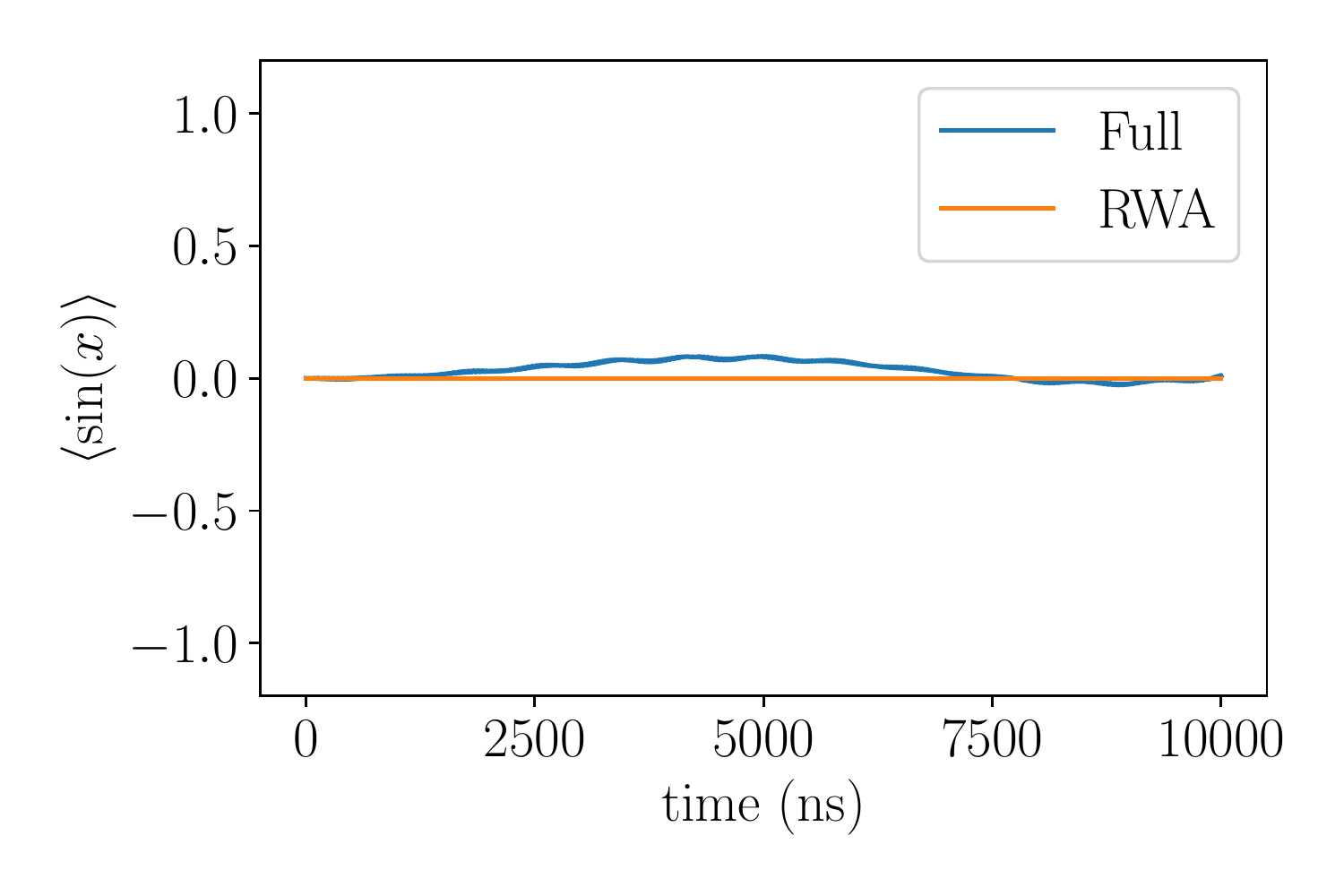}
		\caption{}
		\label{subfig:ns_6_avg_pos_sin_theta_0_1}
	\end{subfigure}
	\hfill
	\begin{subfigure}[ht]{0.32\textwidth}
		\centering
		\includegraphics[width=\textwidth]{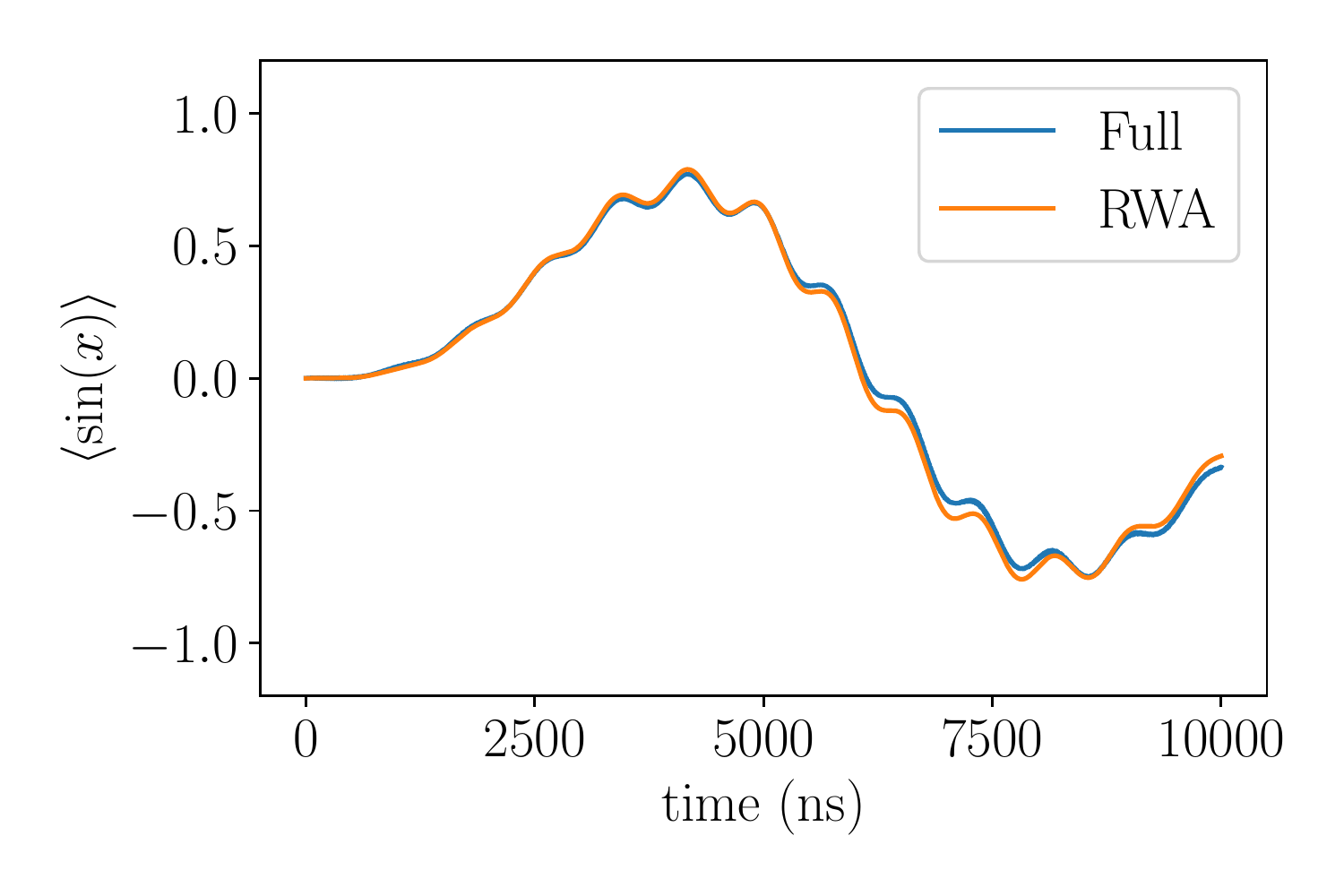}
		\caption{}
		\label{subfig:ns_6_avg_pos_sin_theta_0.5pi_1}
	\end{subfigure}
	\hfill
	\begin{subfigure}[ht]{0.32\textwidth}
		\centering
		\includegraphics[width=\textwidth]{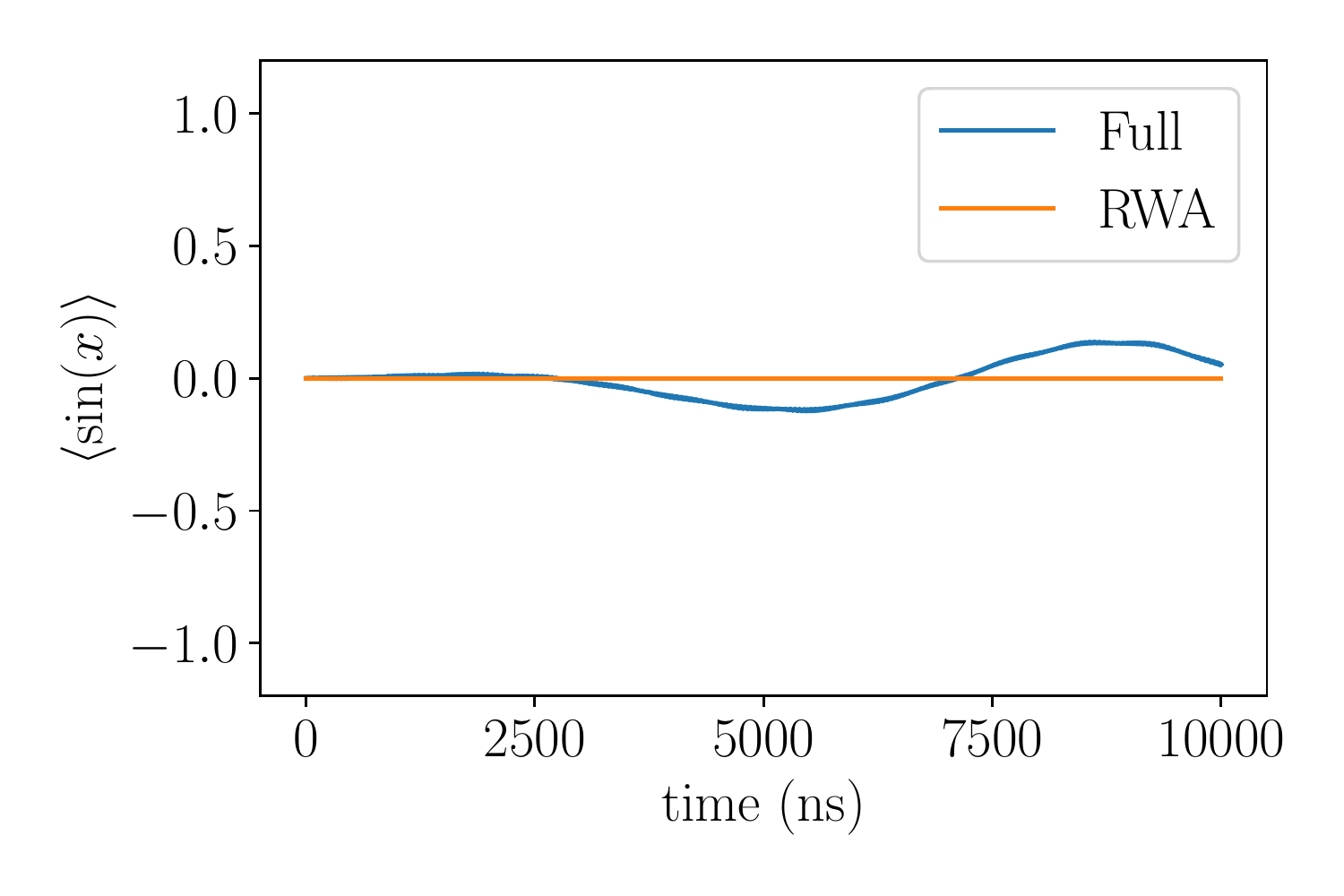}
		\caption{}
		\label{subfig:ns_6_avg_pos_sin_theta_pi_1}
	\end{subfigure}
\caption{$n=3$, $n_s = 6$. Sites 0, 2, and 4 are centers of the three potential wells.
There is a global inconsistency between (\subref{subfig:ns_6_prob_exp_theta_0_1})(\subref{subfig:ns_6_avg_pos_cos_theta_0_1})(\subref{subfig:ns_6_avg_pos_sin_theta_0_1}) $\theta = 0$ and (\subref{subfig:ns_6_prob_exp_theta_pi_1})(\subref{subfig:ns_6_avg_pos_cos_theta_pi_1})(\subref{subfig:ns_6_avg_pos_sin_theta_pi_1}) $\theta = \pi$.
Parameters are $\tilde{\omega} = 0.01~\mathrm{ns}^{-1}$, $\omega = 3.0$, i.e., $\Delta = 0.00333~\mathrm{ns}^{-1} = 2\pi \times 0.531~\mathrm{MHz}$, $\Omega = 0.00152~\mathrm{ns}^{-1} = 2\pi \times 0.242~\mathrm{MHz}$ at (\subref{subfig:ns_6_prob_exp_theta_0_1})(\subref{subfig:ns_6_avg_pos_cos_theta_0_1})(\subref{subfig:ns_6_avg_pos_sin_theta_0_1}) $\theta = 0$, (\subref{subfig:ns_6_prob_exp_theta_0.5pi_1})(\subref{subfig:ns_6_avg_pos_cos_theta_0.5pi_1})(\subref{subfig:ns_6_avg_pos_sin_theta_0.5pi_1}) $\theta = \pi / 2$,  (\subref{subfig:ns_6_prob_exp_theta_pi_1})(\subref{subfig:ns_6_avg_pos_cos_theta_pi_1})(\subref{subfig:ns_6_avg_pos_sin_theta_pi_1}) $\theta = \pi$.  The magnetic field is $B = 45~\mathrm{Gauss}$. $\theta = 0$ and $\theta = \pi$ exhibit similar dynamics in terms of position probabilities, $\expval{\cos(x)}$, and $\expval{\sin(x)}$, consistent with our expectation from the DIGA. $\theta = \pi / 2$ is a more generic value of $\theta$ which has less symmetry and its results are shown here for comparison.}
\label{fig:comparison_exp_ns_6}
\end{figure}

For $n = 3$, $n_s = 6$, the simulation in Fig.~\ref{fig:comparison_exp_ns_6} shows good agreement between the full simulation of 126 Rydberg states and the 6-site tight-binding model from the RWA. The simulated experimental time is 10 microseconds, again within a reasonable estimate of the Rydberg lifetime and coherence time. 
Continuum semiclassical calculation shows the position-space probabilities are approximately equal for $\theta = 0$ and $\theta = \pi$ at fixed $\tilde{\omega}$ and $\omega$, and this is reflected in  Figs.~\ref{subfig:ns_6_prob_exp_theta_0_1} and~\ref{subfig:ns_6_prob_exp_theta_pi_1}. 
The first and third columns of Fig.~\ref{fig:comparison_exp_ns_6} also demonstrate the similarity between $\theta = 0$ and $\theta = \pi$ via the time-dependent expectation values $\expval{\cos(x)}$ and $\expval{\sin(x)}$.
A more generic value $\theta = \pi / 2$ is  shown in the second column of Fig.~\ref{subfig:ns_6_prob_exp_theta_0.5pi_1}  as a comparison.
$\expval{\sin(x)}$ vanishes at $\theta = \pi/2$ but does not vanish at $\theta = 0$ or $\theta = \pi$.
Tunneling frequencies obtained by fitting to the time dependence of $\expval{\cos(x)}$ are shown in Table~\ref{table:params_n_3_ns_6}. The relative ratios of these frequencies at $\theta = 0, \pi / 2, \pi$ are approximately $\sqrt{3}:1:\sqrt{3}$, consistent with the DIGA prediction in Eqs.~(\ref{eq:cosx_0_pi})(\ref{eq:cosx_0.5pi}).
Unlike in the case of even $n$, tunneling is not forbidden at $\theta=\pi$.

\begin{figure}[!ht]
\centering
	\begin{subfigure}[ht]{0.32\textwidth}
		\centering
		\includegraphics[width=\textwidth]{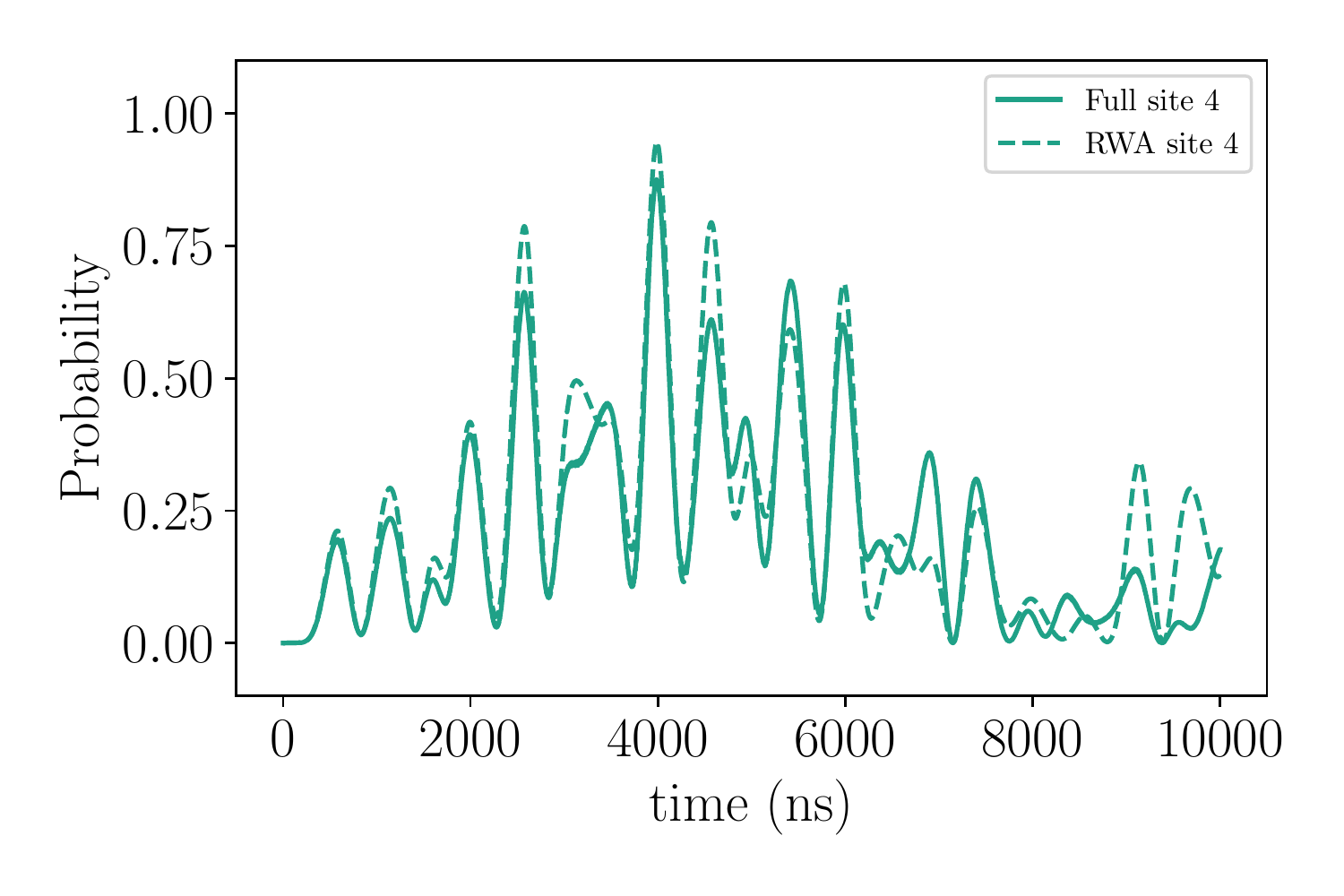}
		\caption{}
		\label{subfig:ns_8_prob_exp_theta_0_1}
	\end{subfigure}
	\hfill
	\begin{subfigure}[ht]{0.32\textwidth}
		\centering
		\includegraphics[width=\textwidth]{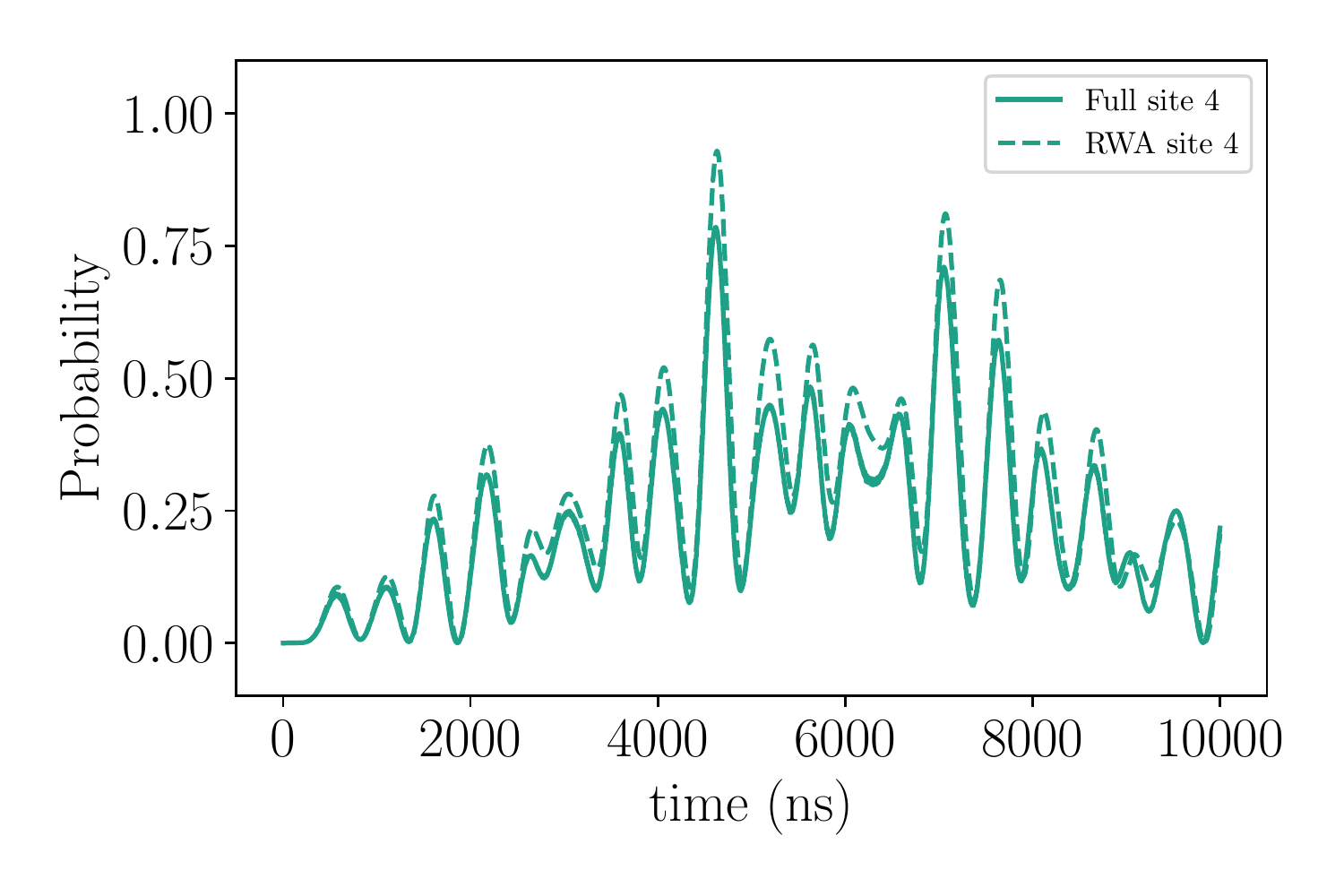}
		\caption{}
		\label{subfig:ns_8_prob_exp_theta_0.5pi_1}
	\end{subfigure}
	\hfill
	\begin{subfigure}[ht]{0.32\textwidth}
		\centering
		\includegraphics[width=\textwidth]{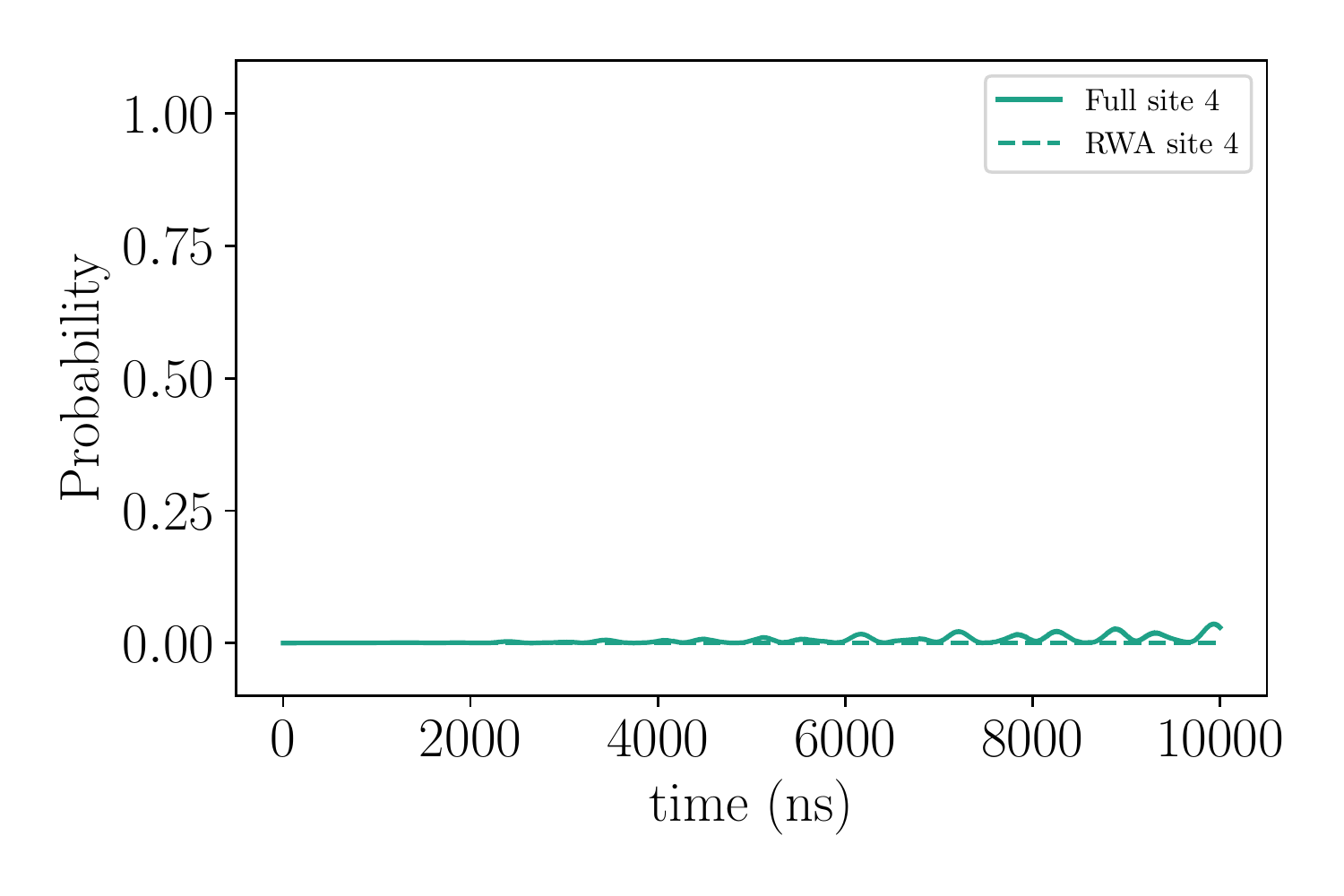}
		\caption{}
		\label{subfig:ns_8_prob_exp_theta_pi_1}
	\end{subfigure}
\caption{For $n=2$, $n_s = 8$, the probability at the potential minimum site 4, antipodal to the initial site 0.
Parameters are $\tilde{\omega} = 0.01~\mathrm{ns}^{-1}$, $\omega = 2.0$, i.e., $\Delta = 0.005~\mathrm{ns}^{-1} = 2\pi \times 0.796~\mathrm{MHz}$, $\Omega = 0.00405~\mathrm{ns}^{-1} = 2\pi \times 0.645~\mathrm{MHz}$ at (\subref{subfig:ns_8_prob_exp_theta_0_1}) $\theta = 0$, (\subref{subfig:ns_8_prob_exp_theta_0.5pi_1}) $\theta = \pi / 2$,  (\subref{subfig:ns_8_prob_exp_theta_pi_1}) $\theta = \pi$.  The magnetic field is $B = 45~\mathrm{Gauss}$. The tunneling angular frequency in the DIGA is expected to be proportional to $\cos (\theta / 2)$ given by Eqs.~(\ref{eq:n_2_p_00})(\ref{eq:n_2_p_01}). The 't Hooft anomaly occurs at (\subref{subfig:ns_8_prob_exp_theta_pi_1}) $\theta = \pi$ where the tunneling frequency vanishes.}
\label{fig:comparison_exp_ns_8}
\end{figure}

\begin{figure}[!ht]
\centering
	\begin{subfigure}[ht]{0.32\textwidth}
		\centering
		\includegraphics[width=\textwidth]{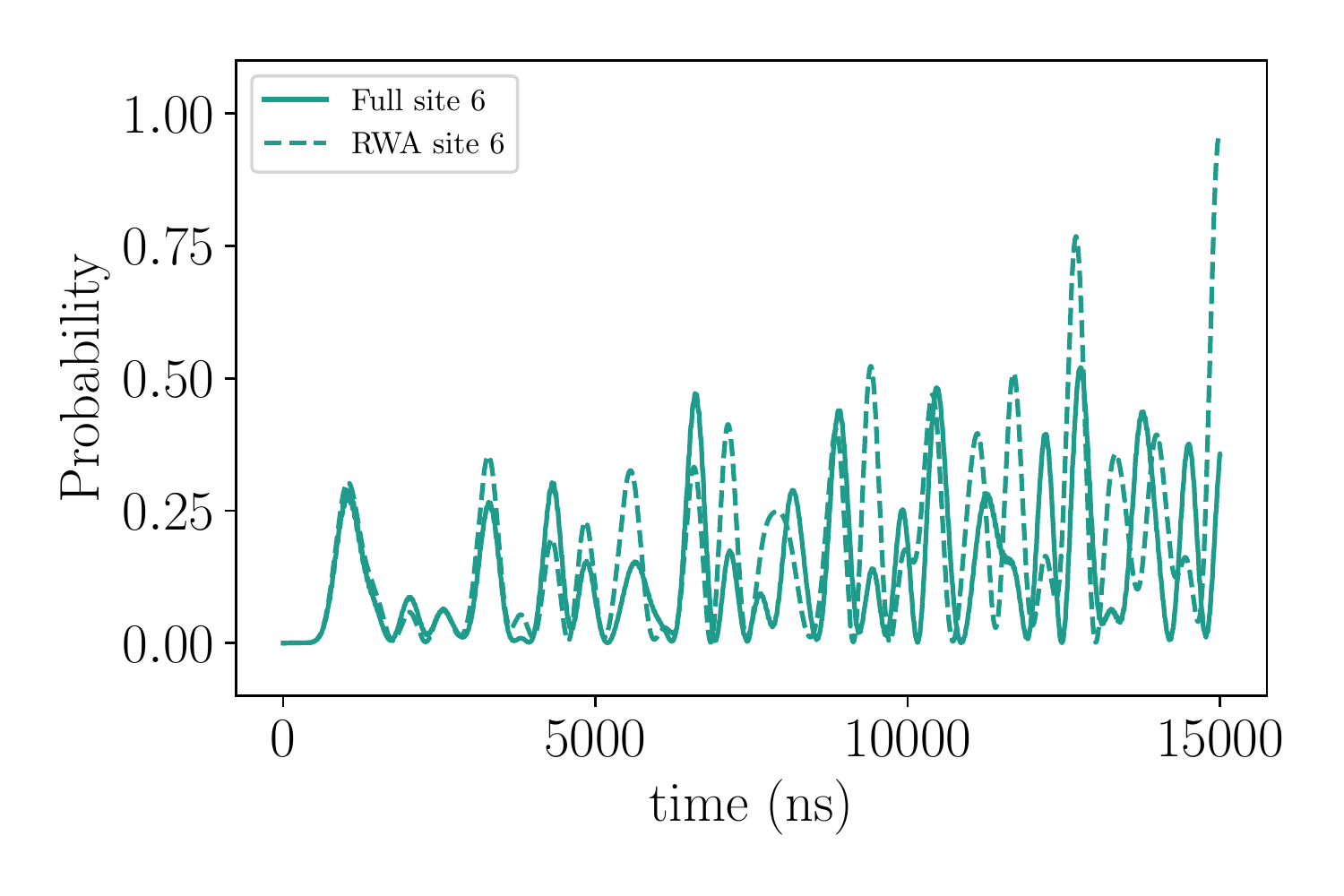}
		\caption{}
		\label{subfig:ns_12_n_2_prob_exp_theta_0_1}
	\end{subfigure}
	\hfill
	\begin{subfigure}[ht]{0.32\textwidth}
		\centering
		\includegraphics[width=\textwidth]{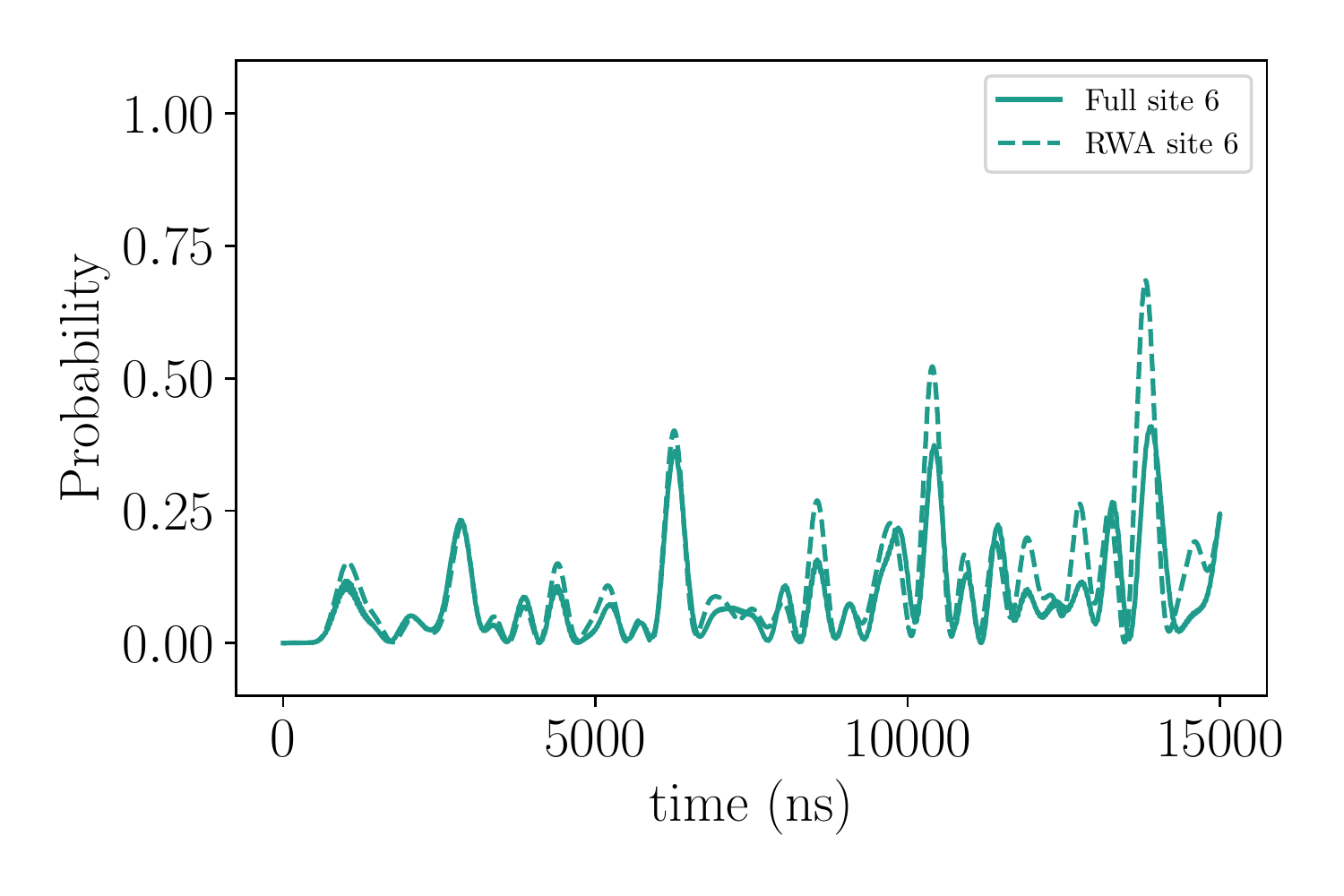}
		\caption{}
		\label{subfig:ns_12_n_2_prob_exp_theta_0.5pi_1}
	\end{subfigure}
	\hfill
	\begin{subfigure}[ht]{0.32\textwidth}
		\centering
		\includegraphics[width=\textwidth]{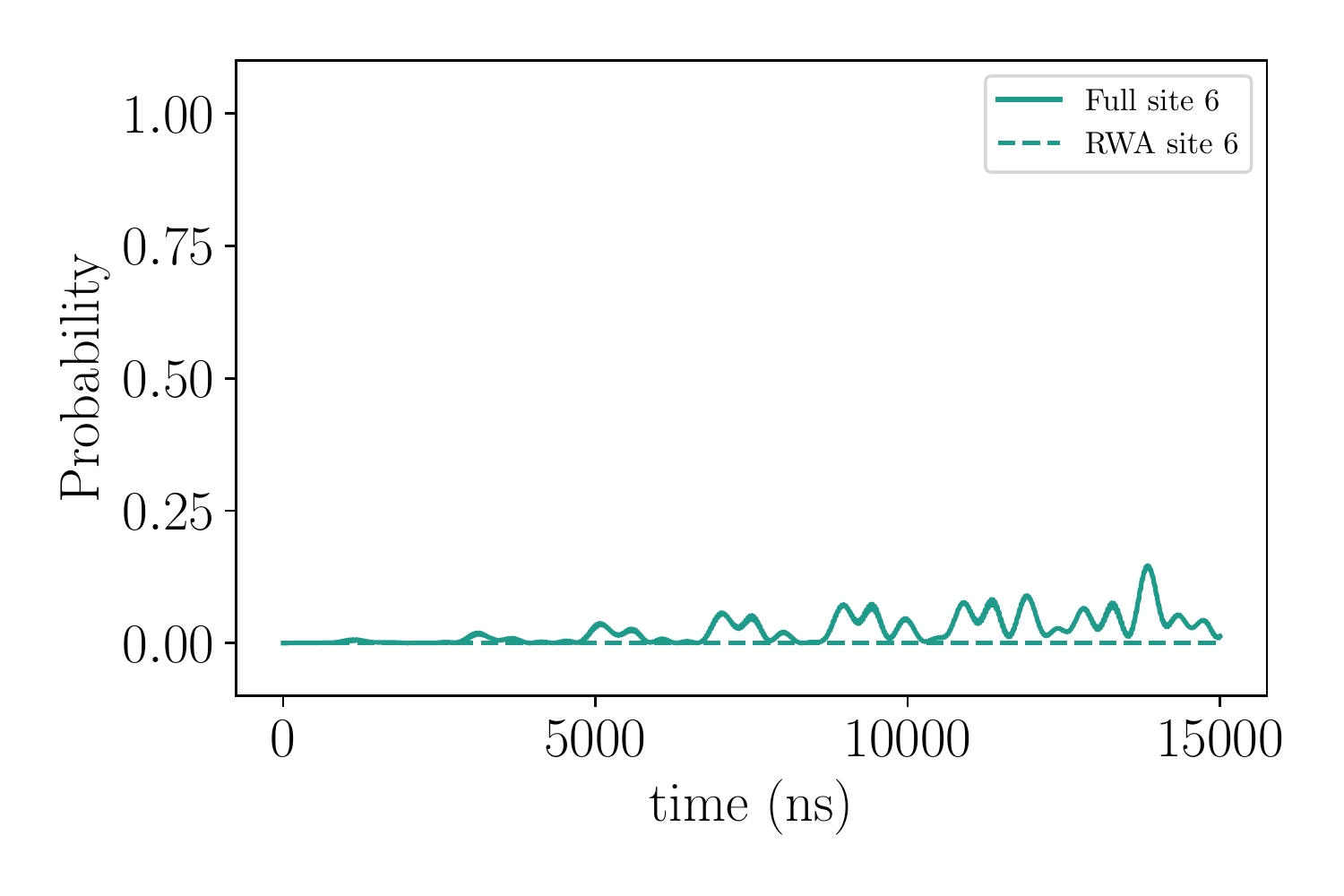}
		\caption{}
		\label{subfig:ns_12_n_2_prob_exp_theta_pi_1}
	\end{subfigure}
\caption{For $n=2$, $n_s = 12$, the probability at the potential minimum site 6, antipodal to the initial site 0.
Parameters are $\tilde{\omega} = 0.005~\mathrm{ns}^{-1}$, $\omega = 2.5$, i.e., $\Delta = 0.003125~\mathrm{ns}^{-1} = 2\pi \times 0.497~\mathrm{MHz}$, $\Omega = 0.00365~\mathrm{ns}^{-1} = 2\pi \times 0.581~\mathrm{MHz}$ at (\subref{subfig:ns_12_n_2_prob_exp_theta_0_1}) $\theta = 0$,  (\subref{subfig:ns_12_n_2_prob_exp_theta_0.5pi_1}) $\theta = \pi / 2$, (\subref{subfig:ns_12_n_2_prob_exp_theta_pi_1}) $\theta = \pi$.  The magnetic field is $B = 45~\mathrm{Gauss}$.
The tunneling angular frequency in the DIGA is expected to be proportional to $\cos (\theta / 2)$ given by Eqs.~(\ref{eq:n_2_p_00})(\ref{eq:n_2_p_01}). The 't Hooft anomaly occurs at (\subref{subfig:ns_12_n_2_prob_exp_theta_pi_1}) $\theta = \pi$ where the tunneling frequency vanishes.}
\label{fig:comparison_exp_ns_12_n_2}
\end{figure}

The continuum limit is approached by increasing $n_s$. In this work, we only consider the experimentally-simplest initial state, a Kronecker delta distribution on the lattice, \textit{i.e.}, a state corresponding to a single site. This has the effect of introducing overlap with higher frequency states as $n_s$ is increased, complicating the dynamics. Nonetheless the basic physical point can still be extracted. Using these single-site initial states, the most relevant results from simulations with $n_s = 8$, $n = 2$  are shown in Fig.~\ref{fig:comparison_exp_ns_8}, and with $n_s = 12$, $n = 2$  in Fig.~\ref{fig:comparison_exp_ns_12_n_2}. The time-dependent probability at the potential minimum antipodal to the initial site shows that the tunneling frequency decreases as $\theta$ increases from 0 to $\pi$, which is qualitatively consistent with the $\cos(\theta / 2)$ factor. 

As mentioned above the Kronecker delta initial state is not the perturbative ground state of the local harmonic-oscillator potential, so its overlap with harmonic-oscillator excited states leads to  fast oscillations which the DIGA does not capture. We show in Appendix~\ref{sec:relations} that a cleaner slow tunneling profile can be seen in the real-time dynamics if the initial state dominantly overlaps with the perturbative ground state of the local potential well. These perturbative ground states are superpositions of many different sites for large $n_s$ and may be more challenging to realize in an experiment.

\begin{table}[t!]
\begin{center}
\small
\begin{tabular}{l|l|l|l|l l l l}
\hline\hline
~ & Parameters & $\theta$ & $\tilde{\omega}_{\mathrm{tun}}$~$(10^{-4}~\mathrm{ns}^{-1})$ & $\tilde{\omega}_{\mathrm{fast}}$~$(\mathrm{ns}^{-1})$ & $3 A_1$ & $A_2$ & $\varphi_{\mathrm{fast}}$\\
\hline
(\subref{subfig:comparison_exp_theta_0_1})(\subref{subfig:ns_6_avg_pos_cos_theta_0_1})(\subref{subfig:ns_6_avg_pos_sin_theta_0_1}) & \multirow{3}{*}{\thead{$\tilde{\omega} = 0.01~\mathrm{ns}^{-1}$, $\omega = 3.0$, \\\textit{i.e.},\\ $\Delta = 0.00333~\mathrm{ns}^{-1} = 2\pi \times 0.531~\mathrm{MHz}$,\\ $\Omega = 0.00152~\mathrm{ns}^{-1} = 2\pi \times 0.242~\mathrm{MHz}$}} & $0$ & $8.20$ & $0.0073$ & $0.92$ & $0.058$ & $-0.003$ \\ \cline{1-1} \cline{3-8} 
(\subref{subfig:comparison_exp_theta_0.5pi_1})(\subref{subfig:ns_6_avg_pos_cos_theta_0.5pi_1})(\subref{subfig:ns_6_avg_pos_sin_theta_0.5pi_1}) &                        & $\pi / 2$ & $4.94 = 8.57 / \sqrt{3}$ & $0.0086$ & $0.90$ & $0.055$ & $-0.085$ \\ \cline{1-1} \cline{3-8} 
(\subref{subfig:comparison_exp_theta_pi_1})(\subref{subfig:ns_6_avg_pos_cos_theta_pi_1})((\subref{subfig:ns_6_avg_pos_sin_theta_pi_1}) &                        & $\pi$ & $8.85$ & $0.0076$ & $0.90$ & $0.070$ & $-0.051$ \\ \hline\hline
\end{tabular}
\end{center}
\caption{Parameters used in Fig.~\ref{fig:comparison_exp_ns_6}, and results of a fit to the real-time evolution of $\expval{\cos(x)}$.
$\expval{\cos(x)}$ at $\theta = 0$ and $\theta = \pi$ is modeled by $\expval{\cos(x)} = A_1 (1 + 2 \cos(\tilde{\omega}_{\mathrm{tun}} \tilde{t})) + A_2 \cos(\tilde{\omega}_{\mathrm{fast}} \tilde{t} + \varphi_{\mathrm{fast}})$, which is Eq.~(\ref{eq:cosx_0_pi}) modified by an additional fast-oscillation term at frequency $\tilde{\omega}_{\mathrm{fast}} \sim 2 \tilde{\lambda} = 2 \Delta$.
$\expval{\cos(x)}$ at $\theta = \pi / 2$ is modeled by $\expval{\cos(x)} = A_1 (2 \cos(\tilde{\omega}_{\mathrm{tun}} \tilde{t}) + \cos(2 \tilde{\omega}_{\mathrm{tun}} \tilde{t})) + A_2 \cos(\tilde{\omega}_{\mathrm{fast}} \tilde{t} + \varphi_{\mathrm{fast}})$, which is Eq.~(\ref{eq:cosx_0.5pi}) modified by an additional fast-oscillation term.
Fit parameters are $\tilde{\omega}_{\mathrm{tun}}$, $\tilde{\omega}_{\mathrm{fast}}$, $A_1$, $A_2$, and $\varphi_{\mathrm{fast}}$.
With the chosen parametrization of fit functions, $\tilde{\omega}_{\mathrm{tun}}$ at $\theta = 0$ and $\theta = \pi$ is approximately $\sqrt{3}$ times $\tilde{\omega}_{\mathrm{tun}}$ at $\theta = \pi / 2$, which is consistent with Eqs.~(\ref{eq:cosx_0_pi})(\ref{eq:cosx_0.5pi}) from DIGA.
}
\label{table:params_n_3_ns_6}
\end{table}	

We also note that in this experimental scheme the topological angle $\theta$ can be tuned continuously through tuning relative phases of microwaves. The tuning of $\theta$ is completely external, so it should be possible to vary $\theta$ adiabatically, \textit{e.g.}, from $\theta = 0$ to $\theta = \pi$ with the observation of a level crossing at $\theta = \pi/2$.
The adiabatic evolution can exhibit the ground-state degeneracy due to the 't Hooft anomaly or  global inconsistency.

\section{Conclusions and Outlook}

Quantum simulations offer an exciting opportunity  to study aspects of gauge theories that are inaccessible to classical Euclidean simulations due to sign problems. In this work we have investigated real-time phenomena connected with an 't Hooft anomaly. The anomaly arises in theories with a finite $\theta$-term, which carries its own sign problem, so these phenomena are associated with two types of sign problems.

We have focused on the real-time  quantum mechanical particle on the circle with $\mathbb{Z}_n$ symmetric potential and a $\theta$-term. This is the low energy, weak coupling limit of the massive charge-$n$ Schwinger model compactified on a small spatial circle, so in a sense it is one of the simplest gauge theory truncation exhibiting the anomaly structure of interest. The anomaly arises for even $n$ and is associated with a slow time scale for tunneling between the perturbative vacua localized in wells on opposite sides of the circle. 
We study the slow dynamics in two regimes: the semiclassical continuum limit, and in discretized models suitable for simple analog simulations.  Continuum semiclassics is useful for qualitative understanding, and also represents a target for experimental realizations. 

The discretized theories are tight-binding models, and we have studied their realization on a synthetic dimension encoded in a Rydberg atom.
Our simulations for Rydberg synthetic lattices with $n_s = 4, 6, 8, 12$ sites show that the idealized tight-binding results can be reproduced reasonably well. For the $\mathbb{Z}_2$-potential, the simulated spectrum of the idealized tight-binding model with $n_s = 4 - 2000$ converges to its continuum limit for $n_s \gtrsim O(10)$. 
Also in the  $\mathbb{Z}_2$ case, the simulated real-time dynamics exhibits a tunneling frequency between the potential minima proportional to $\cos\paren{\theta / 2}$, consistent with the semiclassical result.
At $\theta = \pi$, the tunneling rate vanishing, a dynamical manifestation of the 't Hooft anomaly.
In contrast, for a $\mathbb{Z}_3$-symmetric potential, the real-time dynamics is similar for $\theta = 0$ and $\theta = \pi$, with nonzero tunneling in both cases. This is again consistent with the semiclassical expectation and is compatible with the global inconsistency symmetry structure, which is milder than an 't Hooft anomaly.

While we have focused on the example of Rydberg synthetic lattices, the  tight-binding models may also  be realized in synthetic lattices based on ground state neutral atoms~\cite{Mancini2015,Stuhl2015,Sugawa1429} or molecules~\cite{sundar2018synthetic}. Alternatively, the 't Hooft anomaly and global inconsistency of the particle on a circle model might also be effectively realized in classical analog experiments based on optics, electrical circuits, or mechanical elements.


More generally, the work presented in this paper can be extended to study other interesting models. On the theoretical side, the dimensional reduction technique can be applied to nonabelian gauge theories in four dimensions to obtain simpler quantum mechanical models while preserving interesting physics. For example, the reduction of SU(2) gauge theory on a spatial torus $T^3$ yields a quantum mechanical model richer than the single particle on the circle~\cite{LUSCHER1983233}, and this model accurately describes the string tension to glueball mass${}^2$ ratio of the full 4D theory~\cite{Koller:1987fq,VAN_BAAL_2001}. However, since the degree of freedom in these models lives on more complicated manifolds than the circle, the realization by synthetic dimensions is less natural. A study of these models with digital quantum simulations is in progress. On the experimental side, a more natural generalization is to consider $N$ coupled quantum mechanical rotors, realized as synthetic dimensions in $N$  separate Rydberg atoms. Such a system may provide an analog simulation of the low energy limit of multiple coupled abelian gauge theories. In general, reducing higher dimensional field theories to lower dimensional theories or quantum mechanics can allow the exploration of nonperturbative phenomena from the parent theory in near-term simulations, aiding the development of observables and techniques useful for reaching longer-term goals.

\section*{Acknowledgements}

This work is supported in part by the U.S.\ Department of Energy, Office of Science, Office of High Energy Physics QuantISED program under an award for the Fermilab Theory Consortium ``Intersections of QIS and Theoretical Particle Physics''. 
C.H. and B.G. acknowledge support from the National Science Foundation under grant No.~1945031.

\begin{appendices}

\section{Discretization of the continuum Hamiltonian to the tight-binding model\label{sec:tbh_derivation}}

In this appendix, we derive the tight-binding Hamiltonian Eq.~(\ref{eq:tight-binding_ham}) by discretizing the continuum Hamiltonian Eq.~(\ref{eq:continuum_ham_1}). We discretize the space of the continuous $x$ by $x_i = i \xi$ ($i = 1, 2, \cdots, n_s$) where $\xi$ is the spacing. The kinetic part of the continuum Hamiltonian is then
\begin{equation}
    K \paren{\theta} \equiv \frac{1}{2} \paren{p - \frac{\theta}{2 \pi}}^2 = \frac{1}{2} \paren{- i \frac{\partial}{\partial x} - \frac{\theta}{2 \pi}}^2
    ,
\end{equation}
where $\theta$ is a parameter.
When $\theta = 0$, the operator $K$ in the representation of $x$ is the Laplace operator, and can be discretized to
\begin{equation}
    \mel{x_i}{K_{\xi}\paren{\theta = 0}}{x_{i'}} = - \frac{1}{2 \xi^2} \paren{\delta_{i, i'+1} - 2 + \delta_{i, i'-1}}
    .
\end{equation}
The two Kronecker-$\delta$'s can be interpreted as nearest-neighbor hopping terms. We can Fourier transform from the discrete $x$ space to the discrete momentum space by
\begin{equation}
    \ket{p_j} = \frac{1}{\sqrt{n_s}} \sum_{i=1}^{n_s} e^{ i p_j x_i} \ket{x_i}
    ,
\end{equation}
where $p_j = 2\pi j / (n_s \xi)$ ($j = 1, 2, \cdots, n_s$).
The operator $K$ in the representation of discrete $p$ is
\begin{equation}
    \mel{p_j}{K_{\xi}\paren{\theta = 0}}{p_{j'}} = \frac{1}{n_s} \sum_{i, i' = 1}^{n_s} e^{- i p_j x_i + i p_{j'} x_{i'}} \mel{x_i}{K_{\xi}\paren{\theta = 0}}{x_{i'}} = - \frac{1}{2 \xi^2} \delta_{j, j'} \paren{ e^{- i p_j \xi} + e^{i p_j \xi} - 2}
    .
\label{eq:momentum_k_op_discrete}
\end{equation}
For the case of nontrivial $\theta$, we substitute $p$ by $p - \theta / (2\pi)$ in Eq.~(\ref{eq:momentum_k_op_discrete}) and thus define $K_{\xi} (\theta)$ by
\begin{equation}
    \mel{p_j}{K_{\xi}\paren{\theta}}{p_{j'}} \equiv - \frac{1}{2 \xi^2} \delta_{j, j'} \paren{ e^{- i p_j \xi + i \frac{\theta}{2 \pi} \xi} + e^{i p_j \xi - i \frac{\theta}{2 \pi} \xi} - 2}
    .
\label{eq:momentum_k_op_discrete_theta}
\end{equation}
Then it can be fourier transformed back to the discrete $x$ representation
\begin{equation}
    \mel{x_i}{K_{\xi}\paren{\theta}}{x_{i'}} = \frac{1}{n_s} \sum_{j, j' = 1}^{n_s} e^{i p_j x_i - i p_{j'} x_{i'}} \mel{p_j}{K_{\xi}\paren{\theta}}{p_{j'}} = - \frac{1}{2 \xi^2} \paren{e^{i \frac{\theta}{2 \pi} \xi} \delta_{i, i'+1} - 2 + e^{ - i \frac{\theta}{2 \pi} \xi} \delta_{i, i'-1}}
    .
\label{eq:configuration_k_op_discrete_theta}
\end{equation}
Our discretization procedure preserves the $2 \pi$-periodicity of $\theta$, whereas directly treating the first-order term of $p$ from expanding $(p - \theta / (2 \pi))^2$ as discrete first-order derivative would break the $2 \pi$-periodicity of $\theta$.
Discretization of the potential part of the Hamiltonian in the discrete $x$ representation is
\begin{equation}
    \mel{x_i}{V_{\xi}}{x_{i'}} \equiv \delta_{i, i'} V(n x_i) = \delta_{i, i'} \lambda \paren{1 - \cos\paren{\frac{n i}{2 \pi}}}
    .
\end{equation}
Finally, we write the Hamiltonian in the form of second quantization and obtain Eq.~(\ref{eq:tight-binding_ham}).

\section{Dilute instanton gas approximation (DIGA)\label{sec:diga}}

\subsection{One-instanton contribution}

For the theory of Eq.~(\ref{eq:theory_1}), the   Euclidean equation of motion is
\begin{equation}
	\frac{d^2}{d \tau^2} x \paren{\tau} - \frac{d}{d x} V \paren{n x} = 0
	.
\end{equation}
There are instanton and anti-instanton solutions
\begin{equation}
	x_c^{\pm} \paren{\tau} = \pm \frac{4}{n} \arctan \paren{e^{n \sqrt{\lambda} \paren{\tau-\tau_0}}}
	,
\end{equation}
with initial and final conditions
\begin{equation}
	x_c^{\pm} \paren{\tau=-\infty} = 0,
\end{equation}
\begin{equation}
	x_c^{\pm} \paren{\tau=+\infty} = \pm \frac{2 \pi}{n}.
\end{equation}
$\tau_0$ is the center of the instanton/anti-instanton. Actually, compared to the instantons of the $V=0$ theory, these are like fractional instantons. 
Their classical action is
\begin{equation}
	S_E \bracket{x_c^{\pm}} = 
	\frac{8}{n} \sqrt{\lambda} \mp \frac{i \theta}{n} = \frac{8 \omega}{n^2} \mp \frac{i \theta}{n}
	,
\end{equation}
also in keeping with the fractional instanton interpretation. Here we have defined the curvature of $V (x)$ at $x = 0$ as
\begin{equation}
	V '' \paren{x = 0} = n^2 \lambda =: \omega^2
	.
\end{equation}
We denote the Euclidean action of the instanton by $S_E \bracket{x_c^{+}} = S_c$.

Quantum fluctuations around the classical stationary solution $x_c^{+}$ can also be included. At the quadratic order, contributions from the  fluctuations can be written as functional determinants and  evaluated using the Gel'fand-Yaglom method~\cite{doi:10.1063/1.1703636,10.2307/2041911,Dunne_2008}.

To regularize the functional determinant, we use the known Euclidean propagator from $x = 0$ to an arbitrary final position $x$ for the harmonic oscillator Hamiltonian $H_{\omega}$ with potential $(1/2) \omega^2 x^2$:
\begin{equation}
	\mel{x}{e^{-H_{\omega} \mathcal{T}}}{x=0} = \sqrt{\frac{\omega}{2 \pi \sinh \paren{\omega \mathcal{T}}}} e^{- \frac{1}{2} \omega x^2 \coth \paren{\omega \mathcal{T}}}
	.
\end{equation}
Then the one-instanton propagator for tunneling is
\begin{equation}
\begin{aligned}
	& \mel{x=\frac{2 \pi}{n}}{e^{-H \mathcal{T}}}{x=0}_{\text{one instanton}} \\
	= & \mel{x=0}{e^{-H_{\omega} \mathcal{T}}}{x=0} \frac{\mel{x=\frac{2 \pi}{n}}{e^{-H \mathcal{T}}}{x=0}_\text{one instanton}}{\mel{x=0}{e^{-H_{\omega} \mathcal{T}}}{x=0}} \\
	= & \sqrt{\frac{\omega}{2 \pi \sinh \paren{\omega \mathcal{T}}}} \times e^{-S_c} \bracee{\frac{\det \bracket{- \paren{d^2/d \tau^2} + V''\paren{x_c}}}{\det \bracket{- \paren{d^2/d \tau^2} + \omega^2}}}^{-1/2} \paren{1 + ...}
	.
\label{eq:one-inst_propgator}
\end{aligned}
\end{equation}
where $...$ denotes 2-loop and higher contributions, which we omit in the rest of this computation.

The NLO factor has a zero mode, while all of the other fluctuation modes have positive eigenvalues, and it should be addressed separately as follows:
\begin{equation}
\begin{aligned}
	& \bracee{\frac{\det \bracket{- \paren{d^2/d \tau^2} + V''\paren{x_c}}}{\det \bracket{- \paren{d^2/d \tau^2} + \omega^2}}}^{-1/2} \\
	\approx & \sqrt{\frac{\Re S_c}{2 \pi}} \bracee{\frac{\det' \bracket{- \paren{d^2/d \tau^2} + V''\paren{x_c}}}{\omega^{-2} \det \bracket{- \paren{d^2/d \tau^2} + \omega^2}}}^{-1/2} \int \omega d \tau_0
	,
\label{eq:NLO_factor}
\end{aligned}
\end{equation}
where the notation $\det'$ refers to the determinant with the zero mode removed, and $\mathcal{T}$ is assumed to be large. The integral $\int \omega d \tau_0$ can yield $\omega \mathcal{T}$ in the one-instanton contribution but we still keep this form for further evaluation of the DIGA result.

Then we use the Gelfand-Yaglom method to compute the ratio of determinants. To simplify the calculation, we use consider the dimensionless form
\begin{equation}
\begin{aligned}
	& \frac{\det' \bracket{- \paren{d^2/d \tau^2} + V''\paren{x_c}}}{\omega^{-2} \det \bracket{- \paren{d^2/d \tau^2} + \omega^2}} \\
	= & \frac{\det' \bracket{- \paren{d^2/d r^2} + V''\paren{x_c} / \omega^2 }}{\omega^{-2} \det \bracket{- \paren{d^2/d r^2} + 1}} \\
	= & \frac{\det' \bracket{- \paren{d^2/d r^2} + \cos\bracket{4 \arctan\paren{e^{r - r_0}}} }}{\det \bracket{- \paren{d^2/d r^2} + 1}} \\
	= & \frac{\det' \mathcal{M}}{\det \mathcal{M}^{\mathrm{free}}}
	,
\label{eq:func_det}
\end{aligned}
\end{equation}
where $r := \omega \tau$, $r_0 := \omega \tau_0$. $r_0$ does not affect the value of the determinant in the large-$\mathcal{T}$ limit and we set $r_0 = 0$ for simplicity in the following calculation. With the definition
\begin{equation}
    W \paren{r} := \cos\bracket{4 \arctan\paren{e^{r}}} - 1
    ,
\end{equation}
the differential operators are
\begin{equation}
	\mathcal{M} := - \frac{d^2}{d r^2}+ 1 + W(r)
	,
\end{equation}
\begin{equation}
	\mathcal{M}^\mathrm{free} := - \frac{d^2}{d r^2} + 1
	.
\end{equation}
Both of these operators acting on odd/even functions still give odd/even functions, so the eigenfunctions of them can be divided into odd and even sectors by appropriate initial conditions at $r = 0$. The ratio of the functional determinants can be written as
\begin{equation}
\begin{aligned}
    \frac{\det' \mathcal{M}}{\det \mathcal{M}^{\mathrm{free}}} = \frac{\det' \mathcal{M}_{\mathrm{even}}}{\det \mathcal{M}^{\mathrm{free}}_{\mathrm{even}}} \frac{\det \mathcal{M}_{\mathrm{odd}}}{\det \mathcal{M}^{\mathrm{free}}_{\mathrm{odd}}}
	,
\label{eq:func_det_even_odd}
\end{aligned}
\end{equation}
The zero mode is an even function and thus belongs to the even sector of $\mathcal{M}$.

We use the Gel'fand-Yaglom method for the odd sector and obtain
\begin{equation}
	\frac{\det \paren{\mathcal{M}_{\mathrm{odd}}}}{\det \paren{\mathcal{M}_{\mathrm{odd}}^\mathrm{free}}} = \frac{1}{2}
	.
\label{eq:func_det_odd}
\end{equation}

We remove the zero mode~\cite{PhysRevD.72.125004} in the even sector and obtain
\begin{equation}
	\frac{\det' \paren{\mathcal{M}_{\mathrm{even}}}}{\det \paren{\mathcal{M}_{\mathrm{even}}^\mathrm{free}}} = \frac{1}{2}
	.
\label{eq:func_det_even}
\end{equation}

Eqs.~(\ref{eq:one-inst_propgator})(\ref{eq:NLO_factor}) (\ref{eq:func_det}) (\ref{eq:func_det_even_odd}) (\ref{eq:func_det_odd}) (\ref{eq:func_det_even}) combined together yield
\begin{equation}
\begin{aligned}
	\mel{x=\frac{2 \pi}{n}}{e^{-H \mathcal{T}}}{x=0}_{\text{one instanton}} 
	\approx \sqrt{\frac{\omega}{2 \pi \sinh \paren{\omega \mathcal{T}}}} \times e^{-S_c} \sqrt{\frac{\Re S_c}{2 \pi}} 2 \int \omega d \tau_0
	.
\label{eq:one-inst_propgator_final}
\end{aligned}
\end{equation}

\subsection{Dilute instanton gas\label{subsec:dilute_instanton_gas}}

Multiple instantons and anti-instantons can contribute to a tunneling process. If the centers of these instantons and anti-instantons in  Euclidean time are separated well enough so that different centers $\tau_i$ and $\tau_j$ satisfy $\abs{\tau_i - \tau_j} \gg \omega^{-1}$, then the dilute instanton gas approximation is valid~\cite{Vainshtein:1981wh}.
We define the instanton density
\begin{equation}
	d := e^{-\Re S_c} \sqrt{\frac{\Re S_c}{2 \pi}} \times 2 = 2 \exp\paren{-\frac{8}{n^2} \omega} \frac{2}{n} \sqrt{\frac{\omega}{\pi}}
	.
\end{equation}
$k$ instantons and $\bar{k}$ anti-instantons with $k - \bar{k} = l \, (\mathrm{mod} \, n)$ contribute to a tunneling from $x = 0$ to $x = 2 \pi l/ n$ as
\begin{equation}
\begin{aligned}
	& \mel{x=\frac{2 \pi l}{n}}{e^{-H \mathcal{T}}}{x=0}_{\text{DIGA}} \\
	\approx & \sqrt{\frac{\omega}{2 \pi \sinh \paren{\omega \mathcal{T}}}} \sum_{\substack{k, \bar{k} \in \mathbb{N}^0 \\ k - \bar{k} = l \, \paren{\mathrm{mod} \, n}}} \frac{ \paren{ \mathcal{I} \mathcal{T} }^{k} }{k!} \frac{ \paren{ \bar{\mathcal{I}} \mathcal{T} }^{\bar{k}} }{\bar{k}!}
\end{aligned}
\label{eq:diga_general_n}
\end{equation}
for $\omega \mathcal{T} \gg 1$, where $l = 0, 1, \cdots, n-1$, $\mathcal{I} := \omega d e^{i \theta / n}$, $\bar{\mathcal{I}} := \omega d e^{-i \theta / n}$.
To evaluate the summation over $k$ and $\bar{k}$, we construct an effective Hamiltonian $H_{\mathrm{wells}}$ for $n$ potential wells, explained further in Appendix~\ref{sec:relations}.

In the semiclassical limit, the potential barriers are high enough so that the $n$ lowest perturbative states are ground states of local harmonic oscillators centered at $x = 2 \pi l/ n$ for any $l = 0, 1, \cdots, n-1$. We denote these local harmonic oscillator ground states as $\ket{l}$. For $\omega \mathcal{T} \gg 1$, we take
\begin{equation}
\begin{aligned}
	\mel{l}{e^{-H_{\mathrm{wells}} \mathcal{T}}}{0}
	= e^{-\omega \mathcal{T} / 2} \sum_{\substack{j, \bar{j} \in \mathbb{N}^0 \\ j - \bar{j} = l \, \paren{\mathrm{mod} \, n}}} \frac{ \paren{ \mathcal{I} \mathcal{T} }^{j} }{j!} \frac{ \paren{ \bar{\mathcal{I}} \mathcal{T} }^{\bar{j}} }{\bar{j}!}
	.
\end{aligned}
\label{eq:eff_propagator_general_n}
\end{equation}
The factor of $e^{- \omega \mathcal{T} / 2}$ is from the perturbative ground-state energy $\omega / 2$ and in the $\omega \mathcal{T} \gg 1$ limit, consistent with the factor $\sinh^{-1/2} (\omega \mathcal{T})$ in Eq.~(\ref{eq:diga_general_n}). We will make this point more rigorous in Appendix~\ref{sec:relations}.
The Hamiltonian matrix that yields Eq.~\ref{eq:eff_propagator_general_n} is
\begin{equation}
	H_{\mathrm{wells}} = H_0 + H_{\mathrm{DIGA}} := \frac{\omega}{2} \mathbb{I}_n + \begin{pmatrix}
	0 & - \bar{\mathcal{I}} & 0 & \cdots & 0 & - \mathcal{I}\\
	- \mathcal{I} & 0 & - \bar{\mathcal{I}} & \cdots & 0 & 0 \\
	\cdots & \cdots & \cdots & \cdots & \cdots & \cdots \\
	- \bar{\mathcal{I}} & 0 & 0 & \cdots & - \mathcal{I} & 0
	\end{pmatrix}
\label{eq:H_wells}
\end{equation}
where $\mathbb{I}_n$ is the $n \times n$ identity matrix. The $(l, 0)$-th matrix element of the $j+\bar{j}$-th power of $H_{\mathrm{DIGA}}$ in the expansion of $\exp(- H_{\mathrm{wells}} \mathcal{T})$ is
\begin{equation}
\begin{aligned}
	& \frac{1}{\paren{j + \bar{j}}!} \mel{l}{\paren{-H_{\mathrm{DIGA}} \mathcal{T}}^{j + \bar{j}} }{0} \\
	= & \frac{\binom{j+\bar{j}} {j}}{\paren{j + \bar{j}}!} 
	\paren{ \mathcal{I} \mathcal{T} }^{j}\paren{ \bar{\mathcal{I}} \mathcal{T} }^{\bar{j}} \\
	= & \frac{ \paren{ \mathcal{I} \mathcal{T} }^{j} }{j!} \frac{ \paren{ \bar{\mathcal{I}} \mathcal{T} }^{\bar{j}} }{\bar{j}!}
	,
\end{aligned}
\end{equation}
where $l = j - \bar{j} \, (\mathrm{mod} \, n)$.

Eq.~(\ref{eq:H_wells}) is a tridiagonal matrix with two additional corners and can be regarded as an $n$-site tight-binding model. To diagonalize $H_{\mathrm{DIGA}}$, we consider the discrete Fourier transform of the basis $\{ \ket{l} \}_{l = 0, 1, \cdots, n-1}$
\begin{equation}
	\ket{k} = \frac{1}{\sqrt{n}} \sum_{l = 0}^{n-1} e^{- i 2 \pi k l / n} \ket{l}
	,\footnote{{This Fourier transform has a minus sign difference relative to the exponent from the definition of Bloch states, so the state $\ket{k}$ has crystal momentum  $-k$.}}
\end{equation}
where $k = 0, 1, 2, \cdots, n-1$. We can check that $\ket{k}$ is an energy eigenstate from
\begin{equation}
	H_{\mathrm{DIGA}} \ket{k} = \paren{- e^{i 2 \pi k / n} \mathcal{I} - e^{- i 2 \pi k / n} \bar{\mathcal{I}} } \ket{k} = - 2 \omega d \cos \paren{\frac{2 \pi k + \theta}{n}} \ket{k}
	.
\end{equation}
The spectrum of $H_0 + H_{\mathrm{DIGA}}$ is
\begin{equation}
	E_k \paren{\theta} = \frac{\omega}{2} - 2 \omega d \cos \paren{\frac{2 \pi k + \theta}{n}}
	.
\end{equation}
This result is Eq.~(\ref{eq:diga_spectrum}).

At a real time $t$, the probability amplitude of hopping from well $l$ to well $l'$ under the DIGA is
\begin{equation}
\begin{aligned}
	& \mel{l'}{e^{- i H_{\mathrm{DIGA}} t}}{l} \\
	= & \frac{1}{n} \sum_{k = 0}^{n-1} e^{i 2 \pi k \paren{l - l'} / n} e^{- i E_k \paren{\theta} t} \\
	= & \frac{1}{n} \sum_{k = 0}^{n-1} \exp \bracket{i \frac{2 \pi k}{n} \paren{l - l'} + i 2 \omega d \cos \paren{\frac{2 \pi k + \theta}{n}} t}
	.
\end{aligned}
\end{equation}

For $n = 2$, the probability from well $0$ to itself is
\begin{equation}
\begin{aligned}
    P_{\theta} \paren{0, 0; t} = \cos^2 \paren{2 \omega d \cos \paren{\frac{\theta}{2}} t }
    ,
\end{aligned}
\end{equation}
and the probability from hopping from well $0$ to $1$ is
\begin{equation}
\begin{aligned}
    P_{\theta} \paren{0, 1; t} = \sin^2 \paren{2 \omega d \cos \paren{\frac{\theta}{2}} t }
    .
\end{aligned}
\end{equation}
The tunneling frequency is $2 \omega d \cos (\theta / 2) = \omega_{\mathrm{DIGA}} \cos (\theta / 2)$. At $\theta = \pi$, the tunneling is highly suppressed.

For $n = 3$, at high-symmetry points $\theta = 0$ and $\theta = \pi$, the probabilities from well $0$ to wells $0$, $1$, $2$ are the same under DIGA:
\begin{equation}
\begin{aligned}
    P_{\theta = 0} \paren{0, 0; t} =  P_{\theta = \pi} \paren{0, 0; t} = 1 - \frac{8}{9} \sin^2 \paren{\frac{3}{2} \omega d t }
    ,
\end{aligned}
\end{equation}
\begin{equation}
\begin{aligned}
    P_{\theta = 0} \paren{0, 1; t} = P_{\theta = \pi} \paren{0, 1; t} = \frac{4}{9} \sin^2 \paren{\frac{3}{2} \omega d t }
    ,
\end{aligned}
\end{equation}
\begin{equation}
\begin{aligned}
    P_{\theta = 0} \paren{0, 2; t} = P_{\theta = \pi} \paren{0, 2; t} = \frac{4}{9} \sin^2 \paren{\frac{3}{2} \omega d t }
    .
\end{aligned}
\end{equation}
Both $\theta = 0$ and $\theta = \pi$ have a symmetry between $P(0, 1; t)$ and $P(0, 2; t)$, \textit{i.e.}, probabilities of tunneling counterclockwise and clockwise, a parity symmetry in dynamics.
However, at a generic value of $\theta$, for example, $\theta = \pi / 2$, this parity symmetry in dynamics is absent, as is shown explicitly by
\begin{equation}
\begin{aligned}
    P_{\theta = \pi/2} \paren{0, 0; t} = \frac{1}{9} \paren{1 + 2 \cos \paren{ \sqrt{3} \omega d t }}^2
    ,
\end{aligned}
\end{equation}
\begin{equation}
\begin{aligned}
    P_{\theta = \pi/2} \paren{0, 1; t} = \frac{16}{9} \sin^2 \paren{\frac{\sqrt{3}}{2} \omega d t } \cos^2 \paren{\frac{\sqrt{3}}{2} \omega d t - \frac{\pi}{6}}
    ,
\end{aligned}
\end{equation}
\begin{equation}
\begin{aligned}
    P_{\theta = \pi/2} \paren{0, 2; t} = \frac{16}{9} \sin^2 \paren{\frac{\sqrt{3}}{2} \omega d t } \cos^2 \paren{\frac{\sqrt{3}}{2} \omega d t + \frac{\pi}{6}}
    .
\end{aligned}
\end{equation}
The expectation values of $\cos(x)$ and $\sin(x)$ in real-time evolution from site $0$ are thus
\begin{equation}
    \expval{\cos\paren{x}}_{\theta = 0} = \expval{\cos\paren{x}}_{\theta = \pi} = \frac{1}{3} \paren{1 + 2 \cos\paren{3 \omega d t}}
    ,
\label{eq:cosx_0_pi}
\end{equation}
\begin{equation}
    \expval{\sin\paren{x}}_{\theta = 0} = \expval{\sin\paren{x}}_{\theta = \pi} = 0
    ,
\label{eq:sinx_0_pi}
\end{equation}
\begin{equation}
    \expval{\cos\paren{x}}_{\theta = \pi/2} = \frac{1}{3} \paren{2 \cos\paren{\sqrt{3} \omega d t} + \cos\paren{2 \sqrt{3} \omega d t}}
    ,
\label{eq:cosx_0.5pi}
\end{equation}
\begin{equation}
    \expval{\sin\paren{x}}_{\theta = \pi/2} = \frac{1}{3} \paren{2 \sin\paren{\sqrt{3} \omega d t} - \sin\paren{2 \sqrt{3} \omega d t}}
    .
\label{eq:sinx_0.5pi}
\end{equation}

\section{Relations between propagators and probability densities\label{sec:relations}}

In this appendix, we describe how to obtain the time-dependent probability density near the potential well $l$ with a given initial state $\ket{\psi}$ from the propagator $\mel{l}{\exp(- H \mathcal{T})}{0}$.

With energy eigenstates denoted as $\ket{\varepsilon_i}$, the spectral decomposition of the Euclidean propagator from $x = 0$ to $x = 2 \pi l / n$ and the transition amplitude from $\ket{\psi}$ to $x = 2 \pi l / n$ are
\begin{equation}
    \mel{x=2 \pi l/n}{e^{-H \mathcal{T}}}{x = 0} = \sum_i e^{- \varepsilon_i \mathcal{T}} \braket{x=2 \pi l/n}{\varepsilon_i} \braket{\varepsilon_i}{x = 0}
    ,
\label{eq:propagator_0_l}
\end{equation}
\begin{equation}
    \mel{x=2 \pi l/n}{e^{-H \mathcal{T}}}{\psi} = \sum_i e^{- \varepsilon_i \mathcal{T}} \braket{x=2 \pi l/n}{\varepsilon_i} \braket{\varepsilon_i}{\psi}
    .
\label{eq:propagator_psi_l}
\end{equation}
We take the DIGA result for the propagator from
Eq.~(\ref{eq:diga_general_n}) and expand the $\sinh^{-1/2}(\omega \mathcal{T})$ function for large $\omega \mathcal{T}$. We obtain
\begin{equation}
    \mel{x=2 \pi l/n}{e^{-H \mathcal{T}}}{x = 0}_{\mathrm{DIGA}} = \sqrt{\frac{\omega}{2 \pi}} \sum_{\substack{k, \bar{k} \in \mathbb{N}^0 \\ k - \bar{k} = l \, \paren{\mathrm{mod} \, n}}} \frac{ \paren{ \mathcal{I} \mathcal{T} }^{k} }{k!} \frac{ \paren{ \bar{\mathcal{I}} \mathcal{T} }^{\bar{k}} }{\bar{k}!} \sum_{j} \sqrt{2} e^{- \omega \mathcal{T} / 2} a_j e^{- 2 j \omega \mathcal{T}}
    ,
\label{eq:propagator_diga_sinh_expanded}
\end{equation}
where
\begin{equation}
    \frac{1}{\sqrt{\sinh \omega \mathcal{T}}} = \sqrt{2} e^{- \omega \mathcal{T} / 2} \sum_{j} a_j e^{- 2 j \omega \mathcal{T}}  = \sqrt{2} e^{- \omega \mathcal{T} / 2} \paren{1 + \frac{1}{2} e^{- 2 \omega \mathcal{T}} + \frac{3}{8} e^{- 4 \omega \mathcal{T}} + \frac{5}{16} e^{- 6 \omega \mathcal{T}} + \frac{35}{128} e^{- 8 \omega \mathcal{T}} + ...}
    ,
\end{equation}
with
\begin{equation}
    a_j = \frac{\paren{2 j}!}{4^j \paren{j!}^2}
    .
\end{equation}
In the semiclassical limit, energy eigenstates are linear combinations of eigenstates with the same perturbative energy of harmonic oscillators centered at all potential wells and satisfy $\abs{\braket{x = 0}{\varepsilon_i}} = \abs{\braket{x = 2 \pi / n}{\varepsilon_i}} = ... = \abs{\braket{x = 2 \pi (n - 1) / n}{\varepsilon_i}}$. By comparing Eq.~(\ref{eq:propagator_0_l}) and Eq.~(\ref{eq:propagator_diga_sinh_expanded}), for the lowest $n$ states $\ket{\varepsilon_i}$ ($i = 0, 1, ..., n - 1$) linear combinations of the the harmonic oscillator ground states at $n$ potential wells, we obtain
\begin{equation}
    \abs{\braket{x = 0}{\varepsilon_i}} = \abs{\braket{x = 2 \pi l / n}{\varepsilon_i}} = \paren{\frac{\omega}{\pi}}^{1/4} \sqrt{\frac{a_0}{n}}
    ,
\label{eq:amplitude_varepsilon}
\end{equation}
where the factor of $1/\sqrt{n}$ is because of $n$ exponential functions from the sum over $k, \bar{k}$ with $n$ frequencies differing by the nonperturbative tunneling scale $O(\abs{\mathcal{I}})$. This result is also consistent with the value of ground state wavefunction evaluated at the center of the quadratic potential.

Using Eqs.~(\ref{eq:propagator_0_l})(\ref{eq:propagator_psi_l})(\ref{eq:propagator_diga_sinh_expanded})(\ref{eq:amplitude_varepsilon}) and choosing $\ket{\psi} \approx \ket{\varepsilon_{\mathrm{HO}, 0, 0}}$, the ground state wavefunction of the harmonic oscillator near $x = 0$, we obtain
\begin{equation}
\begin{aligned}
    & \mel{x=2 \pi l/n}{e^{-H \mathcal{T}}}{\psi} \\
    \approx & \sum_i e^{- \varepsilon_i \mathcal{T}} \braket{x=2 \pi l/n}{\varepsilon_i} \braket{\varepsilon_i}{\varepsilon_{\mathrm{HO}, 0, 0}} \\
    = & \paren{\frac{\omega}{\pi}}^{1/4} \sum_{\substack{k, \bar{k} \in \mathbb{N}^0 \\ k - \bar{k} = l \, \paren{\mathrm{mod} \, n}}} \frac{ \paren{ \mathcal{I} \mathcal{T} }^{k} }{k!} \frac{ \paren{ \bar{\mathcal{I}} \mathcal{T} }^{\bar{k}} }{\bar{k}!} \sqrt{n} e^{- \omega \mathcal{T} / 2}
    ,
\end{aligned}
\end{equation}
where we have used $\braket{\varepsilon_i}{x = 0} \approx (\omega / \pi)^{1/4} \sqrt{a_0 / n} \braket{\varepsilon_i}{\psi}$ for $\ket{\varepsilon_i}$ that is a linear combination of harmonic oscillator ground states.

The probability density from $\ket{\psi}$ to $x = 2 \pi l / n$ for a real time $T$ is then
\begin{equation}
    \rho \paren{\psi, \pi; T} = \bracket{\mel{x=2 \pi l/n}{e^{-H \mathcal{T}}}{\psi} \mel{x=2 \pi l/n}{e^{H \mathcal{T}}}{\psi}}_{\mathcal{T} \rightarrow i T}
\end{equation}
which yields the same $T$-dependence as from using $H_{\mathrm{wells}}$ (also with $\mathcal{T} \rightarrow i T$) in Appendix~\ref{subsec:dilute_instanton_gas}.

In addition to the nonperturbative frequency at order $O(\abs{\mathcal{I}})$ in real-time dynamics, there are also perturbative frequencies approximately at order $O(\omega)$ in the semiclassical limit due to the separations of the harmonic oscillator energy levels. When the initial wavefunction takes the form $\propto ((1 + \cos (x_i))/2) ^ {2 \alpha} \approx 1 - \alpha x_i^2 / 2$ and $\alpha \approx \omega$, then this wavefunction gets close to the ground state wavefunction $\propto \exp (- \omega x_i^2 / 2) \approx 1 - \omega x_i^2 / 2$ of the harmonic oscillator with frequency $\omega$. When $\alpha$ gets close to $\omega$, we would expect that the initial wavefunction is approximately the same as a ground eigenstate in one of the potential wells, and then the dominant phenomenon in real-time dynamics will be the nonperturbative tunneling described by the DIGA in the semiclassical limit.

\begin{figure}[!t]
\centering
	\begin{subfigure}[ht]{0.48\textwidth}
		\centering
		\includegraphics[width=\textwidth]{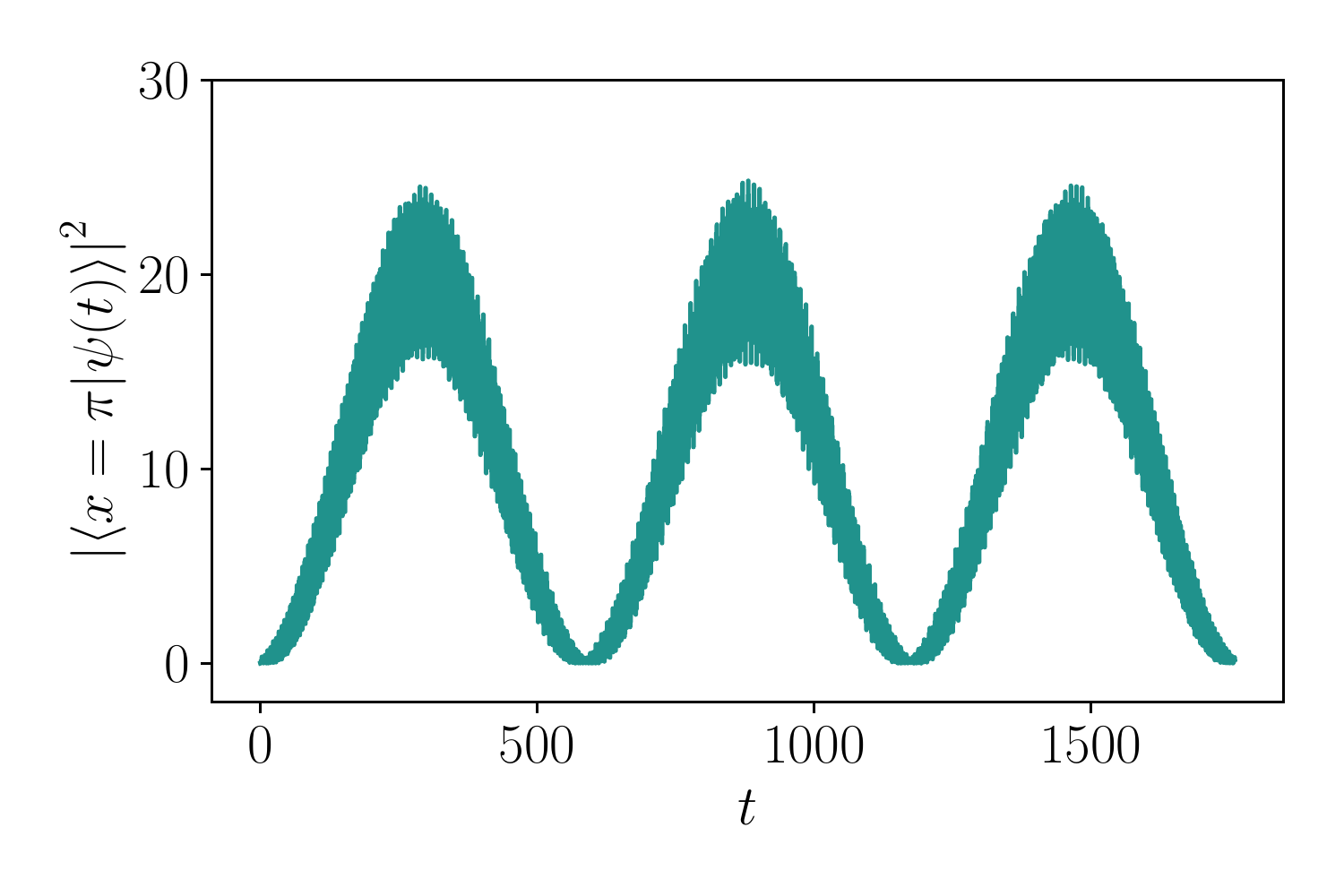}
		\caption{}
		\label{subfig:vary_gauss_width_1}
	\end{subfigure}
	\hfill
	\begin{subfigure}[ht]{0.48\textwidth}
		\centering
		\includegraphics[width=\textwidth]{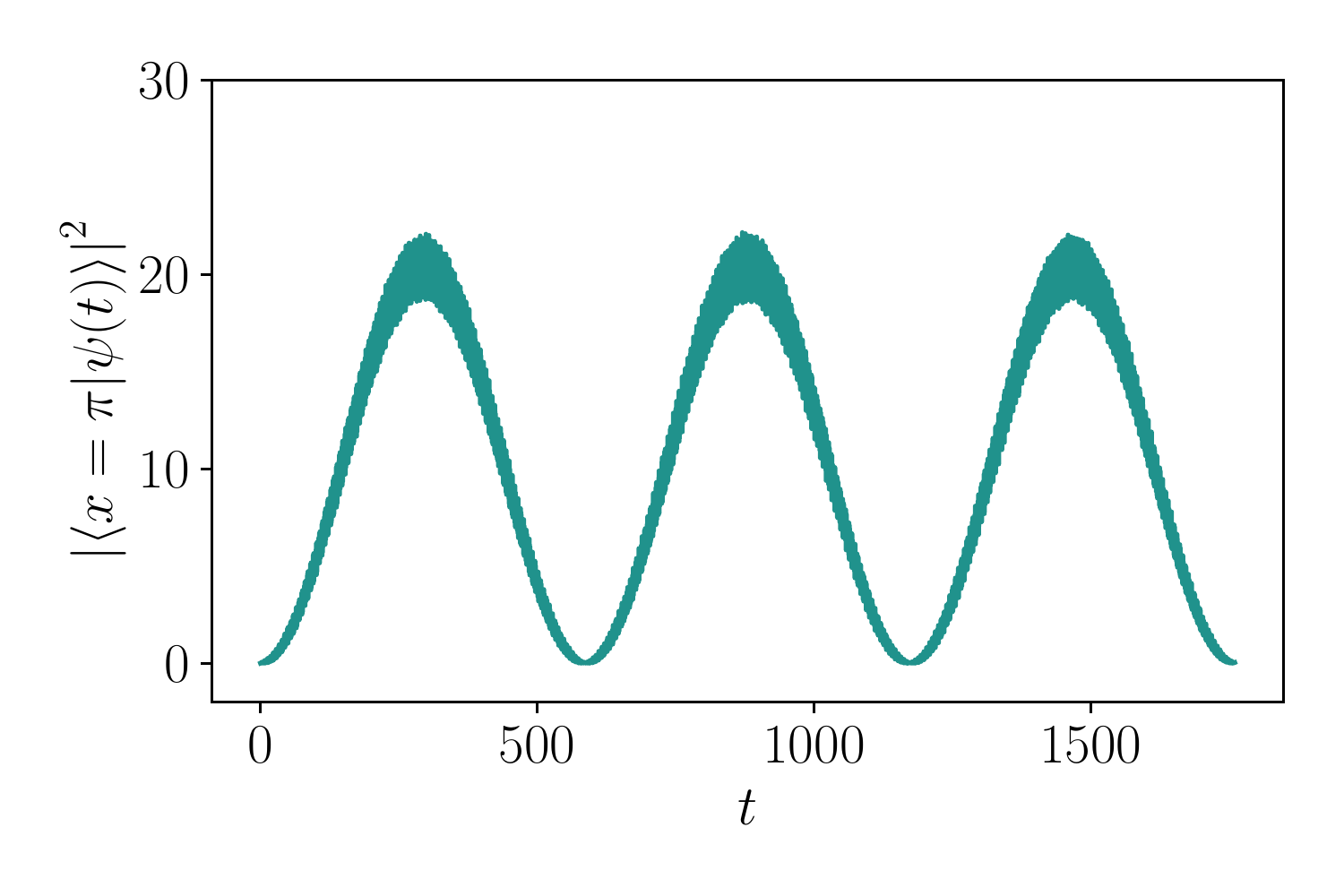}
		\caption{}
		\label{subfig:vary_gauss_width_2}
	\end{subfigure}
	\hfill
	\begin{subfigure}[ht]{0.48\textwidth}
		\centering
		\includegraphics[width=\textwidth]{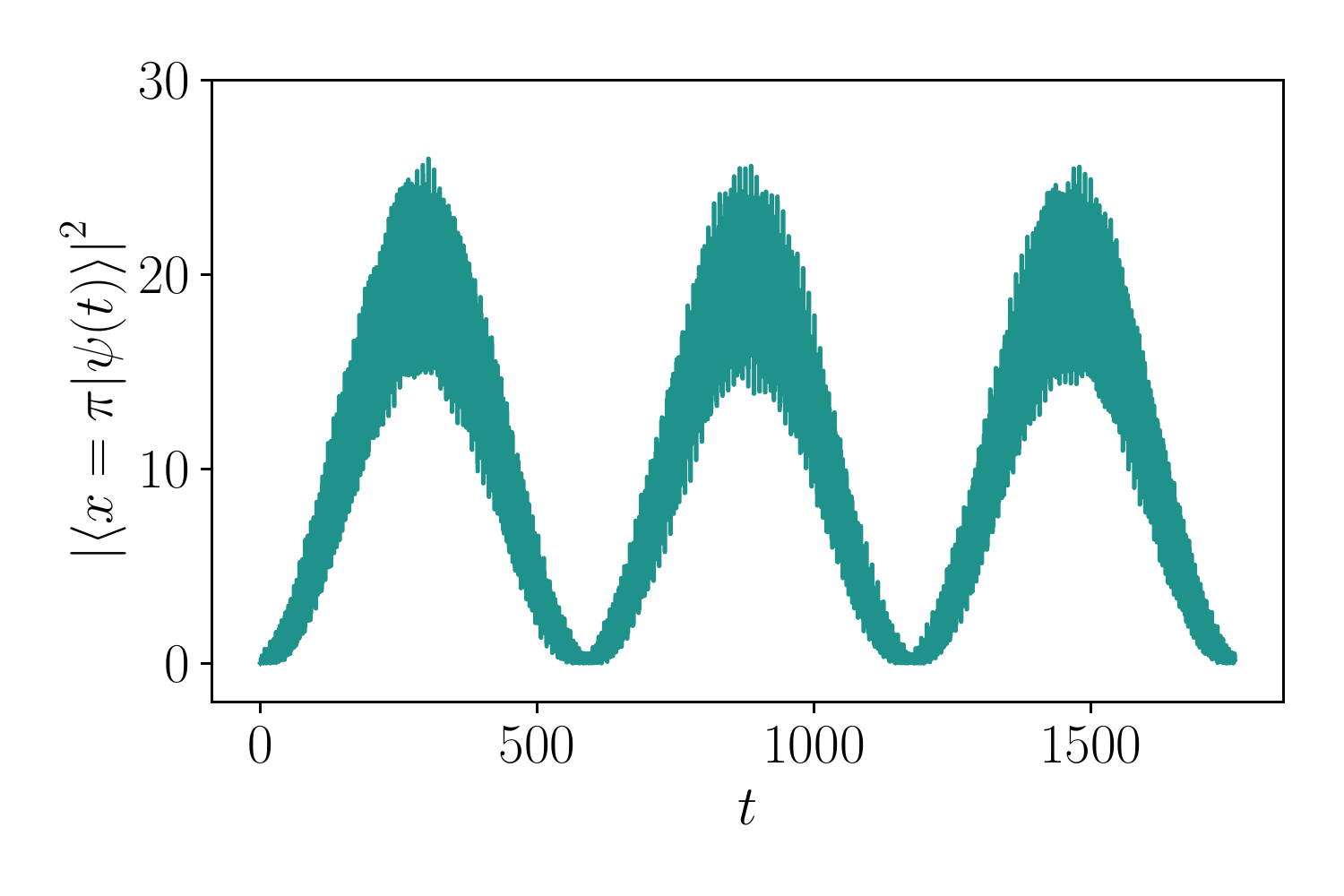}
		\caption{}
		\label{subfig:vary_gauss_width_3}
	\end{subfigure}
	\hfill
	\begin{subfigure}[ht]{0.48\textwidth}
		\centering
		\includegraphics[width=\textwidth]{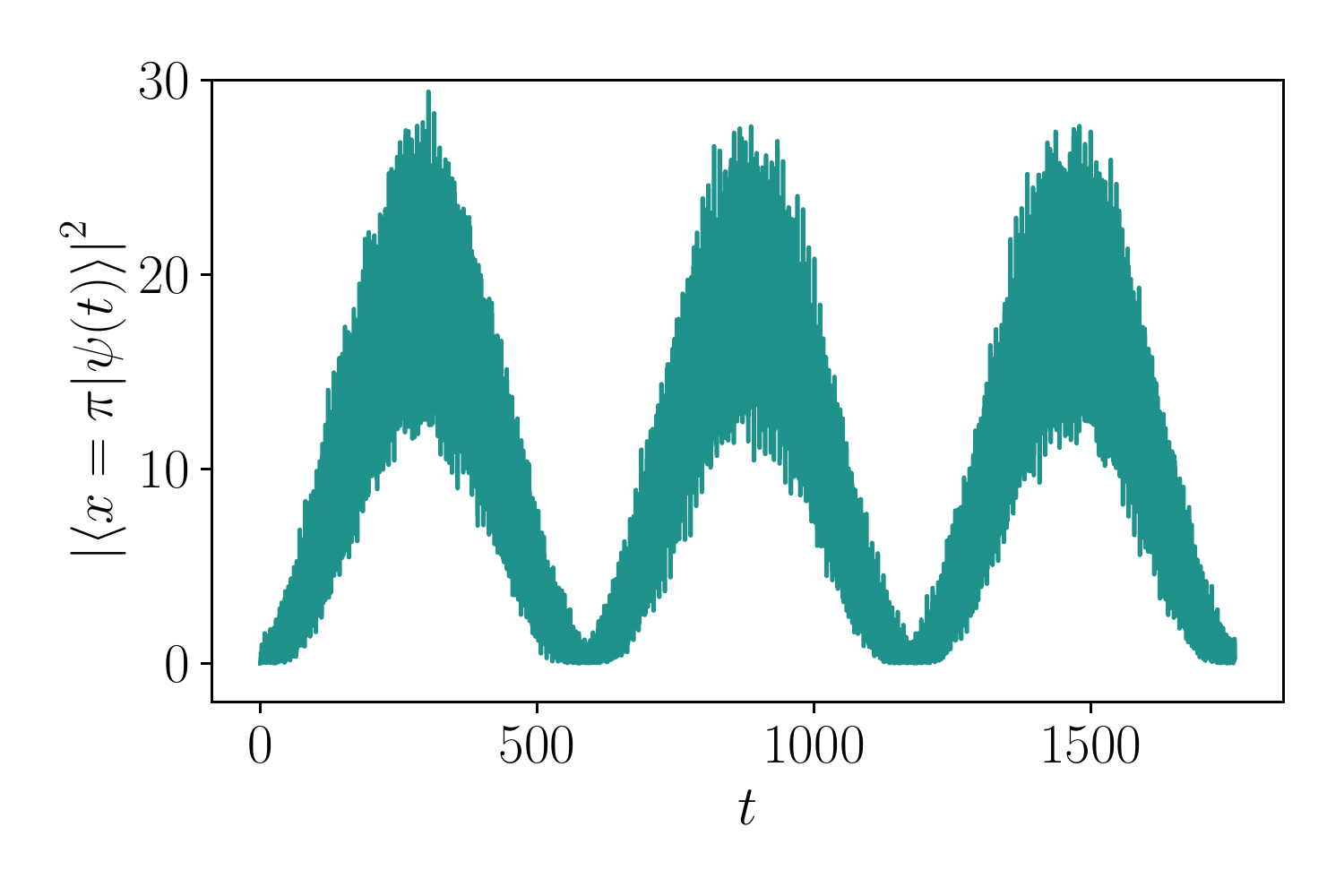}
		\caption{}
		\label{subfig:vary_gauss_width_4}
	\end{subfigure}
\caption{{Time evolution of the probability densities at $x = \pi$ from various  localized initial states centered at $x = 0$. In this example we take  $n=2$ wells, $n_s = 120$ sites, and a frequency near the semiclassical limit, $\omega = 4.0$. The initial states $\propto (1 + \cos(x)/2)^{2\alpha}$ take parameters (\subref{subfig:vary_gauss_width_1}) $\alpha = 2.0$, (\subref{subfig:vary_gauss_width_2}) $\alpha = 4.0$, (\subref{subfig:vary_gauss_width_3}) $\alpha = 6.0$, (\subref{subfig:vary_gauss_width_4}) $\alpha = 8.0$. The fuzziness of the density curves arises from overlap between the initial state and excited states, and is minimized near $\alpha = \omega$ where the width matches that of the perturbative Gaussian ground state in a single well.}}
\label{fig:vary_gauss_width}
\end{figure}

We numerically demonstrate that this claim is qualitatively correct. As shown in Fig.~\ref{fig:vary_gauss_width}, the fuzziness in the density as a function of real time diminishes near $\alpha \approx \omega$.

We observe clean slow profiles in Fig.~\ref{fig:vary_gauss_width} for a large range of the Gaussian width and also in Figs.~\ref{fig:comparison_exp_1},~\ref{fig:comparison_exp_ns_6}, but do not observe as clean profiles in Figs.~\ref{fig:comparison_exp_ns_8},~\ref{fig:comparison_exp_ns_12_n_2}. A few factors contributing can contribute to this. (1) The Kronecker delta distribution has sharp edges so can have significant overlap with more excited states than Gaussian initial states used in Fig.~\ref{fig:vary_gauss_width}. (2) With the greater numbers of sites in Figs.~\ref{fig:comparison_exp_ns_8},~\ref{fig:comparison_exp_ns_12_n_2} than in Figs.~\ref{fig:comparison_exp_1}
\ref{fig:comparison_exp_ns_6}, the slow profiles would have been still clean if $\omega$ (with $I$ being approximately fixed in the dimensionful form) were chosen greater because then the separation between the lowest $n$ states and higher states at order $\omega$ will be greater and reduce the effects from these higher states. However, greater $\omega$ would make the frequency of tunneling exponentially slower. With a finite window of the Rydberg decay time, to observe at least a significant fraction of a full tunneling period, $\omega$ cannot be chosen too large.

\end{appendices}

\renewcommand\refname{References}
\bibliographystyle{unsrt}
\bibliography{main}

\end{document}